%% file: Standard_Model.tex
\documentclass[11pt]{article}
\usepackage{amsmath,amssymb,amsthm}

\usepackage{hyperref}
\hypersetup{
    colorlinks=true,
    citecolor=blue,
    linkcolor=purple,
    filecolor=magenta,      
    urlcolor=cyan,
    breaklinks=true
    }
\usepackage[nameinlink]{cleveref}

\Crefname{fact}{Fact}{Facts}
\Crefname{claim}{Claim}{Claims}
\usepackage[margin=1in]{geometry}
\usepackage{thmtools} 
\usepackage{thm-restate}
\usepackage{math-thm}
\usepackage[dvipsnames]{xcolor}
\usepackage{algorithm, algorithmic}
\usepackage{enumitem}
\usepackage[inline]{asymptote}
\usepackage{caption}
\usepackage[title]{appendix}
\usepackage{quiver}
\usepackage[export]{adjustbox}
\usepackage{booktabs}
\usepackage{multirow}

\title{Boolean\vspace{0.05cm} function monotonicity testing\\ requires (almost) $n^{1/2}$ queries}

\date{\today}
\author{
\ifx\version\public
    Mark Chen
    \\
    yc3879@columbia.edu
    \\
    Columbia University
    \and
    Xi Chen
    \\
    xichen@cs.columbia.edu
    \\
    Columbia University
    \and
    Hao Cui
    \\
    hc3377@columbia.edu
    \\
    Columbia University
    \and
    William Pires
    \\
    wp2294@columbia.edu
    \\
    Columbia University
    \and
    Jonah Stockwell
    \\
    js5384@columbia.edu
    \\
    Columbia University
\fi
}

\newcommand{\Prob}{\Pr}

\newcommand{\rmO}{\mathsf{O}}
\newcommand{\brmO}{\boldsymbol{\mathsf{O}}}
\newcommand{\parent}{\mathrm{par}}
\def\bD{{\boldsymbol{D}}}
\def\bO{\brmO}
\newcommand{\eps}{\epsilon}

\def\bA{\boldsymbol{A}}
\def\bu{\boldsymbol{u}}
\def\brho{\boldsymbol{\rho}}

\def\bP{\boldsymbol{P}}
\def\bR{\boldsymbol{R}}
\def\ff{\boldsymbol{f}}
\def\bf{\boldsymbol{f}}

\def\bT{\boldsymbol{T}}
\def\bC{\boldsymbol{C}}
\def\bX{\boldsymbol{X}}

\def\Dyes{\mathcal{D}_{\text{yes}}}
\def\Dno{\mathcal{D}_{\text{no}}}

\def\bS{\boldsymbol{S}}
\def\bs{\boldsymbol{s}}

\def\bh{\boldsymbol{h}}
\def\bH{\boldsymbol{H}}
\def\dist{\mathsf{dist}}
\def\ALG{\textsf{ALG}}

\def\zo{\{0,1\}}
\def\bi{\boldsymbol{i}}
\def\bH{\boldsymbol{H}}
\def\bM{\boldsymbol{M}}

\def\Hyes{\mathcal{H}_{\text{yes}}}
\def\Hno{\mathcal{H}_{\text{no}}}
\def\calM{\mathcal{M}}

\def\bx{\boldsymbol{x}}

\def\Lyes{\mathcal{O}_{\text{yes}}}
\def\Lno{\mathcal{O}_{\text{no}}}
\def\calD{\mathcal{D}}

\def\version{public}
\def\submission{submission}
\def\public{public}

\begin{document}
\maketitle

\begin{abstract}
\ifx\version\submission
We show that for any constant $c>0$, any (two-sided error) adaptive algorithm for testing monotonicity of Boolean functions  must have query complexity $\Omega(n^{1/2-c})$. This improves the $\tilde\Omega(n^{1/3})$ lower bound  of \cite{chen2017beyond} and almost matches the $\tilde{O}(\sqrt{n})$ upper bound of \cite{khotminzerSafraOptUpperBound}.
\fi

\ifx\version\public
We show that for any constant $c>0$, any (two-sided error) adaptive algorithm for testing monotonicity of Boolean functions  must have query complexity $\Omega(n^{1/2-c})$. This improves the $\tilde\Omega(n^{1/3})$ lower bound  of \cite{chen2017beyond} and almost matches the $\tilde{O}(\sqrt{n})$ upper bound of \cite{khotminzerSafraOptUpperBound}.





\fi
\end{abstract}

\newpage
\hypersetup{linktoc=page}
{\small \tableofcontents}

\newpage

\section{Introduction}
The goal of research in property testing is to understand abilities and limitations of randomized algorithms that can determine whether an unknown ``massive object'' has a particular property or is far from having the property (see \cite{Goldreich_2017,bhattacharyya2022property} for overviews of contemporary property testing research).
A cornerstone problem in property testing of Boolean functions  has been that of monotonicity testing, i.e., to determine whether an unknown Boolean function $f: \zo^n \to \zo$  is monotone or $\eps$-far from monotone. Recall that $f$ is monotone if  $f(x) \leq f(y)$ for all $x \prec y$\footnote{We write $x\prec y$ to denote $x_i \leq y_i$ for all $i \in [n]=\{1,  \ldots, n\}$.}, and is $\epsilon$-far from
monotone if for every monotone function $g:\zo^n \to \zo$, the number of points $x\in \{0,1\}^n$ on which $f$ and $g$ disagree is at least  $\epsilon 2^n$. An $\eps$-tester for monotonicity is a randomized algorithm that can make membership queries to $f$, and should accept with probability at least $2/3$ when $f$ is monotone, and reject with probability at least $2/3$ when $f$ is $\eps$-far from monotone.   

For more than two decades, there has been a line of work that aims to pin down the number of membership queries needed for monotonicity testing \cite{goldreich1998testing,fischer2002monotonicity,ChakrabartySeshadhri,ChenServedioTan, ChenDeServedioTan,belovs2016polynomial, chen2017beyond, khotminzerSafraOptUpperBound,CS19}.
We review them later in \Cref{sec:previouswork}.
In summary, despite significant progress, there remains an intriguing gap between the best upper bound of $\tilde{O}(\sqrt{n})$ \cite{khotminzerSafraOptUpperBound}, which is achieved by a nonadaptive algorithm,  and the 
best lower bound of $\tilde{\Omega}(n^{1/3})$ \cite{chen2017beyond} for adaptive algorithms.
It also remains an open question whether adaptivity can help test monotonicity with query complexity below $\sqrt{n}$.

\subsection{Our Results}

In this paper, we close this gap by proving a  nearly tight lower bound for monotonicity testing:

\begin{theorem}\label{thm:2l-layer-2s-adaptive-lb}
For any constant $c>0$, there exists a constant $\eps_c$ such that any two-sided, adaptive algorithm for testing whether an unknown Boolean function $f:\{0,1\}^n\rightarrow \{0,1\}$ is monotone or $\eps_c$-far from monotone must make ${\Omega}(n^{0.5-c})$ queries.
\end{theorem}

Our lower bound  
proof of \Cref{thm:2l-layer-2s-adaptive-lb} builds on a new construction of Boolean functions called  \emph{multilevel Talagrand functions}, which we discuss in depth in \Cref{sec:overview}.
These functions allow~us to  
prove additional lower bounds for testing monotonicity of Boolean functions as well as its closely related problem of testing unateness\footnote{A function $f:\{0,1\}^n\rightarrow \{0,1\}$ is said to be unate iff there exists an $a\in \{0,1\}^n$ such that $f(x\oplus a)$ is monotone, where $\oplus$ denotes  the bitwise XOR.}.

First, we give a tight $\tilde{\Omega}(\sqrt{n})$ lower bound for the query complexity of any monotonicity testing algorithm that are only allowed a constant number of rounds of adaptivity\footnote{Introduced in \cite{Adaptivity}, an algorithm is \emph{$r$-round-adaptive} if 
  it makes $r+1$ batches of queries, where queries in the $i$-th batch can depend on results from the previous $i-1$ batches only.
Under this definition, a nonadaptive algorithm is $0$-round-adaptive. See \Cref{def:rounds} for the formal definition.}:

\begin{restatable}{theorem}{adaptivityhierarchy}
\label{thm:adaptivity-hierarchy}
For any constant $r\in \mathbb{N}$,
  there exists a constant $\eps_r$ such that any two-sided, {$r$-round}-adaptive algorithm for testing whether 
  an unknown Boolean function $f:\{0,1\}^n\rightarrow \{0,1\}$ is monotone or $\eps_r$-far from monotone must make $\tilde{\Omega}(\sqrt{n})$ queries. 
\end{restatable}
Finally, we work on the \emph{relative-error} testing framework recently proposed \cite{chen2025relative} to study the testability of sparse Boolean functions (see \Cref{sec:relativebackground} for the definition of the~model).~We show that relative-error   monotonicity testing (and unateness testing as well)   require $\Omega((\log N)^{1-c})$ queries for any constant $c>0$, where $N:=|f^{-1}(1)|$ denotes the sparsity of
  $f$. This nearly matches the upper bounds of 
  $\tilde{O}(\log N)$   for testing  both monotonicity \cite{chen2025relative} and unateness  \cite{chen2025relativeerrorunatenesstesting} in this model,
  improving the best known lower bounds of 
  $\tilde{\Omega}((\log N)^{2/3})$ \cite{chen2025relative,chen2025relativeerrorunatenesstesting}.

\begin{restatable}{theorem}{relativeerror}\label{thm:relativeerror}
    For any constants $c,\alpha >0$, there exists a constant $\eps_{c,\alpha}$ such that any two-sided, adaptive algorithm for testing whether an unknown Boolean function $f:\{0,1\}^n\rightarrow \{0,1\}$ satisfying $|f^{-1}(1)|=\Theta(N)$ for some given parameter $N \leq 2^{\alpha n}$ is monotone (unate) or $\eps_{c,\alpha}$-far from monotone (unate)
    in relative distance must make $\tilde{\Omega}\left((\log N\right)^{1-c})$ queries.   
\end{restatable}

\subsection{Technical Overview}\label{sec:overview}

We give a high-level overview of our proofs of \Cref{thm:2l-layer-2s-adaptive-lb} and \Cref{thm:adaptivity-hierarchy}.
Following Yao’s minimax principle, our goal
 is to build a pair of distributions 
 $\Dyes$ and $\Dno$ over Boolean functions $f:\{0,1\}^n$ $\rightarrow \{0,1\}$ such that (a) $\bf\sim \Dyes$ is always monotone; (b) $\bf\sim \Dno$ is $\Omega(1)$-far from monotone with probability $\Omega(1)$; and (c) no deterministic algorithm $\ALG$ with $\ll \sqrt{n}$ queries can distinguish $\Dyes$ from $\Dno$, which means that 
$$
\Pr_{\bf\sim \Dyes}\big[\text{$\ALG$ accepts } \bf\big]\le \Pr_{\bf\sim \Dno}\big[\text{$\ALG$ accepts } \bf\big]+o_n(1).
$$
As mentioned earlier, our construction of $\Dyes$ and $\Dno$ is based on multilevel Talagrand functions. To introduce them properly, we start by reviewing constructions of \cite{belovs2016polynomial} and 
\cite{chen2017beyond}.

\def\bj{\boldsymbol{j}}

\subsubsection{\cite{belovs2016polynomial}: Talagrand Functions}

The first polynomial query lower bound for adaptive monotonicity testing algorithms was obtained by \cite{belovs2016polynomial}.
Their construction of $\Dyes,\Dno$ modifies the \emph{Talagrand functions} (or Talagrand random DNFs) \cite{Talagrand}.\footnote{The distributions sketched here are slightly different from those actually used in \cite{belovs2016polynomial}. These modifications are made to align them more closely with the construction of \cite{chen2017beyond} and our new multilevel construction.}
Let $\smash{N:=2^{\sqrt{n}}}$.\footnote{The technical overview will focus on \Cref{thm:2l-layer-2s-adaptive-lb,thm:adaptivity-hierarchy} under the standard testing model. We always use $N$ to denote $2^{\sqrt{n}}$. Later in \Cref{sec:relativeerrorsec} we use $N$ to denote $|f^{-1}(1)|$ when we work on the relative-error model there.}
To draw a function $\bf\sim \Dno$, one first draws   $N$ (positive) size-$\sqrt{n}$ terms $\bT_1,\ldots, \bT_N$, where each term $\bT :\{0,1\}^n\rightarrow \{0,1\}$ is of the form 
$$
\bT (x)=x_{\bj_1}\land\cdots \land x_{\bj_{\sqrt{n}}} 
$$
with each variable drawn independently and uniformly at random 
  from $[n]$.
Together, they ``partition'' middle layers\footnote{We say $x\in \{0,1\}^n$ is in middle layers if it satisfies $(n/2)-\sqrt{n}\le |x|\le (n/2)+\sqrt{n}$.
 Throughout the overview the reader should only consider points in middle layers; all lower bound constructions, including those of \cite{belovs2016polynomial} and \cite{chen2017beyond}, apply a standard truncation so that an algorithm would never query any point  outside of middle layers.} of $\{0,1\}^n$ into $H_1,\ldots,H_N$, where $H_i$ contains every $x\in \{0,1\}^n$ that satisfies $\bT_i$ but not any other term (which we will refer to as $x$ \emph{uniquely} satisfying $\bT_i$).
Note that, formally speaking, this is not a partition because there are points that do not satisfy any terms or satisfy~at least two terms.
By standard calculations, most likely $H_1,\ldots,H_N$ together cover  $\Omega(1)$-fraction of points in middle layers.
For convenience, we refer to each $H_i$ as a 
  \emph{subcube} in the \emph{partition};  formally, they are not due to the removal of
  overlaps and the restriction to middle layers.

Next, we draw a random anti-dictatorship function
  $\bh_i:\{0,1\}^n\rightarrow \{0,1\}$ for each subcube $H_i$,  by drawing a random secret variable $\bs_i\sim [n]$ independently and setting $\bh_i(x)=\overline{x_{\bs_i}}$.
Finally, the function $\bf(x)$ is set to be $\bh_i(x)$ if $x\in H_i$; $0$ if $x$ does not satisfy any terms;
or $1$ if $x$ satisfies at least two terms.
Given that $H_1,\ldots,H_N$ together cover $\Omega(1)$ fraction of middle layers (and the folklore that the middle layers consist of $\Omega(1)$-fraction of 
  the $2^n$ points in $\{0,1\}^n$), one can show that the anti-dictatorship function $\bh_i$'s will lead to many violations to monotonicity in each subcube $H_i$ and thus, $\bf\sim \Dno$ is $\Omega(1)$-far from monotone with probability at least $\Omega(1)$.

On the other hand, to draw a function $\bf\sim \Dyes$, the only difference 
  is that each $\bh_i$ is~a~random dictatorship function: $\bh_i(x)=x_{\bs_i}$ with each secret variable $\bs_i\sim [n]$ uniformly and independently.
It can be shown that $\bf\sim \Dyes$ is always monotone. (This uses the observation that the number of terms satisfied by an $x$ is monotonically non-decreasing as bits of $x$ are flipped from $0$'s to $1$'s.)

Given that the only difference between $\Dyes$ and $\Dno$ lies in the dictatorship vs anti-dictatorship functions $\bh_i$, it is not surprising that a deterministic algorithm $\ALG$ can only tell them apart by \emph{flipping the secret variable $\bs_i$ of some subcube $H_i$:}
We will repeatedly use this phrase to mean that 
  $\ALG$ queried two points $x,y$ in middle layers such that $x,y\in H_i$ for some $i\in [N]$ and $x_{\bs_i}\ne y_{\bs_i}$. 
  
To see why achieving this requires many queries, 
  consider the scenario where $\ALG$ just queried 
  a point $x$ satisfying $x\in H_i$ for some $i\in [N]$.
Next, $\ALG$ hopes to query $y$, by flipping variables in $x$, such that $y\in H_i$ and 
  $y_{\bs_i}\ne x_{\bs_i}$ with a good probability.
Given that $\bs_i$ is distributed uniformly, naturally
  $\ALG$ would like to flip as many variables of $x$
  as possible.
However, $\ALG$ cannot flip more than $O(\sqrt{n}\log n)$ 
  variables of $x$ from $1$'s to $0$' because doing so
  would move the point outside of $ H_i$ with high probability.
(Recall that $\bT_i$ is a random term of size $\sqrt{n}$; if all we know about it is that $\bT_i(x)=1$, flipping more than $O(\sqrt{n}\log n)$ many $1$'s to $0$'s would falsify $\bT_i$ with high probability.)
On the other hand, while $\bT_i$ does not post any
  constraint on how many $0$'s can be flipped to $1$'s,
we cannot flip more than $O(\sqrt{n}\log n)$ because $y$ needs to remain in middle layers.

This is the high-level intuition behind the  $\tilde{\Omega}(n^{1/4})$ lower bound of  \cite{belovs2016polynomial}.
Given the discussion above, it is natural to wonder whether the construction can lead to a tight $\tilde{\Omega}(\sqrt{n})$ lower bound:
It seems that the set of variables flipped in a subcube $H_i$, which we will refer to as the \emph{dangerous set} of $H_i$\footnote{Formally, a variable $i \in [n]$ is in the dangerous set of $H_i$ if $\ALG$ queried two points $x,y\in H_i$ with $x_i \neq y_i$.}, grows only by $O(\sqrt{n}\log n)$ for each additional query that lands in $H_i$. If this is indeed the case, then $\tilde{\Omega}(\sqrt{n})$ queries are needed for its size to grow to $\Omega(n)$ and only by then $\ALG$ has a good chance of flipping the secret variable $\bs_i\sim [n]$.


However, as pointed out in  \cite{chen2017beyond}, there is   a more efficient way to grow the dangerous set of $H_i$, quadratically (rather than linearly) in the number of queries, which then leads to an $\tilde{O}(n^{1/4})$-query algorithm to distinguish the two distributions. 
We briefly review this strategy, which we~will refer to as the \emph{quadratic-speedup strategy}. Looking ahead, both the two-level construction of \cite{chen2017beyond} and our new construction are designed to mitigate the quadratic-speedup strategy.

For ease of exposition, we assume that the algorithm has access to the following stronger oracle: upon a query $x\in \{0,1\}^n$, the oracle returns not only $f(x)$ but also the (unique) index $i\in [N]$ such that $x\in H_i$, or ``none'' if it does not satisfy any terms, or ``at least two'' if it satisfies at least two terms
(indeed, all our lower bounds are established against such an oracle; see \Cref{sec:information-maintained}).\medskip

\def\by{\boldsymbol{y}}

\noindent\textbf{The quadratic speedup strategy:}
 The algorithm starts with a point $x\in H_i$ for some $i\in [N]$.
 (Given that $H_1,\ldots,H_N$ consist of $\Omega(1)$-fraction of middle points, this occurs for a random $x$ with probability $\Omega(1)$.)
It makes $n^{1/4}$ queries to find $n^{3/4}$ variables $\bS\subseteq [n]$ that do not appear in $\bT_i$:
\begin{flushleft}\begin{quote}
Set $\bS=\emptyset$ and repeat the following $n^{1/4}$ times: Flip $\sqrt{n}$ many random $1$'s in $x$ to $0$'s to obtain $\by$; query $\by$; add the variables flipped to $\bS$ if $\by$ remains in $H_i$.
\end{quote}\end{flushleft}
Because $\bT_i$ is of size $\sqrt{n}$, a constant 
  fraction of $\by$'s stay in $H_i$ and for each such $\by$, the $\sqrt{n}$ variables flipped do not appear in $\bT_i$, leading to an $\bS$ of $\Omega(n^{3/4})$ variables that do not appear in $\bT_i$.
\def\bz{\boldsymbol{z}}

After this preprocessing step, the algorithm can grow the dangerous set much more efficiently. 
In each round, it can (1) flip $n^{3/4}$ many random $0$'s of $x$ to $1$'s and (2) to move it back into middle layers, flip variables in $\bS$ from $1$'s to $0$'s to obtain $\bz$ from $x$.
After querying $\bz$, if $\bz\in H_i$ (which can be shown to happen with $\Omega(1)$ probability), the dangerous set grows by $n^{3/4}$ because of (1).
Now to summarize, if the algorithm is allowed $q$ queries, then it spends $q/2$ queries during preprocessing to build $\bS$ of size $\Omega(q\sqrt{n}))$. After another $q/2$ queries, it can grow a dangerous set of size $\Omega(q^2\sqrt{n})$.

\subsubsection{\cite{chen2017beyond}: Two-level Talagrand Functions}\label{sec:two-level-Talagrand}

To obtain a $\tilde{\Omega}(n^{1/3})$ lower bound, 
  \cite{chen2017beyond} extended the Talagrand construction 
  of \cite{belovs2016polynomial}~into a \emph{two-level construction.}
To draw $\bf\sim \Dyes$, one first draws  $N$  \mbox{size-$\sqrt{n}$} {random} terms $\bT_1,\ldots,\bT_N$ and then for each $i\in [N]$, draws $N$ size-$\sqrt{n}$ random clauses $\bC_{i,1},\ldots,$ $\bC_{i,N}$, each $\bC$ of the form
$$
\bC(x)=x_{\bj_1}\lor \cdots \lor x_{\bj_{\sqrt{n}}},
$$
where $\bj_1,\ldots,\bj_{\sqrt{n}}$ are variables picked independently and uniformly at random from $[n]$.
Similarly they together partition the middle layers  into $H_{i,j}$'s, $i,j\in [N]$, where $x\in H_{i,j}$ if it uniquely satisfies $\bT_i$ (among $\bT_1,\ldots,\bT_N$) and then uniquely falsifies $\bC_{i,j}$ (among $\bC_{i,1},\ldots,\bC_{i,N}$).
It can be shown that $H_{i,j}$'s together again cover an $\Omega(1)$-fraction of middle layers.
To finish the construction, we draw a random dictatorship function $\bh_{i,j}$ for each $i,j\in [N]$ by setting $h_{i,j}(x)=x_{\bs_{i,j}}$ for a uniformly and independently chosen secret variable $\bs_{i,j}\sim [n]$. 
Finally, $\bf(x)$ is set to be $\bh_{i,j}(x)$ if $x\in H_{i,j}$, and we skip details about how to set $\bf(x)$ when $x$ does not belong to any $H_{i,j}$; this needs to be done properly to make sure that $\bf\sim \Dyes$ is always monotone (see \Cref{subsec:Talagrand}).

\begin{figure}[t!]
  \centering
  \[\begin{tikzcd}
	&&& f \\
	& \bullet && \cdots && \bullet \\
	\bullet & \cdots & \bullet && \bullet & \cdots & \bullet \\
	{h_{1,1}} & \cdots & {h_{1,N}} && {h_{N,1}} & \cdots & {h_{N,N}}
	\arrow["{T_1}"{description}, from=1-4, to=2-2]
	\arrow["{T_{N}}"{description}, from=1-4, to=2-6]
	\arrow["{C_{1,1}}"{description}, from=2-2, to=3-1]
	\arrow["{C_{1,N}}"{description}, from=2-2, to=3-3]
	\arrow["{C_{N,1}}"{description}, from=2-6, to=3-5]
	\arrow["{C_{N,N}}"{description}, from=2-6, to=3-7]
	\arrow[dashed, from=3-1, to=4-1]
	\arrow[dashed, from=3-3, to=4-3]
	\arrow[dashed, from=3-5, to=4-5]
	\arrow[dashed, from=3-7, to=4-7]
\end{tikzcd}\]
\caption{A picture of the two-level Talagrand construction from \cite{chen2017beyond}.}
  \label{fig:twolayerconstruction}
\end{figure}
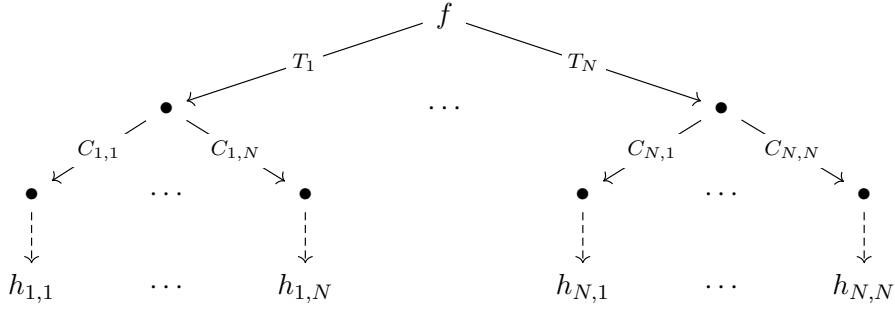

To draw $\bf\sim \Dno$, the only difference is that each $\bh_{i,j}$ is set to be the anti-dictatorship function with the secret variable $\bs_{i,j}$. Similarly one can show that $\bf\sim \Dno$ is $\Omega(1)$-far from monotone  $\Omega(1)$.
See \Cref{fig:twolayerconstruction} for an illustration of the two-level construction.

Intuitively the two-level construction  \cite{chen2017beyond} seeks to mitigate the quadratic-speedup strategy by slowing down the growth of the dangerous set of a given subcube $H_{i,j}$. Given any $x\in H_{i,j}$, on the one hand, flipping more than $O(\sqrt{n}\log n)$ many $1$'s of $x$ to $0$'s would most likely falsifies  $\bT_i$ in the first level; on the other hand, flipping more than $O(\sqrt{n}\log n)$ many $0$'s of $x$ to $1$'s would most likely satisfies $\bC_{i,j}$ in the second level, moving the point outside of $H_{i,j}$ in both cases. 

Using the two-level construction, \cite{chen2017beyond} obtained an $\tilde{\Omega}(n^{1/3})$ lower bound for monotonicity testing.
Their analysis of the two-level construction also turned out to be tight.
Indeed, they showed  how to apply    the quadratic-speedup strategy in a slightly more sophisticated way to distinguish $\Dyes$ from $\Dno$ using $\tilde{O}(n^{1/3})$ queries.
To this end, 
  the algorithm first spends
  $n^{1/3}$ queries on some term $\bT_i$ to build a set $\bS$ of $n^{5/6}=n^{1/3}\cdot \sqrt{n}$ variables that do not appear in $\bT_i$; this can be done in a way similar to the preprocessing step described early in the quadratic-speedup strategy.~Instead of focusing on a single subcube $H_{i,j}$ below $\bT_i$,
the algorithm reuses $\bS$ and applies the quadratic-speedup strategy to 
  grow the dangerous sets of $n^{1/6}$ subcubes  $H_{i,j}$, each up to size
  $n^{5/6}$, using $n^{1/6}$ queries on each subcube.
Basically, the algorithm spends $n^{1/6}$ queries on each subcube  to flip almost all variables in
  $\bS$ from $1$'s to $0$'s.
The quadratic-speedup strategy makes this possible because (1) roughly speaking, $(n^{1/6})^2\cdot \sqrt{n}=n^{5/6}$ and (2) flipping variables in $\bS$ from $1$'s from $0$'s never falsifies $\bT_i$ given that $\bS$ was built to avoid $\bT_i$.
 Given that the secret variable $\bs_{i,j}$ for each subcube $H_{i,j}$ is drawn independently and that the algorithm flipped $\Omega(n)$ variables in the $n^{1/6}$ subcubes altogether, it is likely that the secret variable of one of these subcubes was flipped during the process.

Note that, even though the quadratic-speedup strategy can still be applied to~attack each $H_{i,j}$, its impact is mitigated because the construction of $\bS$ remains linear.
This is the high-level intuition why a better~lower bound can be obtained using the two-level construction.
Given this thought,
it is only natural to  conjecture that once more levels (or equivalently, more alternations between terms and clauses) are added, any algorithm that distinguishes $\Dyes$ from $\Dno$ may have to penetrate the construction level by level and demand more queries overall. 
However, as pointed out in \cite{chen2017beyond}, this was not the case (we spell out more details for why in \Cref{ssec:algo-can-skip-level-cwx17}).

At a high level, one can similarly define $\Dyes,\Dno$ by drawing terms $\bT_i$, clauses $\bC_{i,j}$, and then terms $\bT_{i,j,k}$ again, with $i,j,k\in [N]$, and then plug in either random dictatorship functions or anti-dictatorship functions $\bh_{i,j,k}$ drawn independently for each subcube $H_{i,j,k}$.
The same properties (a) and (b) still hold for $\Dyes$ and $\Dno$. 
But, despite the three levels in the construction, an algorithm  can cheat by jumping directly onto a term $\bT_i$ and work on the two-level construction rooted at $\bT_i$.

\subsubsection{Multilevel Talagrand Functions}

The key observation we make in this paper is the following:
Even though the three-level Talagrand construction described above can be defeated by only $\tilde{O}(n^{1/3})$ queries, this is achieved by making the \emph{sum} of sizes of dangerous sets of the subcubes $H_{i,j,k}$ to be $\Omega(n)$;
their \emph{union}  is much smaller.

The intuition was already hinted in the sketch of the algorithm for the two-level construction.\footnote{For readers who are familiar with \cite{chen2017beyond}, the sketch of the algorithm for the two-level construction given here is slightly different, modified to highlight the idea to be discussed next in this paragraph. In hindsight, this issue of \emph{sum} vs \emph{union} is very subtle because it only shows up when one extends the algorithm~of \cite{chen2017beyond} to the three-level Talagrand construction.
For the two-level construction, the  original algorithm given in \cite{chen2017beyond} actually does make 
  the union of dangerous sets of size $\Omega(n)$.}
To grow the dangerous sets of $H_{i,j}$'s fast, one first secures a large set $\bS$ of variables that do not appear in $\bT_i$; this  $\bS$ is also the same set of variables on which the algorithm applies the quadratic-speedup strategy to flip quickly in $n^{1/6}$ subcubes $H_{i,j}$ below $\bT_i$.
As a result, when taking the union of these $n^{1/6}$ dangerous sets, the majority of variables in it come from $\bS$, which
  is of size roughly $n^{5/6}$ instead of $n$ because $\bS$ only grows linearly.

Inspired by this, we now give our construction of
  multilevel Talagrand functions (see \Cref{sec:MTF} for the formal description). 
To draw a $(2\ell)$-level Talagrand function $\bf\sim\Dyes$, one first builds a complete $2\ell$-level tree of arity $N$, in which every edge $e$ is labeled either a random $\sqrt{n}$-size term $\bT_e$ or a random $\sqrt{n}$-size clause $\bC_e$, depending on the parity of its level. (Edges of the root are labeled terms and then they alternate as we go down the tree.) In a similar fashion, this tree partitions middle layers into ``subcubes'' $H_u$, one for each leaf $u$ of the tree: For an $x$ to be added to $H_u$, it needs to uniquely satisfy a term $\bT_e$ among all those incident to the root to move down along $e$ from the root to a level-$1$ node $u_1$, and then uniquely falsify a clause $\bC_{e'}$ among all those incident to $u_1$ to move down from $u_1$ along $e'$ to a level-$2$ node $u_2$, and repeats this for the $2\ell$ total levels to finally reach the leaf $u$.
This path of $x$ is what we refer to later as the \emph{unique activation path} of $x$; we use the word activation because terms need to be satisfied while clauses need to be falsified. To finish the construction,
for each leaf $u$, the function $\bh_u:\{0,1\}^n\rightarrow \{0,1\}$ is set to be either the constant-$0$ or the constant-$1$ function with probability $1/2$.
The final function $\bf$ sets $\bf(x)$ according to $\bh_u(x)$ if $x\in H_u$ and carefully sets $\bf(x)$ when $x\notin H_u$ for any leaf $u$ so that $\bf\in \Dyes$ is always monotone.

Note that, other than using constant functions in $\bh_u$ instead of random dictatorship functions (which is mainly for the ease of proof), the construction of $\Dyes$ is a fairly standard generalization of the two-level construction of \cite{chen2017beyond}.
The key difference lies in the construction of $\Dno$.

To draw $\bf\sim \Dno$, one first draws the same tree of random terms and clauses as in $\Dyes$ and define in the same way the subcubes $H_{u}$ for each leaf $u$. To draw the $\bh_u$ functions, we first draw a \emph{global} secret variable $\bs\sim [n]$ uniformly and then set each $\bh_u$ independently to be the dictatorship function $x_{\bs}$ or the anti-dictatorship function $\overline{x_{\bs}}$ with probability $1/2$. Similarly, one can show that $\bf\sim \Dno$ is $\Omega(1)$-far from monotone with probability at least $\Omega(1)$.\footnote{As the reader may expect, both constants hidden in $\Omega(1)$ go down exponentially as $\ell$ goes up; this is the reason why our lower bound in \Cref{thm:2l-layer-2s-adaptive-lb} needs the constant $c>0$. See more discussion on this in \Cref{sec:conclusion}.}

Our construction raises new obstacles for algorithms that aim to distinguish $\Dyes$ from $\Dno$. It is no longer sufficient for $\ALG$ to spend queries to build dangerous sets at leaves that have total sizes summing to $\Omega(n)$. Indeed, their union {now} has to have size $\Omega(n)$. Intuitively, if the union of dangerous sets is only $o(n)$, then with probability $1-o_n(1)$, the secret variable $\bs$ is never flipped and thus, information theoretically $\ALG$ cannot tell whether each $\bh_u$ is a constant function or a dictatorship/anti-dictatorship function about $\overline{x_{\bs}}$.
With this in mind, we say the ``knowledge'' of $\ALG$ is \emph{safe} if (roughly speaking) the union of  dangerous sets of all leaves is of size $o(n)$.
(Formally, the ``knowledge'' of an algorithm is what we define as an \emph{outcome} in \Cref{sec:information-maintained} given that, as mentioned earlier, our lower bounds are established against a stronger oracle that returns more information than $f(x)$ for a query $x$, such as its unique activation path; the definition of safe outcomes is in \Cref{sec:safeoutcomes}.)
\Cref{lem:hardtotell}~in \Cref{sec:safeoutcomes} formally shows that if the current knowledge of $\ALG$ is safe, then the underlying function is (almost) equally likely to be drawn from $\Dyes$ or $\Dno$.
So all that is left is to show that, as one runs $\ALG$ on $\bf\sim \Dyes$, the final knowledge of $\ALG$ is safe with probability at least $1-o_n(1)$.
This is proved using different strategies for \Cref{thm:2l-layer-2s-adaptive-lb} and \Cref{thm:adaptivity-hierarchy}, which we sketch below.



\subsubsection{Proof Overview  
  of \Cref{thm:2l-layer-2s-adaptive-lb}}

To prove \Cref{thm:2l-layer-2s-adaptive-lb}, we show that for every deterministic, adaptive  $\ALG$ with $q=\tilde{O}(n^{0.5-  {1}/({4\ell + 2})})$ queries, its final knowledge when running on $\bf\sim\Dyes$ is safe with high probability. To this end, we define for each node {in the tree underlying the $(2\ell)$-level Talagrand function} (not necessarily a leaf) $u$ a set $P_u$ as the set of points queried so far whose unique activation path  contains $u$. This set $P_u$ can be intuitively considered as the set of queries made by $\ALG$ to attack the term or clause labeled on the edge above $u$. (As mentioned earlier,~$\ALG$ knows the unique activation path of every query made so far since this is revealed by the stronger oracle after each query.) Given $P_u$, we define $A_{u,b}$, for each $b\in \{0,1\}$, to be
  the set of variables $i\in [n]$ that all points in $P_u$ set to be $b$.
In particular, the dangerous set of a leaf node $u$ is just the complement of $A_{u,0}\cup A_{u,1}$.

Given the definition of $P_u$'s above using unique activation paths, they naturally have the following nested structure: $P_u\subseteq P_v$ if $v$ is an ancestor of $u$. This structure  carries over to sets $A_{u,b}$ as well: $A_{v,b}\subseteq A_{u,b}$ if $v$ is an ancestor of $u$.
Looking ahead, this nested structure will play a crucial role in our lower bound proofs. 

The proof that the union of dangerous sets of all leaves is of size $o(n)$
  consists of two parts:
\begin{flushleft}\begin{enumerate}
\item First we show in \Cref{lem:main2} that with high probability (as running
  $\ALG$ on $\bf\sim \Dyes$), $A_{u,1}$ is of size at least $(n/2)-|P_u|\cdot O(\sqrt{n}\log n)$ if the edge $e$ above $u$ is labeled with a term $\bT_e$ or $A_{u,0}$ is of size at least $(n/2)-|P_u|\cdot O(\sqrt{n}\log n)$ if $e$ is labeled with a clause $\bC_e$. (This condition is referred to in \Cref{lem:main2} as so-called \emph{good} outcomes.) \Cref{lem:main2} should not come as a surprise because, following earlier discussions, when $e$ has a term (or clause), it is unlikely for $\ALG$ to make a new query to satisfy
  it (or falsify it) and at the same time flip more than $O(\sqrt{n}\log n)$ bits from $1$ to $0$ (or from $0$ to $1$).
\item The more challenging part is \Cref{lem:main1}: Assuming the condition above holds  for all $A_{u,0}$ and $A_{u,1}$ (i.e. that the outcome is good), one can show that it is always the case that the union of dangerous sets of all leaves is of size only $o(n)$.
\end{enumerate}\end{flushleft}
The proof of \Cref{lem:main1} is based on a sequence of inequalities that $A_{u,0}$ and $A_{u,1}$ satisfy, taking advantage of the nested structure. These inequalities can be viewed as obstacles each level of terms or clauses in the construction sets against $\ALG$ when it tries to penetrate down the tree. Once these inequalities are in place, the proof of \Cref{lem:main1} finishes with an induction, using these inequalities, to upperbound the union of dangerous sets by $o(n)$ in size.

\subsubsection{Proof Overview of \Cref{thm:adaptivity-hierarchy}}

To prove \Cref{thm:adaptivity-hierarchy},
  we show that for any $r$-round-adaptive algorithm $\ALG$ with $r=2\ell-1$ and $q=\tilde{O}(\sqrt{n})$ queries, its final knowledge when running on $\bf\in \Dyes$ is safe with high probability.~To this end we define similarly a dangerous set for each node (that is not necessarily a leaf) and prove \Cref{lem:pppp}:
  After the algorithm has made its $t$-th batch of queries, the union of the dangerous sets of nodes at level $t$ contains (with high probability) only $o(n)$ coordinates which were not already in a dangerous set of some node at level $t-1$ before querying this batch.
Given that initially the dangerous set at the root is $\emptyset$, what we need follows by repeatedly applying this lemma $r+1$ times.





\subsection{Previous Work}\label{sec:previouswork}

We review previous work on Boolean function monotonicity testing.

The work of Goldreich, Goldwasser, Lehman and  Ron \cite{goldreich1998testing} initiated the study of monotonicity testing. 
They showed that the nonadaptive ``edge tester''
  can achieve an upper bound of $O(n/\epsilon)$. 
Later Fischer, Lehman, Newman, Raskhodnikova, Rubinfeld and
 Samorodnitsky \cite{fischer2002monotonicity} obtained the first lower bounds for monotonicity testing, showing that 
  $\Omega(\sqrt{n})$ queries are needed for any nonadaptive, one-sided error algorithm, and $\Omega(\log n)$ queries are needed for any nonadaptive, two-sided error algorithm for monotonicity testing.

More than a decade later, Chakrabarty
and Seshadhri \cite{ChakrabartySeshadhri} improved the linear upper bound of \cite{goldreich1998testing} by giving a nonadaptive ``pair tester'' that uses $\tilde{O}(n^{7/8}/\epsilon^2)$ queries. 
Chen, Servedio~and Tan \cite{ChenServedioTan} improved their analysis to obtain an $\tilde{O}(n^{5/6}/\epsilon^4)$ upper bound. Finally, Khot, Minzer and Safra \cite{khotminzerSafraOptUpperBound} proved a directed version of Talagrand’s isoperimetric inequality and used it to give a tight analysis of the pair tester with query complexity $\tilde{O}(\sqrt{n}/\epsilon^2)$,
  which remains the best upper bound for monotonicity testing to date.

Turning to lower bounds (and assuming $\eps$ is a constant), \cite{ChenServedioTan} showed that  $\tilde{\Omega}(n^{1/5})$ queries are needed for any two-sided error, nonadaptive algorithm. This was later improved by   Chen,~De, Servedio and Tan \cite{ChenDeServedioTan}, giving an almost tight 
lower bound of $\Omega(n^{1/2-c})$ for two-sided error, nonadaptive algorithms for any constant $c>0$.
For adaptive algorithms,  Belovs and Blais \cite{belovs2016polynomial} were the first to obtain a polynomial  lower bound. They showed that any two-sided error, adaptive algorithm
needs $\tilde\Omega(n^{1/4})$ queries.
After \cite{belovs2016polynomial}, Chen, Waingarten and Xie \cite{chen2017beyond} improved the adaptive lower bound to $ \tilde{\Omega}(n^{1/3})$; they also removed the constant $c$ in the nonadaptive lower bound of\cite{ChenDeServedioTan}. 
In summary, for nonadaptive algorithm, the query complexity of monotonicity testing is pinned down at $\tilde{\Theta}(\sqrt{n})$ \cite{khotminzerSafraOptUpperBound,chen2017beyond};
for adaptive algorithms, before this work, there remained a gap between the best  bounds $\tilde{O}(\sqrt{n})$ \cite{khotminzerSafraOptUpperBound} and  $\tilde{\Omega}(n^{1/3})$ \cite{chen2017beyond}.

While we are only interested in Boolean functions $f:\zo^n \to \zo$, there has also been an extensive line of work studying monotonicity on the hypergrid and real-valued functions \cite{fischer2002monotonicity, halevy2007distribution, ailon2006information, halevy2008testing, saks2008parallel, fattal2010approximating, bhattacharyya2012lower, chakrabarty2013optimal, ChakrabartySeshadhri, berman2014lp, blais2014lower, black2018d, black2020domain, black2024isoperimetric, harms2020downsampling, braverman2022improved, black2023directed, black2025monotonicity}.
\medskip


\noindent \textbf{Organization.} After preliminaries in \Cref{sec:preliminaries}, we give the formal definition 
  of multilevel Talagrand functions in \Cref{sec:MTF}
  and describe the stronger oracle that we will work with
  in the rest of the paper.
We also introduce the notion of outcomes as a concise way
  of encoding all information an algorithm receives from
  the oracle after a number of queries are made.
We prove \Cref{thm:2l-layer-2s-adaptive-lb} in \Cref{sec:mainlowerbound1}, \Cref{thm:adaptivity-hierarchy} in \Cref{sec:mainlowerbound2} and \Cref{thm:relativeerror} in \Cref{sec:relativeerrorsec}.
We discuss the tightness of our lower bound given in  \Cref{thm:2l-layer-2s-adaptive-lb,thm:adaptivity-hierarchy} in \Cref{sec:upper-bound},
  and conclude in \Cref{sec:conclusion}.

\section{Preliminaries}\label{sec:preliminaries}

 We use bold font letters such as ${\bT}$ and ${\bC}$ for random variables and use calligraphic letters such as $\calD$ and $\calM$ for probability distributions.
Given a finite set $A$, we write $\bx\sim A$ to denote that $\bx$ is an element of $A$ drawn uniformly at random. 
 
 We write $[n]$ to denote $\{1,\ldots , n\}$, and $[i:j]$ to denote integers between $i$ and $j$, inclusive. 
 For a string $x \in \zo^n$, we write $|x|$ to denote its Hamming weight (the number of $1$’s). 
 Given a tuple $u=(u_1,\ldots,u_k)\in \mathbb{Z}^k$ for some $k\ge 0$ and $a\in \mathbb{Z}$, we write $u\circ a$ to denote $(u_1,\ldots,u_k,a)\in \mathbb{Z}^{k+1}$. Given $x\in \{0,1\}^n$ and $S\subset [n]$, we write $x^{S}$ to denote the string $x$ with the bits in $S$ flipped. 


A (positive) size-$s$ term $T$ over $n$ variables $x_1,\ldots,x_n$ is a Boolean function of the form
$$
T(x)=x_{i_1}\land \cdots\land x_{i_s}.
$$
For convenience, we require $i_1,\ldots,i_s$ only to be in $[n]$ and not necessarily distinct.
We write $\frak T_{n,s}$ to denote the set  of all size-$s$ terms 
  over $n$ variables (so $|\frak T_{n,s}|=n^s$).
Drawing $\bT\sim \frak T_{n,s}$ essentially just draws a 
  tuple  $\smash{(\bi_1,\ldots,\bi_s)\sim [n]^{s}}$.
Similarly we write $\frak C_{n,s}$ to denote the set of all $n^s$
  (positive) size-$s$ clauses over $n$ variables $x_1,\ldots,x_n$, where 
  each clause is of the form
$$
C(x)=x_{i_1}\lor \cdots \lor x_{i_s}.
$$

We say a point $x\in \{0,1\}^n$ is in \emph{middle layers} if $|x|$ satisfies 
$$(n/2)-\sqrt{n}\le |x|\le (n/2)+ \sqrt{n}.$$
The following fact is folklore: 

\begin{fact}
The number of points in middle layers is $\Theta(2^n)$.
\end{fact}

{We recall  that a tester for monotonicity is a randomized algorithm that takes as input a parameter $\eps>0$ and black-box access to a function $f:\{0,1\}^n\rightarrow \{0,1\}$. The algorithm should accept $f$ with probability at least $2/3$ when $f$ is monotone, and reject $f$ with probability at least $2/3$ when $f$ is $\eps$-far from monotone. The latter means that $\dist(f,\textsf{monotone})\ge \eps$, where 
$$
\dist\big(f,\textsf{monotone}\big):= \min_{g}
\dist(f,g)\quad\text{with}\quad
\hspace{0.08cm}\dist(f,g):=\Pr_{\bx\sim \{0,1\}^n}\big[f(\bx)\ne g(\bx)\big]
$$
and the minimum is taken over all monotone functions $g$.
We say a tester is \emph{one-sided} (error) if it accepts with probability $1$ when $f$ is monotone. Otherwise we say it is \emph{two-sided} (error).  }


We will use the following lemma of \cite{fischer2002monotonicity} to lowerbound $\dist(f,\textsf{monotone})$:

\begin{lemma}[Lemma 4 in \cite{fischer2002monotonicity}]\label{FLN Lemma}
Let $f:\zo^n \to \zo$ be a Boolean function. Then, $$
\dist\big(f, \textsf{monotone}\big) \geq \max_{\mathfrak{S}}
\left(\frac{|\mathfrak{S}|}{2^n}\right),
$$
where the max is over all sets $\mathfrak{S}$ of pairwise vertex-disjoint pairs $(x,x')$ that violate monotonicity, i.e., $x\prec x'$ but $f(x)>f(x')$. 
\end{lemma}

We now define what we mean by ``rounds of adaptivity'':

\begin{definition}[\cite{Adaptivity}]\label{def:rounds}
A randomized algorithm is said to be \emph{$r$-round-adaptive} if it proceeds in $r+1$ rounds: At the beginning of each round $t\in [0:r]$, it produces a set of queries $Q_t \subseteq \{0,1\}^n$ only based on its own randomness and answers to the previous sets of queries $Q_0, \ldots, Q_{t-1}$. 
At the end of round $t$, the algorithm receives $f(x)$ of all $x\in Q_t$ and either moves to round $t+1$ or terminates if $t=r$.
In particular, a nonadaptive algorithm can also be referred to as a $0$-round-adaptive algorithm.

\end{definition}


\section{Multilevel Talagrand Functions}\label{sec:MTF}
\label{sec:new-construction}

In this section, we introduce \emph{multilevel Talagrand functions} and 
  use them to obtain the two distributions of functions, $\Dyes$ and $\Dno$,
  that will be used to prove our lower bounds for monotonicity testing in
  \Cref{sec:mainlowerbound1} and \Cref{sec:mainlowerbound2}.
Later in \Cref{sec:information-maintained}, we introduce a stronger oracle that returns more information than the membership oracle; both lower bounds in \Cref{sec:mainlowerbound1} and \Cref{sec:mainlowerbound2} are proved against this stronger oracle.
Finally, we define the \emph{outcome} of a multilevel Talagrand function on a given set $Q$ of query points, which is a concise way to organize information obtained from the stronger oracle after making queries in $Q$. 

In this and the next two sections, we always assume that $\sqrt{n}$ is an integer, let $\ell$ be a positive integer constant, and let $N:=2^{\sqrt{n}}$. 
We write $\frak T$ for $\frak T_{n,\sqrt{n}}$ and $\frak C$ for $\frak C_{n,\sqrt{n}}$ for convenience.

\subsection{Multiplexer Trees and Maps}\label{sec:MTM}

We start with the definition of multiplexer trees, which generalizes the $2$-level construction given in \cite{chen2017beyond}.
%
To build a $2\ell$-level \emph{multiplexer tree} $M$, we start with a complete $N$-ary tree of $2\ell$ levels,  with the root at level $0$ and leaves at level $2\ell$.
So there are $N^j$ nodes on each level~$j$ and $N^{2\ell}$ leaves in total.
We refer to the root of the tree by the empty tuple $\varepsilon$ and each node at level $j\in [2\ell]$ by a tuple $u=(u_1,\ldots,u_j)\in [N]^j$,
with the parent node of $u$ being $\parent(u)=(u_1,\ldots,u_{j-1})\in [N]^{j-1}$ and its sibling nodes being $(u_1,\ldots,u_{j-1},u_j')\in [N]^j$ with $u_j'\ne u_j$.
A node $u$ is said to be   an~\emph{internal} node if it is not a leaf, an \emph{odd-level} node if it is on level $j$ for some odd $j$ and an \emph{even-level} node if it is on level $j$ for some even $j$. 

Every edge  $e=(u,v)$ of the tree is directed and goes down the tree. So  $\smash{u\in [N]^{j-1}}$ and $\smash{v\in [N]^{j}}$ for some $j\in [2\ell]$ and $u=\parent(v)$. We will refer to an edge $e=(u,v)$ as the $(v_j)$-th (outgoing) edge of $u$. Two edges $(u,v)$ and $ (u',v')$ are \textit{sibling edges} if $u = u'$.
We call $e=(u,v)$ an odd-level edge if $v$ is an odd-level node, and an even-level edge otherwise.

To finish building the multiplexer tree $M$, we associate   each odd-level edge $e$ with a  size-$\sqrt{n}$ term $T_e\in \frak T$, and each even edge $e$ with a size-$\sqrt{n}$ clause
  $C_e\in \frak C$.
Formally, a $(2\ell)$-level multiplexer tree is a map $M$ from edges to $\frak T\cup \frak C$, such that $M(e)$ is the term $T_e$ of $e$ if it is an odd-level edge and the clause $C_e$ of $e$ if it is an even-level edge.

Every $(2\ell)$-level multiplexer tree $M$ 
  defines a \emph{multiplexer map}
$$\Gamma_{M}:\{0,1\}^n\rightarrow [N]^{2\ell}\cup \{0^*,1^*\},$$
which maps every $x\in \{0,1\}^n$ to either a leaf $u\in [N]^{2\ell}$ of the  tree or one of the two special labels $\{0^*,1^*\}$. 
The definition of $\Gamma_M(x)$ is crucially based on the following notion of \emph{unique activations}:

\begin{definition}\label{def:unique-act}
Given a $(2\ell)$-level multiplexer tree $M$ and a string $x\in \{0,1\}^n$, we say an edge $e$ in the tree is \emph{activated} by $x$ if either (1) $e$ is an odd-level edge and $T_e(x)=1$ (i.e., the term $T_e$ on $e$ is satisfied by $x$) or (2) $e$ is an even-level edge and $C_e(x)=0$ (i.e., the clause $C_e$  is falsified by $x$).

Moreover, we say an edge $e=(u,v)$ is \emph{uniquely activated} by $x$ if it is the only edge activated by $x$  among all its sibling edges, in which case we also say that the node $u$ is \emph{uniquely activated} by $x$ and $(u,v)$ is its uniquely activated edge. (So a node $u$ is \emph{not} uniquely activated by $x$ if either (1) none of its edges is activated, or (2) at least two of its outgoing edges are activated.)

Given $M$ and $x\in \{0,1\}^n$, the \emph{unique activation path} of $x$ is defined to be the path $u^0\cdots u^k$ in the tree, for some $k\in [0:2\ell]$, such that (1) $u^0$ is the root; (2) every edge along the path is uniquely activated; and 
(3) the end $u^k$ of the path is either a leaf or is not uniquely activated. 
\end{definition}

We are now ready to define the multiplexer map $\Gamma_M$.
For each $x\in \{0,1\}^n$, let $u^0\cdots u^k$ be its unique activation path in the tree. 
We set $\Gamma_M(x)=u^{{k}}$ if $u^{k}\in [N]^{2\ell}$ is a leaf; otherwise, we know that $k<2\ell$ and $u^k$ is \emph{not} uniquely activated, in which case we 
  have the following two cases:
\begin{flushleft}\begin{itemize}
    \item Case 1: $k$ is even: Set $\Gamma_M(x)=0^*$ if no edges of $u^k$ is activated (i.e., no terms on edges of $u^k$ is satisfied) and set $\Gamma_M(x)=1^*$ if at least two edges of $u^k$ are activated (i.e., at least two terms on edges of $u^k$ are satisfied).
    \item Case 2: $k$ is odd: Set $\Gamma_M(x)=1^*$ if no edges of $u^k$ is activated (i.e., no   clauses on edges  of $u^k$ is falsified) and set $\Gamma_M(x)=0^*$ if at least two edges of $u^k$ are activated (i.e., at least two clauses on edges of $u^k$ are falsified).
\end{itemize}\end{flushleft}

Before using it to define multilevel Talagrand functions, we record the following simple lemma: 

\begin{lemma}\label{lem:verysimple}
Let $M$ be a $(2\ell)$-level multiplexer tree and $\Gamma_M$ be the multiplexer map it defines. 
Given any $x\in \{0,1\}^n$ and $i\in [n]$ with $x_i=0$, we have
\begin{itemize}
\item If $\Gamma_M(x)=u\in [N]^{2\ell}$, then $\Gamma_M(x^{\{i\}})$ is either $u$ or $1^*$.
\item If $\Gamma_M(x)=1^*$, then $\Gamma_M(x^{\{i\}})=1^*$.
\end{itemize}
\end{lemma}
\begin{proof}
For any even-level node $u$ (whose edges are labelled with terms), note that if $u$ is uniquely activated by $x$, then either it is still uniquely activated by $x^{\{i\}}$ along the same edge, or it has more than one activated edges. 
For any odd-level node $u$ (whose edges are labelled with clauses), if $u$ is uniquely activated by $x$, then either it is still uniquely activated by $x^{\{i\}}$ along the same edge, or none of its edges is activated. 
The lemma follows directly from these two observations.
\end{proof}
\subsection{Multilevel Talagrand Functions}\label{subsec:Talagrand}


Let $M$ be a $(2\ell)$-level multiplexer tree and $H=(h_u)$ be a tuple of  functions  
  $h_u:\{0,1\}^n\rightarrow \{0,1\}$, one for each leaf $u\in [N]^{2\ell}$ of the tree. 
(So $H$ consists of $N^{2\ell}$ functions.)
Together they define the following  \emph{$(2\ell)$-level Talagrand function}  $f_{M,H}:\{0,1\}^n\rightarrow \{0,1\}$.
For each string $x\in \{0,1\}^n$, we set $f_{M,H}(x)=1$ if $|x|> (n/2)+\sqrt{n}$; $f_{M,H}(x)=0$ if $|x|< (n/2)-\sqrt{n}$; and 
$$
f_{M,H}(x)=\begin{cases} 0 & \text{if $\Gamma_M(x)=0^*$}\vspace{0.02cm}\\[0.3ex]
1 & \text{if $\Gamma_M(x)=1^*$}\vspace{0.02cm}\\[0.3ex]
h_u(x) & \text{if $\Gamma_M(x)=u\in [N]^{2\ell} $}
\end{cases},
$$
if $x$ is in middle layers.

\subsection{Distributions $\smash{\Dyes}$ and $\smash{\Dno}$}\label{sec:dist}


We describe the two distributions $\Dyes$ and $\Dno$ over $(2\ell)$-level Talagrand functions $f_{M,H}$ that will be used in our lower bound proofs in \Cref{sec:mainlowerbound1} and 
\Cref{sec:mainlowerbound2}.

To draw $\ff\sim \Dyes$, we 
  first draw a multiplexer tree $\bM$ and a tuple of functions $\bH$ as follows: 
\begin{flushleft}\begin{enumerate}
\item We draw $\bM\sim \calM$ as follows: Start with a $(2\ell)$-level complete $N$-ary tree.
Then  we draw a term $\bT_e\sim \frak T$ for each odd-level edge $e$ (i.e., set $M(e)=\bT_e$) and draw   a clause $\bC_e\sim \frak C$ for each even-level edge (i.e., set $M(e)=\bC_e$), both independently and uniformly at random. 
\item We draw $\bH=(\bh_u)\sim \Hyes$ as follows: For each leaf $u$, $\bh_u$ is set to be the constant-$0$ function with probability $1/2$ and the constant-$1$ function with probability $1/2$, independently. 
\end{enumerate}\end{flushleft} 
Given $\bM\sim\calM$ and $\bH\sim \Hyes$, $\bf$ is set to be the $(2\ell)$-level Talagrand function $\bf=f_{\bM,\bH}$.

To draw $\bf\sim \Dno$, we draw   $\bM\sim \calM$ in the same way as in $\Dyes$. 
On the other hand, the tuple of functions $\bH$ is drawn as follows:
\begin{flushleft}\begin{itemize}
\item[$2'$.] We draw $\bH\sim \Hno$ as follows: First we draw a ``\emph{secret variable}'' $\bs\sim [n]$ uniformly at random.
For each leaf $u$,  $\bh_u$ is set to 
  the dictator function $\bh_u(x)=x_{\bs}$ with probability $1/2$ and set to be
  the anti-dictatorship function $\bh_u(x)=\overline{x_{\bs}}$ with probability $1/2$, independently. 
\end{itemize}\end{flushleft}
Given $\bM\sim \calM$ and $\bH\sim \Hno$, $\bf$ is set to be the $(2\ell)$-level Talagrand function $\bf=f_{\bM,\bH}$.

We prove two lemmas about $\Dyes$ and $\Dno$, respectively. 
\Cref{lem:monotone} shows that every function in the support of $\Dyes$ is monotone; \Cref{lem:farmonotone} shows that $\bf\sim \Dno$ is $\Omega(1)$-far from monotone with probability $\Omega(1)$.
(We note that both hidden constants are exponentially small in $\ell$. As discussed in \Cref{sec:conclusion},
  this is the obstacle for the current
  construction to obtain an $\tilde{\Omega}(\sqrt{n})$  lower bound.)

\begin{lemma}\label{lem:monotone}
Every function in the support of $\Dyes$ is monotone.
\end{lemma}

\begin{proof} 
Fix any multiplexer tree $M$ and any tuple of functions $H$ such that every $h_u$ in $H$ is either the constant-$0$ or -$1$ function, and let $f:=f_{M,H}$.
It suffices to show that for all $x \in \{0,1\}^n$ and $i \in [n]$ with $f(x) = 1$ and $x_i = 0$, we have that $f(y) = 1$, where $y:= x^{\{i\}}$. 

If $|x|\ge  (n/2)+ \sqrt n$, then
  we have $|y|=|x|+1>(n/2)+\sqrt{n}$ and thus,  $f(y) = 1$. 
On the other hand, if $|x|<(n/2)-\sqrt{n}$, then $f(x)=0$, contradicting with the assumption.
So below we assume that $(n/2)-\sqrt{n}\le |x|<(n/2)+\sqrt{n}$ and thus,
  both $x$ and $y$ are in middle layers.

Given that $x$ is in middle layers and  $f(x)=1$, either (1) $\Gamma_M(x)=1^*$; or (2) $\Gamma_M(x)=u$ for some leaf $u$ and $h_u$ is the constant-$1$ function.
For (1), we have by \Cref{lem:verysimple} that $\Gamma_M(y)=1^*$ as well and thus, $f(y)=1$.
For (2), we have by \Cref{lem:verysimple} that $\Gamma_M(y)$ is either the same $u$, in which case $f(y)=h_u(y)=1$, or $\Gamma_M(y)=1^*$, in which case we also have $f(y)=1$.
%
%
%
\end{proof}



\begin{lemma}\label{lem:farmonotone}
A function $\bf\sim \Dno$ satisfies $\dist(f, \textsf{monotone}) = \Omega(1)$ with probability at least $\Omega(1)$.
\end{lemma}



\begin{proof}
Fix an $s\in [n]$. 
We write $\Hno^s$ to denote this distribution of $\bH$ conditioning on $\bs=s$, i.e., each $\bh_u$ is $x_s$ with probability $1/2$ and $\overline{x_s}$ with probability $1/2$. 
It suffices to show that $\bf= f_{\bM,\bH}$ with $\bM\sim\calM$ and $\bH\sim \Hno^s$  has distance $\Omega(1)$ to monotonicity with probability $\Omega(1)$.

Given $\bM\sim \calM$ and $\bH\sim \Hno^s$, 
we write $\bX$ to denote the set of edges $(x,x^\ast)$ in $\{0,1\}^n$ 
such that the following three conditions holds:
\begin{enumerate}
  \item $x_{s}=0$, $x^*=x^{\{s\}}$ and $x$ satisfies $(n/2)-\sqrt{n}\le |x|\le (n/2)+\sqrt{n}-1$; 
    \item $\Gamma_{\bM}(x) = \Gamma_{\bM}(x^*) = \bu$ for some leaf $\bu\in [N]^{2\ell}$; and 
    \item $\bh_{\bu}(x)$ is the anti-dictatorship function $\overline{x_s}$.
\end{enumerate} 
Clearly, all strings in edges of $\bX$ are distinct, 
and every edge in $\bX$ is a violation to monotonicity. 
As a result, by \Cref{FLN Lemma}, it suffices to show that $|\bX|\ge \Omega(2^n)$ with probability $\Omega(1)$.
Given that the number of edges that satisfy the first condition is $\Omega(2^n)$, by linearity of expectation and Markov's inequality,
  it suffices to show that for each 
  edge $(x,x^*)$ satisfying the first condition, we have 
$$\Prob_{\bM\sim \calM,\bH\sim \Hno^s}\big[(x,x^*)\in \bX\big ] = \Omega(1).$$
To this end, we note that the second condition is about $\bM\sim \calM$ and the third condition, conditioning on the second condition, is only about $\bH\sim \Hno^s$ and always holds with probability $1/2$.
So below we show that the second condition holds with probability $\Omega(1)$ when $\bM\sim \calM$.

We partition the above event into $N^{2\ell}$ disjoint sub-events, indexed by leaves $u\in [N]^{2\ell}$:
\begin{equation*} 
\sum_{u\in [N]^{2\ell}}\Pr_{\bM\sim \calM} \big[\Gamma_{\bM}(x)=\Gamma_{\bM}(x^*)=u\big].
\end{equation*}
For each $u\in [N]^{2\ell}$, letting $u^0\cdots u^{2\ell}$ denote the path from the root $u^0$ to $u=u^{2\ell}$, the sub-event of $u$ above corresponds to the following $2\ell$ independent conditions: \begin{itemize}
    \item For each $j\in [0:2\ell-1]$, edge $(u^j,u^{j+1})$ is uniquely activated by both $x$ and $x^*$.
\end{itemize}
In particular, the probability of the condition for $j=0$ is at least
$$
\left(\frac{|x|}{n}\right)^{\sqrt n}
\left(1-\left(\frac{n-|x^*|}{n}\right)^{\sqrt n}\right )^{N-1},$$
where the first factor is the probability of the term $\bT_e\sim \frak T$, where $e=(u^0,u^1)$, is satisfied by $x$ (which implies that it is satisfied by $x^*$ as well); the second factor is the probability of $\bT_{e'}\sim\frak T$ of every other edge $e'$ of $u^0$ is not satisfied by $x^*$ (which implies that they are also not satisfied by $x$). 
Given that both $x$ and $x^*$ are in middle layers, the probability is at least 
$$
\left(\frac{(n/2)-\sqrt{n}}{n}\right)^{\sqrt{n}}\left(1-\left(\frac{(n/2)+\sqrt{n}}{n}\right)^{\sqrt{n}}\right)^{N-1}
=\frac{1}{N} \left(1-\frac{2}{\sqrt{n}}\right)^{\sqrt{n}}
\left(1-\frac{1}{N}\left(1+\frac{2}{\sqrt{n}}\right)^{\sqrt{n}}\right)^{N-1}.
$$
Using $(1\pm 2/\sqrt{n})^{\sqrt{n}}=\Theta(1)$ and $(1-\Theta(1/N))^{N-1}=\Theta(1)$, the probability is $\Omega(1/N)$.

Similarly, the probability of each of the $2\ell$ conditions can be shown to be $\Omega(1/N)$. As a result, 
$$
\sum_{u\in [N]^{2\ell}}\Pr_{\bM\sim \calM} \big[\Gamma_{\bM}(x)=\Gamma_{\bM}(x^*)=u\big] \ge N^{2\ell}\cdot \left(\Omega\left(\frac{1}{N}\right)\right)^{2\ell}=\Omega(1)$$ as desired, given that $\ell$ is a constant. 
\end{proof}

\subsection{Outcomes of Query Points}\label{sec:information-maintained}

In \Cref{sec:mainlowerbound1,sec:mainlowerbound2}, we apply Yao's minimax principle and prove our lower bounds for monotonicity testing  by showing that any deterministic, adaptive (or $r$-round adaptive) algorithm $\ALG$ cannot distinguish $\Dyes$ from $\Dno$ when its query complexity is too low. 
Given that $\ALG$ only needs to work on $(2\ell)$-level Talagrand functions and every such function is truncated outside of middle layers, we may assume without loss of generality that every query made by $\ALG$ lies in middle layers. 
Indeed we assume this is the case throughout this and the next two sections.

In our lower bound proofs, we further assume that $\ALG$ has access to a ``stronger'' oracle for the unknown $(2\ell)$-level Talagrand function $f_{M,H}$ that returns more information about a query point $x$ than just the bit $b=f(x)$.
Roughly speaking, the stronger oracle returns not only the bit $b\in \{0,1\}$ but also the minimal  information about terms\hspace{0.04cm}/\hspace{0.04cm}clauses in $M$ and functions in $H$ needed to infer that $f(x)=b$.
Formally, on a query $x\in \{0,1\}^n$, the oracle returns the following information:
\begin{flushleft}\begin{enumerate}
\item First, the oracle returns the unique activation path
  $u^0\cdots u^k$ of $x$ in $M$, where $k\in [0:2\ell]$. The oracle returns additional information depending on the following three cases.
\item \textbf{Case 1:} $k=2\ell$ (and thus, $u^k$ is a leaf and 
  $\Gamma_M(x)=u^k$). In this case, the oracle also returns $h_{u^k}(x)\in \{0,1\}$, and $\ALG$ knows that $f(x)$ is the bit $h_{u^k}(x)$ returned. 

\item \textbf{Case 2:} $k<2\ell$ (so $u^k$ is not uniquely activated) and no edges of $u^k$ is activated by $x$. \\ The oracle just lets $\ALG$ know that $x$ is in Case 2. In this case, $\ALG$ knows that $f(x)=0$\\ if $k$ is even and $f(x)=1$ if $k$ is odd.

\item \textbf{Case 3:} $k<2\ell$ and $u^k$ has at least two edges activated by $x$. The oracle lets $\ALG$ know that $x$ is in Case 3 and return the two smallest indices $a_1<a_2\in [N]$ such that 
  $(u,u\circ a_1)$ and $(u,u\circ a_2)$ are activated by $x$.
{If $k=2\ell-1$, the oracle also returns both $h_{u\circ a_1}(x)$ and $h_{u\circ a_2}(x)$.}\footnote{Technically the oracle does not need to return these two bits; returning these two bits will make the definition of \emph{outcomes} below a bit more concise, where we can make each $\rho_u$ a map over $P_u$ instead of some subset of $P_u$.}
In this case, $\ALG$ knows that $f(x)=1$ if $k$ is even and $f(x)=0$ if $k$ is odd.
\end{enumerate}\end{flushleft}
It is clear from the discussion above that this oracle reveals more information than just $f(x)$ and thus, any lower bound proved for it carries over to the standard membership oracle.

To help organize information collected
  by $\ALG$ after making a number of queries, we define the 
  \emph{outcome} of a $(2\ell)$-level Talagrand function $f_{M,H}$ on a set $Q$ of query points as follows.
Given any $Q\subseteq \{0,1\}^n$ (in middle layers) and $f_{M,H}$ for some $M$ and $H$, the outcome $\rmO = (Q,P,R,\rho)$ of $f_{M,H}$ on $Q$ is a $4$-tuple in which $P,R$ and $\rho$ are tuples with the following components:
\begin{align*}
P&=\big(P_u\subseteq Q: u\ \text{is a node in the tree that is not the root}\big),\\[0.6ex] R&=\big(R_e\subseteq Q: e\ \text{is an edge in the tree}\big)\quad\text{and}\\[0.6ex]
\rho&=\big(\rho_u:\text{$u$ is a leaf in the tree}\big),\quad\text{where $\rho_u:P_u\rightarrow \{0,1\}$ for each leaf $u$.}
\end{align*}
As it becomes clear below, each $P_u$ contains all  $x\in Q$ that are known (by information returned~by the oracle) to activate the edge $(\parent(u),u)$ in $M$;
each~$R_e$ contains all points $x\in Q$ that are known to \emph{not} activate $e$ in $M$;
each $\rho_u$ contains all information revealed so far about the function $h_u$ in $H$.
(The reader may notice that we index the $P$-sets and $R$-sets differently, using nodes and edges, respectively. One reason for this is to emphasize that, given how the oracle works, every time an $x$ is known to activate an edge $e$, it must  activate every edge along the root-to-$e$ path as well. In contrast, knowing an edge $e$  not activated by $x$ does not imply that edges along the root-to-$e$ path are not activated.)

The outcome $\rmO$ is built as follows. Start by setting every set in $P$ and $R$ to be the empty set, and every $\rho_u$ in $\rho$ to be the function with an empty domain.  Then for each point $x\in Q$, 
\begin{flushleft}\begin{enumerate}
\item Let $u^0\cdots u^k$ be the unique activation path of $x$ in $M$.
First we add $x$ to $\smash{P_{u^j}}$ for every  $j\in [k]$. For each $j\in [0:k-1]$, add $x$ to $R_e$ for all {sibling edges} $e$ of $(u^j,u^{j+1})$. Then\\ we consider the same three cases used 
  in the description of the oracle.

\item \textbf{Case 1:} $k=2\ell$. In this case, we just set $\smash{\rho_{u^{k}}(x)=h_{u^k}(x)}.$

\item \textbf{Case 2:} $k<2\ell$ and no edges of $u^k$ is activated. 
Add $x$ to $R_e$ for every  edge $e$ of $u^k$.

\item \textbf{Case 3:} $k<2\ell$ and at least two  edges of $u^k$ are activated, with $a_1<a_2\in [N]$ being the two smallest indices such that $\smash{(u^k,u^k\circ a_1)}$ and $\smash{(u^k,u^k\circ a_2)}$ are activated.
In this case, add $x$ to   $\smash{P_{u^k\circ a_1}},\smash{P_{u^k\circ a_2}}$,
 and to $\smash{R_{e}}$ for every  $\smash{e=(u^k,u^k\circ a)}$ with $a<a_2$ and $a\ne a_1$. {If $k=2\ell-1$, set} $$\rho_{u^k\circ a_1}(x)=h_{u^k\circ a_1}(x)\quad\text{and}\quad \rho_{u^k\circ a_2}(x)=h_{u^k\circ a_2}(x).$$
\end{enumerate}\end{flushleft}

While our definitions of the stronger oracle and its outcomes on a query set are quite involved, the motivation behind them is to have the following fact which gives a characterization of all $f_{M,H}$ that are consistent with an outcome on a set of query points:

\begin{fact}\label{fact:simple1}
Let $\rmO=(Q,P,R,\rho)$ be the outcome of some $(2\ell)$-level Talagrand function on $Q\subseteq \{0,1\}^n$.
Then~it is the outcome of a $(2\ell)$-level Talagrand function $f_{M,H}$ on $Q$ iff $M$ and $H$ satisfy \begin{enumerate}
\item For each odd-level edge $e=(u,v)$, $T_e$ in $M$ satisfies $T_e(x)=1$ for $x\in P_v$ and $T_e(x)=0$ for $x\in R_e$;
\item For each even-level edge $e=(u,v)$, $C_e$ in $M$ satisfies 
$C_e(x)=0$ for  $x\in P_v$ and $C_e(x)=1$ for $x\in R_e$;
\item
 For every leaf $u$, $h_u$ agrees with $\rho_u$ on every $x\in P_u$.
\end{enumerate}
\end{fact}

So by having the  oracle give away more information, the characterization of what an algorithm knows about the hidden multilevel Talagrand function $f_{M,H}$ behind the oracle now has a product structure, which consists of independent conditions on the term $T_e$ or clause $C_e$ of each edge $e$ and on the function $h_u$ of each leaf $u$. 
This will significantly simplify our analysis of $\ALG$ later.

Before moving on, we record the following fact about outcomes that follows directly from the definition.
Looking ahead, we mention that the nested structure of sets $P_u$ along a path given in the first item below will play a crucial role in our lower bound proofs:

\begin{fact}\label{fact:simple2}
Let $\rmO=(Q,P,R,\rho)$ be the outcome of a $(2\ell)$-level Talagrand function on $Q$. We have 
\begin{flushleft}\begin{enumerate}
\item For any two nodes $u,v$ such that $u$ is an ancestor of $v$ and $u$ isn't the root, we have $P_{v}\subseteq P_u$.    
\item For every internal node $u$ other than the root, we have
$$
\sum_{a\in [N]} \big|P_{u\circ a}\big|\le 2 \big|P_{u}\big|.
$$

For the root $\varepsilon$, we have
$\sum_{a\in [N]} \big|P_{a}\big|\le 2|Q|.
$

\item {The number of sets $P_u$ that are nonempty in $\rmO$ is at most $(2\ell+1)|Q|.$}
\end{enumerate}\end{flushleft}
\end{fact}


\subsection{Safe Outcomes}\label{sec:safeoutcomes}

Both lower bound proofs in \Cref{sec:mainlowerbound1,sec:mainlowerbound2} revolve around the notion of  
  ``safe'' outcomes that we define next.
Roughly speaking, if an outcome $\rmO$ is safe, then it is hard for an algorithm to tell based on $\rmO$ whether it comes from functions  from $\Dyes$ or functions   from $\Dno$ (see \Cref{lem:hardtotell} below).

We start with a definition of
  dangerous sets (of variables) in a given outcome:


\begin{definition}[Dangerous Sets]
Let $\rmO=(Q,P,R,\rho)$ be the outcome of some $(2\ell)$-level Talagrand function on $Q$. 
For each leaf $u$, we define the \emph{dangerous set} $D_u$ at $u$ to be $D_u=\emptyset$ if $P_u=\emptyset$ and
$$
D_u:=\big\{i\in [n]:\exists\hspace{0.02cm}x,y\in P_u\ \text{such that}\ x_i\ne y_i\big\}\subseteq [n],\quad\text{if $P_u\ne \emptyset.$}
$$
\end{definition}

We are now ready to define safe outcomes:

\begin{definition}[Safe Outcomes]\label{def:safeoutcomes}
Let $\rmO=(Q,P,R,\rho)$ be the outcome of some $(2\ell)$-level Talagrand function on $Q$.
We say $\rmO$ is safe if the following two conditions are satisfied:
\begin{flushleft}\begin{enumerate}
\item For each leaf $u$ with $P_u\ne \emptyset$, we have $\rho_u(x)=\rho_u(y)$ for all $x,y\in P_u$; and
\item The union of dangerous sets $D_u$ over all leaves $u$ has size $o(n)$.
\end{enumerate}\end{flushleft}
\end{definition}

The first condition above should be expected given that we want safe outcomes to confuse~an algorithm:
given the construction of $\Dyes$, $\rho_u$ is always a constant function when the hidden $f_{M,H}$ is in the support of $\Dyes$ (because $h_u$'s are constant functions). So when this condition is violated, the algorithm already knows that the function must come from $\Dno$. 

We prove the following lemma about safe outcomes:

\begin{lemma}\label{lem:hardtotell}
Let $\rmO=(Q,P,R,\rho)$ be a safe outcome.
Let $\alpha$ (or $\beta$) denote the probability of $\rmO$ being the outcome of $\bf\sim \Dyes$ (or $\bf\sim \Dno$, respectively) on $Q$. Then  we have
$\alpha\le (1+o_n(1))\cdot\beta.
$
\end{lemma}
\begin{proof}
Given any $(2\ell)$-level multiplexer tree $M$, let $\Dyes^M$ denote the distribution $\Dyes$ conditioning on $\bM=M$ (or equivalently, $\smash{\bf\sim\Dyes^M}$ is drawn by drawing $\bH\sim \Hyes$ and setting $\bf=f_{M,\bH}$),
and let $\Dno^M$ denote the distribution $\Dno$ conditioning on $\bM=M$. 
Given that in both $\Dyes$ and $\Dno$, $\bM$ is drawn from the same distribution $\calM$, it suffices to show for every $M$ that $\smash{\alpha_M\le (1+o_n(1))\cdot \beta_M}$, where
$\alpha_M$ ($\beta_M$) denotes the probability of $\rmO$ being the outcome of $\smash{\bf\sim \Dyes^M}$ (or $\smash{\bf\sim \Dno^M}$) on $Q$.

To this end, we may further assume that $M$ satisfies the first two conditions of \Cref{fact:simple1}; since otherwise, we have $\alpha_M=\beta_M=0$ and the inequality holds trivially.
Assuming that $M$ satisfies the first two condition of \Cref{fact:simple1}, $\rmO$ is the outcome of $\smash{\bf\sim \Dyes^M}$ iff the $\bH=(\bh_u)\sim \Hyes$ has  $\bh_u$ agree with $\rho_u$ on $P_u$ for every leaf $u$ with $P_u\ne \emptyset$.
Given that $\rmO$ is safe, every $\rho_u$ is a constant function. As every $\bh_u$ in $\smash{\bH\sim \Hyes}$ is set independently to be the constant-$1$ function with probability $1/2$ and the constant-$0$ function with probability $1/2$, we have 
$\alpha_M=1/2^m$, where $m$ is the number of leaves $u$ with $P_u\ne \emptyset$.

Similarly,  $\rmO$ is the  outcome of $\bf\sim \Dno^M$ on $Q$ iff the $\bH=(\bh_u)\sim \Hno$ has $\bh_u$ agree with $\rho_u$ on $P_u$ for every leaf $u$ with $P_u\ne \emptyset$.
Recall that $\bH\sim \Hno$ starts by drawing a secret variable $\bs\sim [n]$ and then sets each $\bh_u$ independently to be either $x_{\bs}$ or $\overline{x_{\bs}}$.
Consider the case when $\bs$ is not in the dangerous set $D_u$ of any leaf $u$, which, by the definition of safe outcomes, occurs with probability at least $1-o_n(1)$.
In this case, for every leaf $u$ with $P_u\ne \emptyset$, we have $\bs\notin D_u$ and thus, $\rho_u$ agrees with $\bh_u$ on $P_u$ with probability $1/2$. To see this is the case, if $\rho_u$ is the constant-$b$ function on $P_u$ and all points in $u$ have $b'$ in coordinate $\bs$ for some $b,b'\in \{0,1\}$, then $\rho_u$ agrees with $\bh_u$ iff $\bh_u$ is set to be the dictator $x_{\bs}$ when $b=b'$, and the anti-dictatorship $\overline{x_{\bs}}$ when $b\ne b'$. 
As a result, we have $\beta_M\ge (1-o_n(1))\cdot (1/2^m)$ and this finishes the proof of the lemma.
\end{proof}

To help our analysis of  dangerous sets in the next two sections, we define
$$
A_{u,0} = \big\{k \in [n]:  x_k = 0\ \text{for all $x\in P_u$}\big\} \quad\text{and}\quad A_{u,1} = \big\{k \in [n] :  x_k = 1\ \text{for all $x\in P_u$}\big\}.$$
for each node $u$ satisfying $P_u\ne \emptyset$, 
which capture common $0$- or $1$-indices
  of points in $P_u$.

We record the following simple fact about these sets:


\begin{fact}
\label{fact:basic-facts-about-coordinate-sets}
Let $\rmO=(Q,P,R,\rho)$ be the outcome of some $(2\ell)$-level Talagrand function on $Q$. Then 
\begin{enumerate}
\item For any node $u$ with $P_u\ne \emptyset$, we have $$A_{u,0}\cap A_{u,1}=\emptyset\quad\text{and}\quad 
\big|A_{u,0}\big|,\big|A_{u,1}\big|\le (n/2)+\sqrt{n}.
$$
\item For any nodes $u,v$ such that $u$ is an ancestor of $v$ and $P_u$ and $P_v$ are nonempty, we have 
$$A_{u,0}\subseteq A_{v,0}\quad\text{and}\quad
A_{u,1}\subseteq A_{v,1}.$$
\end{enumerate}
\end{fact}
The second part of the first item used the assumption that 
  query points lie in middle layers.



\section{Lower Bounds for Adaptive Monotonicity Testing}
\label{sec:mainlowerbound1}
We prove the following theorem in this section, from which \Cref{thm:2l-layer-2s-adaptive-lb} follows directly:

\begin{theorem}\label{theo:main100}
Fix any integer constant $\ell$. There exists a constant $\epsilon_\ell>0$ such that any \mbox{two-sided}, adaptive algorithm for testing whether an unknown Boolean function $f: \{0,1\}^n \to \{0,1\}$ is monotone or $\epsilon_{\ell}$-far from monotone must make $\tilde{\Omega}({n^{0.5- c}})$ queries with $c={1}/({4\ell + 2})$. 
\end{theorem}

Let $\Dyes$ and $\Dno$ be the two distributions over $(2\ell)$-level Talagrand functions described in \Cref{sec:dist}.
Let $q$ be the following parameter:
\begin{equation}\label{eq:setq}
q = \frac{n^{\frac{1}{2} - \frac{1}{4\ell + 2}}}{\log n}.
\end{equation} 
We prove that no $q$-query, deterministic algorithm $\ALG$ can distinguish $\Dyes$ from $\Dno$ under the  stronger  oracle described in \Cref{sec:information-maintained}.

To this end, we view $\ALG$ as a depth-$q$ tree\footnote{To help distinguish the $\ALG$ tree from the multiplexer tree, we will refer to nodes in the $\ALG$ tree as vertices.}, in which each vertex is labeled with an outcome $\rmO$ (as the outcome of the hidden function $f_{M,H}$ on the queries made so far; the root in particular is labeled the empty outcome in which all components are empty). Each internal vertex of $\ALG$ is also labeled with a point $x\in \{0,1\}^n$ as the next point to query. After the query $x$ is made, $\ALG$ uses the information returned by the oracle to update the outcome and move to the child vertex labeled with the updated outcome. (So the fan-out of the tree can be large.)
Each leaf vertex of the tree, in addition to the current outcome $\rmO$, is also labeled either ``accept'' or ``reject,'' meaning that $\ALG$ either accepts or rejects when this leaf vertex is reached. 

As mentioned before, $\ALG$ only needs to work on functions $f$ in the support of $\Dyes$ and $\Dno$.
For these functions, we always have  $f(x)=1$ if $|x|>n/2+\sqrt{n}$ and $f(x)=0$ if $|x|<n/2-\sqrt{n}$. Hence we may assume without loss of generality that every query $x\in \{0,1\}^n$ made by $\ALG$ lies in middle layers, as otherwise $\ALG$ already knows the value of $f(x)$. 

Looking ahead, \Cref{theo:main100} follows from two main  lemmas, \Cref{lem:main1,lem:main2}, combined with \Cref{lem:hardtotell} for safe outcomes proved in \Cref{sec:safeoutcomes}. 
Both of them are based on the following notion of \emph{good} outcomes:

\begin{definition}\label{good leaves condition}
Let $\rmO=(Q,P,R,\rho)$ be the outcome of some $(2\ell)$-level Talagrand function on a query set $Q$.
We say $\rmO$ is a \emph{good outcome} if it satisfies the following conditions:
\begin{flushleft}\begin{enumerate}
\item For every odd-level node $u$ with $P_u\ne \emptyset$, we have 
$$
\big|A_{u,1}\big| \geq \frac{n}{2}-\big|P_{u}\big|\cdot 100\sqrt{n}\log n.$$
\item For every even-level non-root node $u$ with  $P_u\ne \emptyset$, we have 
$$
\big|A_{u,0}\big| \geq \frac{n}{2}-\big|P_{u}\big|\cdot 100\sqrt{n}\log n.$$
\item For every leaf $u$ such that  $P_{u}\ne \emptyset$, we have $\rho_{u}(x)=\rho_{u}(y)$ for all $x,y\in P_{u}$. (Note that this is the same condition as in the definition of safe outcomes.)
\end{enumerate}\end{flushleft}
\end{definition}

\Cref{lem:main1} shows that every good outcome must be safe as well:
\begin{lemma}\label{lem:main1}
Every good outcome $\rmO=(Q,P,R,\rho)$ with $|Q|\le q$ is also safe. 
\end{lemma}

To state  \Cref{lem:main2}, 
we consider the following distribution $\Lyes$ over outcomes labeled at leaves of the $\ALG$ tree. To draw $\brmO\sim \Lyes$, we first draw $\bf\sim \Dyes$ then we run $\ALG$ on $\bf$ and set $\brmO$ to be the  outcome labeled at the leaf reached  at the end. 
Similarly we define $\Lno$, where $\bf\sim \Dno$.


\begin{restatable}{lemma}{maintwo}\label{lem:main2}
We have 
$$
\Pr_{\brmO\sim \Lyes}\big[\brmO\ \text{is good}\big]\ge 1-o_n(1).
$$
\end{restatable}

\Cref{theo:main100} follows immediately from   \Cref{lem:main1} and \Cref{lem:main2}:
\begin{proof}[Proof of \Cref{theo:main100} Assuming \Cref{lem:main1} and \Cref{lem:main2}]Fix any integer constant $\ell$. Let $\eps_\ell$ and $c_\ell$ be the two hidden constants in \Cref{lem:farmonotone} such that a function $\bf\sim \Dno$ is $\eps_\ell$-far from~monotone with probability   at least $c_\ell$.
We show below that no randomized $q$-query algorithm can test whether a function is monotone or $\eps_\ell$-far from monotone with error probability at most $c_\ell/4$.
The theorem follows via standard amplification arguments.

Assume for a contradiction that such an algorithm exists. Then on $\bf\sim \Dyes$, by \Cref{lem:monotone} this algorithm should accept with probability at least $1-c_\ell/4$; on $\bf\sim \Dno$, by \Cref{lem:farmonotone}, this algorithm should reject with probability at least $c_\ell(1-c_\ell/4)\ge 3c_\ell/4$.
Given that a randomized algorithm is a distribution over deterministic algorithms,
  there must be a $q$-query deterministic  $\ALG$ such that
\begin{equation}\label{eq:hehe9}
\Pr_{\bf\sim \Dyes}\big[\ALG\ \text{accepts}\ \bf\big]-
\Pr_{\bf\sim \Dno} \big[\ALG\ \text{accepts}\ \bf\big] 
\ge (1-c_\ell/4)-(1-3c_\ell/4)=c_\ell/2.
\end{equation}

On the other hand,
let ${\frak O}_\text{acc}$ be the set of outcomes on leaves of $\ALG$ at which $\ALG$ accepts and let ${\frak O}_\text{acc}^\star\subseteq {\frak O}_\text{acc}$ be those that are good.
Then
\begin{align*}
\Pr_{\bf\sim \Dyes}\big[\text{$\ALG$ accepts } \bf\big] &= \sum_{\rmO \in {\frak O}_\text{acc}} \Pr_{\brmO\sim \Lyes}\big[\brmO = \rmO\big]  
\leq \sum_{\rmO  \in {\frak O}_\text{acc}^{\star}} \Pr_{\brmO\sim \Lyes}\big[\brmO = \rmO\big] + o_n(1)
\end{align*}
using \Cref{lem:main2}.
By \Cref{lem:main1}, every outcome $\rmO=(Q,P,R,\rho)\in {\frak O}^*$ is safe. Moreover, note that the probability of $\brmO\sim \Lyes$ (or $\brmO\sim \Lno$) satisfying $\brmO=\rmO$ is exactly the same as the probability that the outcome of $\bf\sim \Dyes$ on $Q$ is $\rmO$ (or that the outcome of $\bf\sim \Dno$ on $Q$ is $\rmO$), it directly follows from \Cref{lem:hardtotell} that the RHS of the inequality above is at most
\begin{align*}
(1 + o_n(1))\sum_{\rmO\in {\frak O}_\text{acc}^\star}  \Pr_{\brmO\sim \Lno}\big[\brmO = \rmO\big] + o_n(1) \leq \Pr_{\bf\sim \Dno}\big[\text{ALG accepts } \bf\big] + o_n(1),
\end{align*}
which contradicts with \Cref{eq:hehe9}. This finishes the proof of \Cref{theo:main100}.
\end{proof}

In the rest of the section, we prove \Cref{lem:main1} in \Cref{sec:main1} and  \Cref{lem:main2} in \Cref{sec:main2}.


\subsection{Proof of \Cref{lem:main1} }\label{sec:main1}


Let $\rmO=(Q,P,R,\rho)$ be a good outcome with $|Q|\le q$. Recall the definition of  $A_{u,0},A_{u,1}$ for each (non-root) node~$u$ with $P_u\ne \emptyset$ from \Cref{sec:information-maintained}.
We start with two bounds on  these sets: 
\begin{claim}
\label{claim:1-good-leaf-size-lower-bound}
For any even-level node $u$ other than the root with $P_u\ne \emptyset$, letting $v=\parent(u)$, we have
$$
\big|A_{u,1}\big| \ge \frac{n}{2} -  \min\left(\big|P_{u}\big|^2, \big|P_{v}\big|\right)
\cdot 150\sqrt{n}\log n.
$$
\end{claim}
\begin{proof}
First, by \Cref{fact:simple2} we have $P_u\subseteq P_v$ so $P_v\ne \emptyset$ as well; by \Cref{fact:basic-facts-about-coordinate-sets} we have $A_{v,1}\subseteq A_{u,1}$.
Then by the definition of good outcomes (and that $v$ is an odd-level node with $P_v\ne \emptyset$), we have 
    $$\big|A_{u,1}\big| \ge \big|A_{v,1}\big| \ge  \frac{n}{2} - \big|P_{v}\big|\cdot 100\sqrt{n}\log n.$$
    
    On the other hand, we also know that for any two strings $x,y\in P_{u}$, we have 
    $$
    \big|\{j\in [n] : x_j = y_j = 0\}\big| \geq \big|A_{u, 0}\big|\geq \frac{n}{2} -  \big|P_{u}\big|\cdot 100\sqrt n \log n,
    $$
    where the second inequality used the definition of good outcomes (and that $u$ is an even-level node other than the root).
    Given that all points are in middle layers, we have  
    \begin{align*}
    \big|\{j\in [n] : x_j = 1, y_j = 0\}\big|=\left(n-|y|\right)- \big|\{j\in [n] : x_j = y_j = 0\}\big|\leq \sqrt{n}+\big|P_{u}\big|\cdot 100\sqrt n \log n.
    \end{align*} 
 As a result, we have 
\begin{align*}
    \big|A_{u, 1}\big| &\geq  |x| - \sum_{y\in P_{u}\setminus \{x\}}\big|\{j: x_j = 1, y_j = 0\}\big|\\ &\geq \frac{n}{2}-\sqrt{n} -\left(\big|P_u\big|-1\right)\left(\sqrt{n}+\big|P_{u}\big|\cdot 100\sqrt n \log n\right)\\[0.8ex] &\ge \frac{n}{2}-\big|P_u\big|^2\cdot 150\sqrt{n}\log n,
\end{align*}
where we used $|P_u|\ge 1$.
Combining the two inequalities for $|A_{u,1}|$ gives the desired claim. 
\end{proof}
The following claim for odd-level nodes can be proved similarly:
\begin{claim}
\label{claim:0-good-leaf-size-lower-bound}
For any odd-level node $u$ at level $k\ge 3$ with $P_u\ne \emptyset$, letting $v=\parent(u)$, we have
$$
\big|A_{u,0}\big| \ge \frac{n}{2} - \min\left(\big|P_{u}\big|^2, \big|P_{v}\big|\right)\cdot 150\sqrt{n}\log n.
$$
\end{claim}


For convenience, we will write $K$ to denote $250\sqrt{n}\log n$ in the rest of this subsection.

Recall that for each leaf $u$, the dangerous set $D_u$ at $u$ is the set of coordinates $i\in [n]$ such that points in $P_u$ don't agree on (and $D_u=\emptyset$ trivially if $P_u=\emptyset$).
When $P_u\ne \emptyset$, we also have  
$$
D_u=\overline{A_{u,0}\cup A_{u,1}}.
$$
To upperbound the union of $D_u$ over all leaves, we introduce the following sets $B_u\subseteq [n]$ for each node (including the root) of the tree: For each node $u$, $B_u$ is the union of dangerous sets $D_w$ over all leaves $w$ in the subtree rooted at $u$. 
So $B_u$ is the same as $D_u$ if $u$ is a leaf, and $B_\eps$ at the root is exactly the union of $D_w$ over all leaves $w$, which we want to bound in size by $o(n)$.
We also have for each internal node $u$ that $B_u=\cup_{a\in [N]} B_{u\circ a}.$
We prove the following fact about these sets:

\begin{restatable}{fact}{Bij}
\label{fact: Bij}
For every node $u$ with $P_u\ne \emptyset$ (so $u$ is not the root), we have $
B_{u} \subseteq \overline{A_{u, 0}\cup A_{u, 1}}.$ 
    
\end{restatable}
\begin{proof}
The case when $u$ is a leaf is trivial given that $B_u=D_u=\overline{A_{u,0}\cup A_{u,1}}$. So we assume below $u$ is an internal, non-root node with $P_u\ne \emptyset$. By definition, for every $i\in B_u$, there must be a leaf $w$ in the subtree rooted at $u$ such that $i\in D_w$ and thus, $i\notin A_{w,0}\cup A_{w,1}$.
Given that $u$ is an ancestor of $w$, it follows from \Cref{fact:basic-facts-about-coordinate-sets} that $A_{u,0}\subseteq A_{w,0}$ and $A_{u,1}\subseteq A_{w,1}$ and thus, $i\in \overline{A_{u,0}\cup A_{u,1}}$.
\end{proof}

As a corollary of \Cref{claim:1-good-leaf-size-lower-bound} and \Cref{claim:0-good-leaf-size-lower-bound}, we have the following inequality for $|B_u|$:


\begin{corollary}\label{Bij Corollary} 
    For each node $u$ at level $k\ge 2$,  letting $v=\parent(u)$, we have
    $|B_{u}| \leq |P_{v}\big|\cdot K$. 
\end{corollary}
\begin{proof}
First, we assume without loss of generality that $P_u\ne \emptyset$.
Otherwise, $P_w=\emptyset$ for  every leaf $w$ in the subtree rooted at $u$ and thus, $D_w=\emptyset$ for every such leaf $w$ and $B_u=\emptyset$ as well.

Assuming that $P_u\ne \emptyset$, we have $P_u\subseteq P_v$ by \Cref{fact:simple2} so $P_v$ is not empty as well. We start with the case when $u$ is an even-level  node (that is not the root).
Using \Cref{fact: Bij}, we have 
   \begin{align*}
    \big|B_{u}\big|
        &\leq n - \big|A_{u, 0} \cup A_{u, 1} |
         = n - \big|A_{u, 0}\big| - \big|A_{u, 1}\big|
        \leq n - \big|A_{u, 0}\big| - \big| A_{v, 1}\big|,
    \end{align*}
    where we used $A_{v,1}\subseteq A_{u,1}$ by \Cref{fact:basic-facts-about-coordinate-sets}.
We also have 
$$
|A_{u,0}|\ge \frac{n}{2}-\big|P_u\big|\cdot 100\sqrt{n}\log n\quad\text{and}\quad
|A_{v,1}|\ge \frac{n}{2}-\big|P_v\big| \cdot 100\sqrt{n}\log n,
$$
where we used the definition of good outcomes (see \Cref{good leaves condition}). 
The statement follows by using $|P_v|\ge  |P_u|$ and $K=250\sqrt{n}\log n$. The case when $u$ is odd-level follows similarly.
\end{proof}

\begin{corollary}
\label{Claim Bij at last layer}
For each node $u$ that is not a leaf and not the root, we have 
    $$\big|B_{u}\big| \leq \sum_{a\in [N]} \min\Big(\big| P_{u\circ a}\big|^2, \big| P_{u}\big|\Big)\cdot K.$$
\end{corollary} 
\begin{proof}
Using $B_u=\cup_{a\in [N]} B_{u\circ a}$,  
we have 
\begin{align*}
\big|B_{u}\big|
    \leq
\sum_{a\in [N]} 
    \big|B_{u\circ a}\big|.
\end{align*}
For each $a\in [N]$, if $P_{u\circ a}=\emptyset$, then $B_{u\circ a}=\emptyset$ because every dangerous set in the subtree rooted at $u\circ a$ is empty.
Combining this with \Cref{fact: Bij}, we have 
$$
\big|B_u\big|\le \sum_{a\in [N]:P_{u\circ a}\ne \emptyset}\Big(n-\big|A_{u\circ a,0}\big|-\big|A_{u\circ a,1}\big|\Big).
$$

For each $a\in [N]$ with $P_{u\circ a}\ne \emptyset$, it follows by combining \Cref{good leaves condition} and \Cref{claim:1-good-leaf-size-lower-bound}, \Cref{claim:0-good-leaf-size-lower-bound} that one of $|A_{u\circ a,0}|$ and $|A_{u\circ a,1}|$ is at least $(n/2)-|P_{u\circ a}|\cdot 100\sqrt{n}\log n$ and the other is at least 
$$
\frac{n}{2}-\min\Big(\big|P_{u\circ a}\big|^2,\big|P_u\big|\Big)\cdot 150\sqrt{n}\log n.
$$
The statement follows by combining these inequalities and that $K=250\sqrt{n}\log n$.
\end{proof}

We just need one more simple technical lemma before proving \Cref{lem:main1}:

\begin{lemma}[Smoothing Lemma]
\label{lem:help-ineq-for-induction}
Let $\alpha$ and $\beta$ be two nonnegative real numbers.
Let  $(p_j)_{j \in [N]}$ be a sequence of nonnegative real numbers that sum to at most $2\beta$. Then we have
$$\sum_{j\in [N]} \min\Big(\beta, p_j^{1+\alpha}\Big) \leq 4\beta^{1+\frac{\alpha}{1+\alpha}}.$$
\end{lemma}
\begin{proof}
Assume without loss of generality that $\beta>0$.
Let $$J_1 := \big\{j: p_j^{1+\alpha} \leq \beta\big\}\quad\text{and}\quad J_2 := \big\{j: p_j^{1+\alpha} > \beta\big\}.$$ For each $j \in J_2$ we have $p_j > \beta^{1/(1+\alpha)}$. 
Using $\sum_j p_j\le 2\beta$, we have 
$ |J_2 | \leq 2{\beta^{\frac{\alpha}{1+\alpha}}}.$
As such, we have 
$$\sum_{j\in [N]} \min\Big(\beta , p_j^{1+\alpha}\Big) 
=
\sum_{j \in J_1} p_j^{1+\alpha} + |J_2| \cdot \beta \leq \sum_{j \in J_1} p_j^{1+\alpha} + 2\beta^{1 + \frac{\alpha}{1+\alpha}}$$ 

On the other hand, for each $j \in J_1$, we have $p_j \leq \beta^{1/(1+\alpha)}$ and thus,
$$
p_j^{1+\alpha} = p_j \cdot p_j^{\alpha} \leq p_j \cdot \left(\beta^{1/(1+\alpha)}\right)^{\alpha} = p_j \cdot \beta^{\alpha/(1+\alpha)}.
$$
As a result, we can bound the sum over $J_1$ by 
$$
\sum_{j \in J_1} p_j^{1+\alpha} \leq \beta^{\alpha/(1+\alpha)} \sum_{j \in J_1} p_j \leq \beta^{\alpha/(1+\alpha)} \cdot 2\beta = 2\beta^{1+\frac{\alpha}{1+\alpha}}
$$
and the desired result follows from summing these two bounds.
\end{proof}

We are now ready to prove \Cref{lem:main1}, i.e., $|B_\eps|=o(n)$:
\begin{proof}[Proof of \Cref{lem:main1}]
First we prove  that every node $u$ at level $k=1,\ldots, 2\ell-1$ satisfies
\begin{equation}\label{eq:IH}
\big|B_{u}\big|  \leq  {4^{2\ell-k}}\big|P_{u}\big|^{1+\frac{1}{2\ell-k+1}}\cdot K. 
\end{equation}
We will proceed by induction on the level of $u$ from $2\ell-1$ to $1$.

For the base case when $u$ is at level $2\ell-1$, we have from \Cref{Claim Bij at last layer} that 
$$
\big|B_u\big|\le \sum_{a\in [N]} \min\Big(\big|P_{u\circ a}\big|^2,\big|P_u\big|\Big)\cdot K.
$$
Using $\sum_{a\in [N]} |P_{u\circ a}|\le 2|P_u|$ from \Cref{fact:simple2} and \Cref{lem:help-ineq-for-induction} (with $\alpha=1$ and $\beta=|P_u|$), we have
$$
\big|B_u\big|\le 4\big|P_u\big|^{3/2}\cdot K.
$$

Next we work on the induction step to prove \Cref{eq:IH} for any node $u$ at some level $k$ that satisfies $1\le k\le 2\ell-2$, assuming \Cref{eq:IH} for nodes at level $k+1$. 
First we have 
\begin{align*}
\big|B_{u}\big| 
    \leq \sum_{a\in [N]} \big|B_{u\circ a}\big|.
\end{align*}
Combining the inductive hypothesis and \Cref{Bij Corollary} on each $|B_{u\circ a}|$ (at level $k+1\ge 2$), we have 
\begin{align*}
  \big|B_u\big|  \leq \sum_{a\in [N]} \min\Big(\big|P_u\big| , {4^{2\ell-k-1}}\big|P_{u\circ a}\big|^{1+\frac{1}{2\ell-k}}\Big)\cdot K\le 4^{2\ell-k-1} \sum_{a\in [N]} \min\Big(\big|P_u\big|,\big|P_{u\circ a}\big|^{1+\frac{1}{2\ell-k}}\Big)\cdot K.
\end{align*}
It then follows from $\sum_{a\in [N]} |P_{u\circ a}|\le 2|P_u|$ and  \Cref{lem:help-ineq-for-induction} that the RHS is at most
$$
4^{2\ell-k-1}\cdot 4\big|P_u\big|^{1+\frac{\frac{1}{2\ell-k}}{1+\frac{1}{2\ell-k}}}\cdot K=4^{2\ell-k} \big|P_u\big|^{1+\frac{1}{2\ell-k+1}}\cdot K.
$$
This finishes the induction and the proof of \Cref{eq:IH}.

Using $|B_\eps|\le \sum_{a\in [N]} B_a$ and then \Cref{eq:IH} on all level-$1$ nodes $a$, we have 
\begin{align*}
\big|B_\eps\big|\le 
\sum_{a\in [N]} \big|B_a\big|
\le 4^{2\ell-1}\sum_{a\in [N]}
\big|P_a\big|^{1+ \frac{1}{2\ell}}\cdot K \le O\Big(q^{1+\frac{1}{2\ell}}K\Big),
\end{align*}
where the last inequality used that $\sum_a |P_a|\le 2|Q|=2q$ from \Cref{fact:simple2}.
Plugging in the choice of $q$ in \Cref{eq:setq} and 
 $K=O(\sqrt{n}\log n)$ finishes the proof that $|B_\eps|=o(n)$.
\end{proof}

\subsection{Proof of \Cref{lem:main2}}\label{sec:main2}
Finally we prove \Cref{lem:main2} which we restate below for convenience.
\maintwo*

Given that $\brmO$ is drawn from $\Lyes$ here, it suffices to prove that $\brmO\sim \Lyes$ satisfies the first two conditions of \Cref{good leaves condition} with probability at least $1-o_n(1)$. This is  because the third condition is always satisfied (see the comment below \Cref{def:safeoutcomes}).

To prove \Cref{lem:main2}, it suffices to prove the following lemma and apply a union bound:

\begin{lemma}\label{lem:usealot}
Let $\rmO=(Q,P,R,\rho)$ be a good outcome labeled at some internal {vertex} of $\ALG$, and let $x\in \{0,1\}^n$ be the next query to make labeled {at this {vertex}}.
Conditioning on $\bf\sim \Dyes$ reaching
  this {vertex} (or equivalently, conditioning on the outcome of $\bf\sim \Dyes$ on $Q$ is being $\rmO$), the probability of $\bf$ reaching a bad outcome after querying $x$ is $o(1/q)$.
\end{lemma}
\begin{proof}
Let $K'=100\sqrt{n}\log n$ in this proof. 

First, the only possibilities for the updated outcome to become bad after querying $x$ are (note that these events below are only necessary but not sufficient for the updated outcome to be bad):
\begin{flushleft}\begin{enumerate}
\item The query point $x$ is added to some $P_u$ which was empty in $\rmO$ for some odd-level node $u$ and the new $|A_{u,1}|$ becomes lower than $(n/2)-K'.$ This cannot happen because the new $|A_{u,1}|$ is just $|x|$ and is at least $(n/2)-\sqrt{n}$ because $x$ is in middle layers\footnote{Recall that we can assume without loss of generality that $\ALG$ only queries points in  middle layers.};
\item The query point $x$ is added to some $P_u$ which was empty in $\rmO$ for some even-level, non-root node $u$ and the new $|A_{u,0}|$ becomes lower than $(n/2)-K'$. This again cannot happen. 
\item The query point $x$ is added to some $P_u$ which was not empty in $\rmO$ for some odd-level node $u$ and the new $|A_{u,1}|$ goes down for more than $K'$. For this to happen, it must be the case that the number of $i\in A_{u,1}$ with $x_i=0$ is at least $K'$.
\item The query point $x$ is added to some $P_u$ which was not empty in $\rmO$ for some even-level, non-root node $u$ and the new $|A_{u,0}|$ goes down for more than $K'$.
For this to happen, it must be the case that the number of $i\in A_{u,0}$ with $x_i=1$ is at least $K'$.
\end{enumerate}\end{flushleft}
We show below that for any odd-level node $u$ such that 
\begin{enumerate}
\item $P_u\ne \emptyset$ in $\rmO$; and  
\item the number of $i\in A_{u,1}$ satisfying $x_i=0$ is at least $K'$,
\end{enumerate}
the probability of $x$ being added to $P_u$ when $\bf\sim \Dyes$ conditioning on reaching $\rmO$ is $o(1/q^2)$. 
The same can be proved, with similar arguments, for even-level nodes (and regarding $A_{u,0}$).
Assuming these, the lemma follows because the number of nonempty $P_u$ in $\rmO$ can be at most $O(\ell |Q|)=O(q)$ by \Cref{fact:simple2} given that $|Q|\le q$ and $\ell$ is a constant.

To this end, fix any odd-level $u$ such that $P_u$ is nonempty and we write $\Delta$ to denote
$$
\Delta:=\big\{i\in A_{u,1}:x_i=0\big\},
$$
with $|\Delta|\ge K'$.
We show that when $\bf\sim \Dyes$ conditioning on it reaching $\rmO$, the probability that~$x$ is added to $P_u$ after it is queried is at most $o(1/q^2)$.
For this purpose, recall from \Cref{fact:simple1} that~the characterization of $\bf\sim \Dyes$ reaching $\rmO$ consists of independent conditions, one condition on the term or clause on each edge and one condition on the function at each leaf.
Regarding~the~term~$\bT_e$ (since $u$ is an odd-level node) at 
  $e=(\parent(u),u)$ in $\bM\sim \calM$:
\begin{flushleft}\begin{enumerate}
\item For $\bf\sim \Dyes$ to reach $\rmO$, the term $\bT_e$ at $e$ can be set to a term $T\in \frak T$ iff (1) $T(y)=1$ for all $y\in P_u$ and (2) $T(y)=0$ for all $y\in R_e$. Let's denote this event $E_1$ for $\bT_e\sim \frak T$.
\item For $\bf\sim \Dyes$ to not only reach $\rmO$ but also have $x$ added to $P_u$ after it is queried, $\bT_e$ can be set to a term $T\in \frak T$ iff (1) $T(y)=1$ for all $y\in P_u\cup \{x\}$ and (2) $T(y)=0$ for all $y\in R_e$. Let's denote this event $E_2$ for $\bT_e\sim \frak T$.
\end{enumerate}\end{flushleft}
With the definition of $E_1$ and $E_2$ above, it suffices to show that
\begin{equation}\label{eq:lastlast}
\Pr_{\bT\sim \frak T} \big[E_2\big]\le 
o\left(\frac{1}{q^2}\right)\cdot \Pr_{\bT\sim \frak T} \big[E_1 \big].
\end{equation}

We prove \Cref{eq:lastlast} in a more generic setting and with looser parameters so that the conclusion can be reused later in the next section. 
Let $A\subseteq [n]$, $\Delta\subseteq A$ with $|\Delta|\ge K'$ and $R\subseteq \{0,1\}^n$ with $|R|\le \sqrt{n}/2$. 
Consider $\bT\sim \frak T$.
Let $E_1^*$ be the event that 
  (1) all variables in $\bT$ come from $A$ and (2) $\bT(y)=0$ for all $y\in R$;
  let $E_2^*$ be the event that (1) all variables in $\bT$ come from $A\setminus \Delta$ and (2) $\bT(y)=0$ for all $y\in R$.

We prove the following claim under this setting, from which \Cref{eq:lastlast} follows directly:

\begin{claim}\label{claim:reuse}
We have 
$$
\Pr_{\bT\sim \frak T}\big[E_2^*\big]\le o\left(\frac{1}{n^{5}}\right)\cdot \Pr_{\bT\sim \frak T} \big[E_1^*\big].
$$
\end{claim}
\begin{proof}We count  ordered tuples $I=(I_1,\ldots,I_{\sqrt{n}})\in [n]^{\sqrt{n}}$ in the following two sets.
\begin{flushleft}\begin{itemize}
\item $U$ contains all $I\in [n]^{\sqrt{n}}$ such that $I_k\in A$ for all $k\in [\sqrt{n}]$ and for every $z\in R$,
  there exists at least one $k\in [\sqrt{n}]$ such that $z_{I_k}=0$; and
\item $V$ contains all $I\in [n]^{\sqrt{n}}$ such that $I_k\in A \setminus \Delta$ for all $k\in [\sqrt{n}]$ and for every $z\in R$, there exists at least one $k\in [\sqrt{n}]$ such that $z_{I_k}=0$.
\end{itemize}\end{flushleft}
It suffices to show that $|V|/|U|\le o(1/n^{5})$. 
To upperbound this ratio, let $t= \log n$ and we use $U'$ to denote the subset of $U$ such that $I\in U$ is in $U'$ if and only if $$\Big|\big\{ k \in [\sqrt n] : I_k \in \Delta \big\}\Big| = t.$$Now it suffices  to show that $|V|/|U'|=o(1/n^{5})$ given that $U'\subseteq U$. We define a bipartite graph $G$ between
  $U'$ and $V$: $I'\in U'$ and $I\in V$ have an edge if and only if
  $I'_k=I_k$ for every $k\in [\sqrt{n}]$ with $I'_k\notin \Delta$.
From the construction, it is clear 
  each $I'\in U'$ has degree at most $|A\setminus \Delta|^t$.

To lowerbound the degree of an $I\in V$,
  letting points in $R$ be $z^1,\ldots,z^{|R|}$, we can fix a set of $|R|$ (not necessarily distinct) indices $k_1, \ldots, k_{|R|}$  in $[\sqrt{n}]$ such that every $z^i$ has  $$\big(z^i\big)_{I_{k_i}}=0.$$ Once these indices are fixed, we can pick any of the $t$ remaining ones and map them to $t$ variables in $\Delta$. 
As a result, the degree of each $I\in V$ is at least:
$${{\sqrt n-|R|} \choose {t} }\cdot |\Delta|^t.
$$
By counting   edges in $G$ in two different ways and using $|A| \leq n$ and $|R| \leq \sqrt{n}/2$, 
we have 
\[ \frac{|U'|}{|V|}\geq {{\sqrt n - |R| }\choose t} \cdot \Bigg(\frac{|\Delta|}{|A\setminus \Delta|}\Bigg)^t \geq \Bigg(\frac{\sqrt{n}/2}{t}\Bigg)^t\cdot \Bigg(\frac{100\sqrt{n}t}{n}
\Bigg)^t > \omega(n^{5}).\]
This finishes the proof of the claim.
\end{proof}
This finishes the proof of \Cref{lem:usealot}.\end{proof}

\section{Tight Lower Bounds for Constant Rounds of Adaptivity}\label{sec:mainlowerbound2}

\label{sec:adaptivity-hierarchy}
In this section we prove the following theorem from which 
\Cref{thm:adaptivity-hierarchy} follows:

\begin{theorem}\label{theo:main200}
For any integer constant $\ell$, there exists a constant $\epsilon_\ell>0$ such that any \mbox{two-sided}, $(2\ell-1)$-round adaptive algorithm for testing whether an unknown Boolean function $f: \{0,1\}^n \to \{0,1\}$ is monotone or $\epsilon_{\ell}$-far from monotone must make $\tilde{\Omega}(\sqrt{n})$ queries. 
\end{theorem}


Fix any integer constant $\ell$, and let $r:=2\ell-1$ be the number of rounds of adaptivity. 
(Recall that an $r$-round adaptive algorithm gets to make $r+1=2\ell$ batches of queries.)
Let $\Dyes$ and $\Dno$ be the distributions over $(2\ell)$-level Talagrand functions described in \Cref{sec:dist}.
Let 
\begin{equation}\label{eq:setq2}
q = \frac{\sqrt{n}}{\log^2 n}.
\end{equation} 
We show that no $q$-query, deterministic, $r$-round adaptive algorithm $\ALG$ can distinguish $\Dyes$ from $\Dno$ under the (stronger) oracle described in \Cref{sec:information-maintained}.

\begin{remark}
In \Cref{sec:upper-bound}, we sketch a $(2\ell+1)$-round-adaptive algorithm spending ${O}(n^{\frac{1}{2}-\frac{1}{4\ell+2}})$ queries that successfully finds a violation in $\bf\sim \Dno$ with probability $\Omega(1)$. This aligns with the intuition if a tester wants to use the ``quadratic-speedup strategy'' (see \Cref{sec:overview}) to flip the secret variable $\bs$, it first needs to attack the $2\ell$-level Talagrand function level by level (each level requiring $1$ round of queries). 
\end{remark}



Given that $\ALG$ is an $r$-round adaptive algorithm, we consider it as a tree of depth $r+1$, with the root at depth $0$ and leaves at depth $r+1$.\footnote{Again we will refer to nodes in the $\ALG$ tree as vertices.}
Each vertex of the $\ALG$ tree is labeled an outcome $\rmO$, 
  as the outcome of the hidden function $f_{M,H}$ on the queries made so far.
Each internal vertex is also labeled a set $S$ of at most $q$ points to be queried in the next batch.  
After the set $S$ of queries is made, $\ALG$ uses the information returned by the oracle to update the outcome to $\rmO'$ and move down to the child vertex labeled with the updated outcome $\rmO'$.  Each leaf of the tree, in addition to the final outcome $\rmO$, is also labeled either ``accept'' or ``reject,'' meaning that $\ALG$ either accepts or rejects when this leaf is reached. 

Given $\ALG$, the two distributions $\Lyes,\Lno$ over final outcomes of $\ALG$ are defined similarly as in the previous section: $\brmO\sim \Lyes$ (or $\brmO\sim \Lno$) is drawn by first drawing $\bf\sim \Dyes$ (or $\bf\sim \Dno$, respectively) and then returning the outcome $\brmO$ of the leaf that $\bf$ reaches in $\ALG$. 


The main technical lemma we prove in this section is the following:

\begin{lemma}\label{lem:mmmm}
We have 
$$
\Pr_{\brmO\sim \Lyes} \big[\brmO\ \text{is safe}\big]\ge 1-o_n(1).
$$
\end{lemma}

\Cref{theo:main200}   follows by combining \Cref{lem:mmmm} with \Cref{lem:hardtotell}, using arguments similar to the proof of \Cref{theo:main100} in the previous section.
We prove \Cref{lem:mmmm} in the rest of this section.

\subsection{Proof of \Cref{lem:mmmm}}

We generalize the definition of dangerous sets $D_u$ to not only leaves but also every non-root node in the multiplexer tree.
Let $\rmO=(Q,P,R,\rho)$ be an outcome. 
For every non-root node $u$, 
  we write $D_u$ to denote the set of coordinates $i\in [n]$ such that points in $P_u$ don't agree on (i.e., $x_i\ne y_i$ for some $x,y\in P_u$);
we set $D_u=\emptyset$ if $P_u=\emptyset$.
Note that for leaves this definition is the same as before,
and we refer to $D_u$ as the \emph{dangerous set} of node $u$.


Let $\rmO=(Q,P,R,\rho)$ be the outcome labeled at an internal vertex in $\ALG$ and let $t\in [0:r]$ be its depth. (In particular, the vertex can be the root with $t=0$.) Let $S$ be the query set of size $|S|\le q$ labeled at this vertex.
We are interested in the updated outcome $\brmO^*=(Q\cup S,\bP^*,\bR^*,\brho^*)$ obtained from $\rmO$ after quering $S$, when $\bf$ is drawn from $\Dyes$  conditioning on $\bf$ reaching $\rmO$ (i.e., conditioning on
  that the outcome of $\bf$ on $Q$ is $\rmO$). 
For clarity, we  use symbols such as $P_u,R_e,\rho_u$ to denote objects defined from $\rmO$, and use $\bP_u^*,\bR_e^*,\brho_u^*$ to denote their counterparts in $\brmO^*$.
The two sets that we will pay special attention to are $D$ (from $\rmO$) and $\bD^*$ (from $\brmO^*$) where: 
\begin{flushleft}\begin{itemize}
\item $D$ is the union of dangerous sets $D_u$ in $\rmO$ over all level-$t$ nodes $u$ (for the special case\\ when $t=0$, the vertex is the root and $\rmO$ is the empty outcome, we set $D=\emptyset$); and
\item $\bD^*$ be the union of dangerous sets $\bD_u$ in $\brmO^*$ over all level-$(t+1)$ nodes $u$.
\end{itemize}  \end{flushleft}

We show that with high probability (over $\bf\sim \Dyes$ conditioning on 
  $\bf$ reaching $\rmO$), 
$\bD^{*}$ can only grow by $o(n)$ in size from $D$ after querying $S$:

\begin{lemma}\label{lem:pppp}
With probability at least $1- o_n(1)$, we have 
$
\left|\bD^*\right|\le \left|D\right|+o(n).
$
\end{lemma}

We delay the proof of   \Cref{lem:pppp} and first use it to prove \Cref{lem:mmmm}: 






\begin{proof}[Proof of \Cref{lem:mmmm} Assuming \Cref{lem:pppp}.]
Let $\brmO^0,\brmO^1,\cdots,\brmO^{r+1}$  denote the sequence of outcomes labeled along the path that $\bf\sim \Dyes$ walks down in $\ALG$, where $\brmO^0$ is the empty outcome labeled at the root and $\brmO^{r+1}$ is the final outcome at the leaf reached. 
Recall that the goal of \Cref{lem:mmmm} is to show that $\brmO=\brmO^{r+1}$ is safe with probability $1-o_n(1)$. 
Given that $\bf$ is drawn from $\Dyes$, the first condition in \Cref{def:safeoutcomes} always holds and thus, it suffices focus on the first condition and show that the union of dangerous sets on leaves in $\brmO$ has size $o(n)$ with probability at least $1-o_n(1)$.

To this end, we write $\bD^t$, for each $t\in [r+1]$, to denote the union of dangerous sets $\bD^t_u$ in $\bO^t$ over all nodes $u$ at level $t$, with $\bD^0$ being the empty set for $t=0$. 
Notice that $\bD^{r+1}$ from $\brmO^{r+1}$ is exactly the set that we would like to bound by $o(n)$ in size.
Then by \Cref{lem:pppp} and a union bound over the $r+1$ rounds, we have that with probability at least $1-(r+1)\cdot o_n(1)=1-o_n(1)$ that  
$$\big|\bD^{t+1}\big|\le \big|\bD^t\big|+o(n),
\quad\text{for each $t\in [0:r]$.}
$$
Given that $\bD^0=\emptyset$ in the empty outcome $\brmO^0$ initially (and that $r$ is a constant), we have $|\bD^{r+1}|=o(n)$, which implies that $\brmO^{t+1}$ is safe with probability at least $1-o_n(1)$.
\end{proof}

\subsection{Proof of \Cref{lem:pppp}}

We assume without loss of generality that $t$ is odd;
  the case when $t$ is even is symmetric.
  
Our plan to upperbound $|\bD^*|-|D|$ uses the following simple inequality:
\begin{align*}
\big|\bD^*\big|-\big|D\big|\le \big|\bD^*\setminus D\big|
=\left| \left(\bigcup_{\substack{\text{level $t+1$}\\ \text{node $u$}}} 
\bD_u^*\right)\Bigg\backslash \left(\bigcup_{\substack{\text{level $t$}\\
\text{node $v$}}}D_v\right)\right|
\le \sum_{\substack{\text{level $t $}\\ \text{node $v$}}} \left|\left(\bigcup_{\substack{\text{child $u$}\\\text{of $v$}}} \bD_u^*\right) \Bigg\backslash D_{v}\right|.
\end{align*}
For each level-$t$ node $v$, we split the terms into those $u$ with 
  $P_u\ne \emptyset$ and those $u$ with $P_u=\emptyset$: 
$$
\left|\left(\bigcup_{\substack{\text{child $u$}\\\text{of $v$}}} \bD_u^*\right) \Bigg\backslash D_{v}\right|\le 
\left|\left(\bigcup_{\substack{\text{child $u$}\\\text{of $v\hspace{-0.05cm}:\hspace{-0.05cm} P_u\ne \emptyset$}}} \bD_u^*\right) \Bigg\backslash D_{v}\right|
+\sum_{\substack{\text{child $u$}\\ \text{of $v\hspace{-0.05cm}:\hspace{-0.05cm}P_u= \emptyset$ }}}\big|\bD_u^* \big|.
$$
For the first term we upperbound it by
\begin{align}\label{eq:hehe99}
\Big|\big\{i\in [n]: i\in A_{v,1}\ \text{but}\ i\notin \bA_{v,1}^*\big\}\Big|
+\sum_{\substack{\text{child $u$}\\ \text{of $v\hspace{-0.05cm}:\hspace{-0.05cm}P_u\ne \emptyset$}}}
\Big|\big\{i\in [n]: i\in A_{u,0}\ \text{but}\ i\notin \bA^*_{u,0}\big\}\Big|,
\end{align}
by showing that it is a subset of the union of all sets in \Cref{eq:hehe99}.
To see this is the case, let $i$ be any coordinate in $\bD_u^*$ for some child $u$ of $v$ with $P_u\ne \emptyset$ but not in $D_v$. 
Then by definition we have
  $i\notin \bA^*_{u,0}\cup \bA^*_{u,1}$ but
  $i\in A_{v,0}\cup A_{v,1}$.
If $i\in A_{v,1}$, then it is in the first set because $\bA_{v,1}^*\subseteq \bA_{u,1}^*$;
If $i\in A_{v,0}$, then it is also in $A_{u,0}$ given that $A_{v,0}\subseteq A_{u,0}$.
So $i$ is in one of the sets in the sum.

As a result, it suffices to upperbound each of the  following three sums by $o(n)$:
\begin{align}\label{eq:sums}
 \sum_{\substack{\text{level $t$}\\ \text{node $v\hspace{-0.05cm}:\hspace{-0.05cm}P_v\ne \emptyset$}}} \big|A_{v,1}\setminus \bA^*_{v,1}\big|; \quad 
 \sum_{\substack{\text{level $t+1$}\\ \text{node $u\hspace{-0.05cm}:\hspace{-0.05cm}P_u\ne \emptyset$}}} \big|A_{u,0}\setminus \bA^*_{u,0}\big|;
 \quad\text{and}\quad\sum_{\substack{\text{level $t+1$}\\ \text{node $u\hspace{-0.05cm}:\hspace{-0.05cm}P_u=\emptyset$}}}
\big|\bD_u^*\big|.
\end{align}
(Notice that for the special case when $t=0$ and $\rmO$ is the empty outcome at the root, it suffices to upperbound the last sum, which is covered by the general case considered here.)

In the rest of the proof we show that each of the three
  sums above is $o(n)$ with probability at least $1-o_n(1)$;
  recall that this is over a draw of $\bf\sim \Dyes$, conditioning
  on $\bf$ reaching $\rmO$.


\subsubsection{First and Second Sums in \Cref{eq:sums}}

Let's focus on the second sum; the first sum follows from similar arguments.

For each node $u$ at level $t+1$ with $P_u\ne \emptyset$,
  we have $|A_{u,0}\setminus \bA_{u,0}^*|>0$ only when 
  at least one new query point $x\in S$ is added to $P_{u}$.
When this happens, we can upperbound $|A_{u,0}\setminus \bA_{u,0}^*|$ by
$$
\sum_{x\in \bP_u^*\setminus P_u} \big|\big\{i\in A_{u,0}:
  x_i=1\big\}\big|.
$$
We prove below that with probability at least $1-o_n(1)$: 
\begin{quote}
\emph{Event $E_1$: No point
  $x\in S$ is added to any $P_u$ with $P_u\ne \emptyset$ and}
\begin{equation}\label{eq:hehe88}
\big|\big\{i\in A_{u,0}:
  x_i=1\big\}\big|\ge  100\sqrt{n}\log n.
\end{equation}  
\end{quote}
When this event occurs, we can upperbound the second sum by
$$
\sum_{\substack{\text{level $t+1$}\\ \text{node $u\hspace{-0.05cm}:\hspace{-0.05cm}P_u\ne \emptyset$}}}
\big|\bP_u^*\setminus P_u\big|\cdot 100\sqrt{n}\log n.
$$
Given that each $x\in S$ can only be added to at most two $P_u$'s 
  on level $t+1$, we have 
$$
\sum_{\substack{\text{level $t+1$}\\ \text{node $u\hspace{-0.05cm}:\hspace{-0.05cm}P_u\ne \emptyset$}}} \big|\bP_u^*\setminus P_u\big|\le 2|S|\le 2q.
$$
As a result, the first sum is at most $O(q\sqrt{n}\log n)=o(n)$ with
  probability at least $1-o_n(1)$.

The proof uses arguments similar to the proof of 
  \Cref{lem:usealot}.
To show that the probability of $E_1$ is $1-o_n(1)$, we work on fixed $u$ and $x\in S$ satisfying $P_u\ne \emptyset$ and \Cref{eq:hehe88}.
It then follows from \Cref{claim:reuse} that when $\bf$ is drawn from $\Dyes$ conditioning on it reaching $\rmO$, the probability of~$x$ being added to $P_u$ is $o(1/n^{5})$. 
(For this, set $A$ to be $A_{u,0}$, $\Delta$ to be the set on LHS of \Cref{eq:hehe88},~and $R$ to be $R_{(\parent(u),u)}$ but after applying bitwise negation on every string in it\footnote{This is because the edge $(\parent(u),u)$ is labeled with a clause and not a term. }.)
The probability of $E_1$ is $1-o_n(1)$ by applying a union bound over the $|S|\le q$ many $x\in S$ and $O(q)$ many nonempty $P_u$'s.

\subsubsection{Third Sum in \Cref{eq:sums}}

To bound the third sum in \Cref{eq:sums}, we show that the following event occurs with probability at least $1-o_n(1)$, when $\bf\sim \Dyes$ is drawn conditioning on reaching $\rmO$:
\begin{flushleft}\begin{quote}
\emph{Event $E_2$: No two points $x,y\in S$ with \begin{equation}\label{eq:66}
\big|\{i:x_i=y_i=0\}\big|\le (n/2)-100\sqrt{n}\log n 
\end{equation}
are added to $P_u$ of some level-$(t+1)$ node $u$ with $P_u=\emptyset$; equivalently, for any level-$(t+1)$ node $u$ with $P_u=\emptyset$ but $\bP^*_u\ne \emptyset$, every two points $x,y\in \bP_u^*$ satisfy
\begin{equation*}
\big|\{i:x_i=y_i=0\}\big|\ge (n/2)-100\sqrt{n}\log n .
\end{equation*}}
\end{quote}\end{flushleft}
We first show that, assuming $E_2$, the third sum can be bounded by $o(n)$. After this we show that $E_2$ occurs with probability at least $1-o_n(1)$.

To bound the third sum, note that assuming $E_2$, every level-$(t+1)$ node $u$ with $P_u=\emptyset$ has 
$$
|\bD_u^*|\le |\bP_u^*| \cdot 300\sqrt{n}\log n.
$$
To see this, we use the following simple fact from  \cite{belovs2016polynomial}:
\begin{fact}
Let $P$ be a set of points from middle layers in $\{0,1\}^n$ and let $x$ be any point in $P$. Then
\begin{align*}
\big|\{i\in [n]: \exists\hspace{0.02cm}y,z\in P\ \text{such that $y_i\ne z_i$}\big\}\big|
&\le \sum_{y\in P} \big|\big\{i\in [n]: x_i\ne y_i\big\}\big|.
\end{align*}
Moreover, each term on the RHS can be  bounded 
from above by
\begin{align*}
\big|\big\{i: x_i=0\big\}\big|-
\big|\big\{i:x_i=y_i=0\big\}\big|
&+\big|\big\{i:y_i=0\big\}\big|-
\big|\big\{i:x_i=y_i=0\big\}\big|\\[0.8ex]
&\ \ \ \ \ \ \ \ \ \ \ \ \ \le n+2\sqrt{n}-2\big|\big\{i:x_i=y_i=0\big\}\big|.
\end{align*}
\end{fact}

As a result, assuming $E_2$, we have 
$$
\sum_{\substack{\text{level $t+1$}\\ \text{node $u\hspace{-0.05cm}:\hspace{-0.05cm}P_u\ne \emptyset$}}}\big|\bD_{u}^*\big|
\le
\sum_{\substack{\text{level $t+1$}\\ \text{node $u\hspace{-0.05cm}:\hspace{-0.05cm}P_u= \emptyset$}}}
\big|\bP_u^*\big|\cdot 300\sqrt{n}\log n.
$$
It follows that the sum is $o(n)$ using that the sum of $|\bP_u^*|$ is at most $2|S|\le 2q$.

The last piece of the puzzle is to show that event $E_2$ occurs with probability at least $1-o_n(1)$.
To this end, we note that there can be up to $N^{t+1}$ nodes at level $t+1$, which is too many to apply a union bound.
Instead, we work on the following event $E_3$ that would imply $E_2$:
\begin{flushleft}\begin{quote}
\emph{Event $E_3$: No two points $x,y\in S$ that satisfy \Cref{eq:66} are added to $P_u$ of some \emph{frontier} node $u$, where a node $u$ (of any level) is called a frontier node if either (1) it is at level $1$ and has $P_u=\emptyset$; or (2) it is at level $>1$, has $P_u=\emptyset$ and $P_{\parent(u)}\ne \emptyset$. }
\end{quote}\end{flushleft}
We note that $E_3$ implies $E_2$ because if there are two points $x,y\in S$ that satisfy \Cref{eq:66} are added to $P_u$ for some level-$(t+1)$ node $u$ with $P_u=\emptyset$, then either $t=0$ and $u$ is at level $1$ so $u$ is a frontier node, or there is an ancestor node $v$ of $u$ that is a frontier node and $x,y$ are added to $P_v$. 
Note that here we used the property stated in \Cref{fact:simple2}, that whenever a point is added to $P_u$ for some node $u$, it must also be added to $P_v$ of all ancestors $v$ of $u$ as well.

On the one hand, the number of frontier nodes in $\rmO$ can be bounded by $O(qN)$. To see this, we note that for a node to be frontier, either it is on level $1$ (no more than $N$ many) or it must be the child of some node $v$ with $P_v\ne \emptyset$.
But there can be at most $O(q)$ many nonempty $P_v$'s. As a result, the number of frontier nodes is at most $O(qN)$.

On the other hand, we show in the claim below that 
  for any $x,y\in S$ that satisfy \Cref{eq:66}  and any frontier node $v$ in $\rmO$,
  the probability of  $x,y\in \bP^*_v$ is tiny (when $\bf\sim \Dyes$ conditioning on reaching $\rmO$). It then follows by a union bound over the $|S|^2 \cdot O(qN)=O(q^3N)$ triples $(x,y,v)$ that $E_3$ occurs with probability at least $1-o_n(1)$:

\begin{claim}
Fix $x,y\in S$ that satisfy \Cref{eq:66} and any frontier node $v$ in $\rmO$.
When $\bf\sim \Dyes$ conditioning on $\bf$ reaching $\rmO$, the probability of $x,y\in \bP_v^*$ in $\brmO^*$ is at most $o(1/(n^{10}N))$.
\end{claim}

\begin{proof}
    Note that even though $v$ is a frontier node and satisfies $P_v=\emptyset$, $R_{e}$ with $e=(\parent(v),v)$ is not necessarily empty (though we do have $|R_e|\le |Q|=O(q)$).
    Assume without loss of generality that $v$ is an odd-level node so $e$ is labeled with a term $\bT_e$ in $\bf$.
     The case when $v$ is an even-level node follows by similar  arguments.
Before the queries in $S$ are made, all we know about the term $\bT_e$ in the unknown function 
  $\bf$ is that $\bT_e(z)=0$ for all $z\in R_e$.
As a result, conditioning on~$\bf\sim \Dyes$ reaching $\rmO$, 
$\bT_e$ is distributed uniformly among all terms in $\frak T$ that satisfy $\bT_e(z)=0$ for all $z \in R_e$. 
Let $E$ be the event of $\bT(z)=0$ for all $z \in R_e$. We want to upperbound:
    \begin{align*}
        \Pr_{\bT\sim\frak T}\big[\bT(x)=\bT(y)=1 \mid E\big]&=\frac{\Pr_{\bT\sim\frak T}[\bT(x)=\bT(y)=1 \land E]}{\Pr_{\bT\sim \frak T}[E]} 
        \leq \frac{\Pr_{\bT\sim \frak T}[\bT(x)=\bT(y)=1]}{1-\Pr_{\bT\sim \frak T}[\overline{E}]}.
    \end{align*}
Given \Cref{eq:66} and that $x,y$ both come from middle layers, we have 
\begin{align*}
\big|\big\{i:x_i=y_i=1\big\}\big|
&= \big|\big\{i:x_i=1\big\}\big|
-\big|\big\{i:x_i=1,y_i=0\big\}\big|\\[0.5ex]&=
\big|\big\{i:x_i=1\big\}\big|
-\big|\big\{i:y_i=0\big\}\big|+
\big|\big\{i:x_i=y_i=0\big\}\big|
\end{align*}
and thus, is at most $(n/2)-98 \sqrt{n}\log n$.
As a result, the probability in the numerator is at most
$$
\left(\frac{(n/2)-98\sqrt{n}\log n}{n}\right)^{\sqrt{n}}=\frac{1}{N}\cdot \left(1-\frac{196\log n}{\sqrt{n}}\right)^{\sqrt{n}}=o\left(\frac{1}{N \cdot n^{10}}\right).
$$
    So what's an upper bound on the probability of $\overline{E}$? 
Given that $|R_e|\le |Q|=O(q)$, we can apply a union bound on the probability of $\bT\sim \frak T$ not falsifying each $z\in R_e$, 
  which is exponentially small in $\sqrt{n}$ given that every $z$ is in middle layers and thus, 
  the probability of $\overline{E}$ is $o_n(1)$.

This finishes the proof of the claim.
\end{proof}

\section{Conclusion}\label{sec:conclusion}

{Using $(2\ell)$-level Talagrand functions, we proved for any constant $c>0$, there exists a constant $\epsilon_c$ such that any adaptive and two-sided error algorithm to test whether a function $f$ is monotone or $\epsilon_c$-far from monotone must make ${\Omega}(n^{1/2-c})$ queries. 
Together with the $\tilde{O}(\sqrt{n}/\epsilon^2)$ upper bound of \cite{khotminzerSafraOptUpperBound}, our result shows that the following conjecture is true, up to any polynomial factor:


\begin{conjecture}[Conjecture 8.1 in \cite{chen2017beyond}]\label{conj:chen}
Adaptivity does not help for monotonicity testing.
\end{conjecture}

In contrast, adaptivity does help for the closely related problem of unateness testing: \mbox{one-sided} nonadaptive unateness testing requires $\tilde{\Omega}(n)$ queries \cite{chen2017beyond} (which is tight by \cite{CS16}), whereas \cite{ChenWaingartenUnate} gave an adaptive tester with $\tilde{O}(n^{2/3}/\eps^2)$ queries 
(which is also tight by \cite{chen2017beyond}).

The major obstacle for our construction to establish a tight $\tilde{\Omega}(\sqrt{n})$ lower bound is that in our $(2\ell)$-level construction, the probability over the draw of a random multiplexer tree $\bM$ that a point $x$ in middle layers has a unique activation path down to a leaf (and thus $\Gamma_{\bM}(x) \not \in \{0^*,1^*\}$) decays exponentially with $\ell$. 
As such, when drawing a function $\bf \sim \Dno$ with hidden variable $\bs$, with high probability only a $2^{-\Omega(\ell)}$-fraction of the edges $(x,x^{\{\bs\}})$ form a violation to monotonicity. Can the construction be adapted so that the distance to monotonicity does not decay exponentially as the number of levels increases? We leave this as an open problem in this work.

\bibliography{the}
\bibliographystyle{alpha}

\appendix

\section{Tightness of \Cref{thm:2l-layer-2s-adaptive-lb,thm:adaptivity-hierarchy}}\label{sec:upper-bound}
In this appendix section, we will give a sketch of an algorithm which demonstrates the tightness of our analysis. Essentially, given the distributions $\Dyes$ and $\Dno$ over $(2\ell)$-level Talagrand functions $f_{M,H}$, there is an algorithm spending $2\ell + 1$ rounds of adaptivity and $O(n^{\frac{1}{2}-\frac{1}{4\ell+2}})$ queries that successfully distinguishes these two distributions with high probability. 

\subsection{A $4$-Round-Adaptive,
$O(n^{\frac{3}{8}})$ Algorithm for Three Levels of Our New Construction}\label{sec:3layers} 
This may seem like a rather weird example to start with, as we now have an odd number of levels. However, this proves to be an effective demonstration of how an algorithm can react to the increase in the number of levels and the secret variable being fixed at the top (in contrast to the distributions described in \cite{chen2017beyond} where a secret variable is picked independently and uniformly at random for every leaf). 

Our algorithm only makes one-sided errors and finds a violation to monotonicity with $\Omega(1)$ probability. Our algorithm will work level by level and then employ the quadratic speedup strategy. Our goal is to find a $w$ which reaches a unique leaf where the function $\bh_{i,j,k}$, at this leaf, is be anti-dictatorship $\overline{x_{\bs}}$ so that $w_{\bs}=0$. The algorithm will then find a point $w' \prec w$ such that $g(w)=0$ but $g(w')=1$ (meaning we flipped $\bs$ in $w$). 

Let $g$ be a function in the support of $\Dno$ with secret variable $\bs$. Without loss of generality, we can assume that we start with a point $x$ with $|x|=n/2$ and $x_{\bs}=0$. Note that $g(x) = 1$.
Furthermore, we assume that $x$ satisfies some term $T_i$ uniquely but doesn't falsify any $C_{i,j}$, as this event happens with constant probability.
\begin{flushleft}\begin{enumerate}
    \item[] \textbf{Round 0:} Similar to Algorithm 7.2 in \cite{chen2017beyond}, we select $n^{3/8}$ random sets $S_1.\cdots, S_{n^{3/8}} \subseteq \{i \in [n] \mid x_i =1\}$ of size $\sqrt{n}$. Let $C_1=\emptyset$ and for each $t\in[n^{3/8}]$, query $g(x^{S_t})$. If the output is $1$, add the elements in $S_t$ to $C_1$. Clearly, such an $S_t$ does not intersect $T_i$, and the total size of $C_1$ is $\Theta(n^{7/8})$ with high probability. 
    \end{enumerate}\end{flushleft}
    We execute the next instructions (Rounds 1 to 4) $n^{1/8}$ times in parallel:
    \begin{enumerate}
    \item[]\textbf{Round 1:} (a) Pick a random random set $C_0 \subseteq \{i \in [n] \mid x_i =0\}$ with $|C_0| = |C_1|$ and query $y:=x^{C_1 \cup C_0}$. Clearly, $|y|=n/2$ and with constant probability we have that $y$ satisfies a unique term $T_i$, falsifies a unique $C_{i,j}$, and does not satisfy any $T_{i,j,k}$ (so $g(y) = 0$).  
    
    
    Note that $\bs$ is included in $C_0$ with probability $\Omega(n^{-1/8})$. Hence, by repeating Rounds $1$ to $3$ $n^{1/8}$ times in parallel we can ensure that $\bs \in C_0$ happens with high probability.  See \Cref{fig:ALGO1} for an illustration.
    \item[] (b) Let $D_1=\emptyset$ and repeat the following $n^{1/4}$ times: Pick a random subset $R \subseteq C_1$ of size $\sqrt n$. Query $g(y^{R})$; if $g(y^R)=0$ (in which case $R$ doesn't intersect with $C_{i,j}$), add the coordinates in $R$ to $D_1$. With high probability, we have $|D_1|=\Theta(n^{3/4})$. Note that $D_1$ doesn't intersect with $T_i$ nor $C_{i,j}$.

    \begin{figure}[t]
  \centering
  \includegraphics[width=0.8\linewidth]{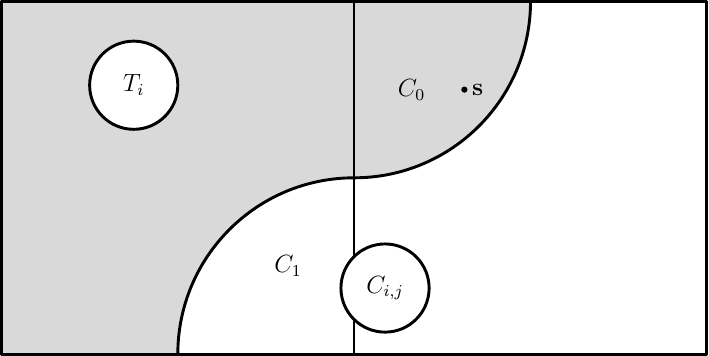}
  \caption{A diagram of the knowledge of the algorithm for the set $[n]$ by the end of Round 1 (a). The whole rectangle represents $[n]$, and the shaded areas (not including $T_i$ or $C_{i,j}$) are the $1$-coordinates. The set $C_0$ contains the anti dictator variable $\mathbf{s}$ is located. It has size $\Theta(n^{7/8})$ and it is disjoint from  $T_i$ and $C_{i,j}$. } 
  \label{fig:ALGO1}
\end{figure}
\end{enumerate}
\noindent
For each execution of Round $1$ we repeat the next instructions (Rounds 2,3 and 4) $n^{1/8}$ times in parallel:
\begin{flushleft}\begin{enumerate}
    \item[]\textbf{Round 2:} (a) Let $D_0$ be a random subset of $C_0$ such that $|D_1| = |D_0|$, let $z=y^{D_1 \cup D_0}$ (so $|z|=n/2$) and query $g(z)$. With constant probability, $z$ satisfies $T_i$ uniquely, falsifies $C_{i,j}$ uniquely, $z$ satisfies a unique term $T_{i,j,k}$, and $h_{i,j,k}$ is the anti-dictatorship $\overline{x_{\bs}}$ meaning $g(z)=1$ (since $D_0$ are $0$'s of $z$). Assuming $\bs \in C_0$, with probability $\Omega(n^{-1/8})$, $\bs\in D_0$, so by repeating Rounds 2 and 3 for $n^{1/8}$ times, we are guaranteed that, with high probability, $\bs\in D_0$ for one of the parallel repetitions. See \Cref{fig:ALGO2} for an illustration.

    (b) Furthermore, let $G_1=\emptyset$ we repeat the following $n^{1/8}$ times:  Pick a random subset $R \subseteq D_1$ of size $\sqrt n$ in $C_1$. Query $g(z^{R})$; if $g(z^{R})=1$ (in which case we know $R$ can't intersect with $T_{i,j,k}$), add the coordinates in $R$ to $G_1$. With high probability we have $|G_1|=\Theta(n^{5/8})$. Note that, we have that $G_1$ is disjoint from $T_{i}, C_{i,j}, T_{i,j,k}$. \footnote{Note that here steps (a) and (b) can all be done in the same round of queries: we don't need to know the outcome of the query in (a) to do (b). }
\end{enumerate}\end{flushleft}
\begin{figure}[t]
  \centering
  \includegraphics[width=0.8\linewidth]{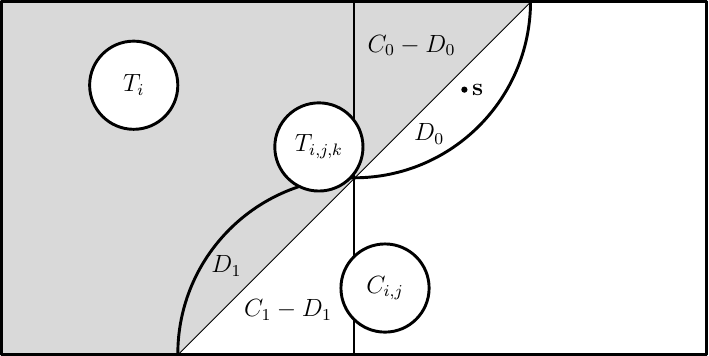}
  \caption{A diagram of the knowledge of the algorithm for the set $[n]$ by the end of Round 2 (a). The whole rectangle represents $[n]$, and the shaded areas (not including $T_i, C_{i,j}$ or $T_{i,j,k}$) are the $1$-coordinates. The set $D_0$, which contains the anti dictator variable $\mathbf{s}$ is located. It has size $\Theta(n^{3/4})$ and it is disjoint from  $T_i, C_{i,j}$ and $T_{i,j,k}$. } 
  \label{fig:ALGO2}
\end{figure}
At this stage, the reader should think of $D_0$ as a set of size $\Theta(n^{3/4})$ containing $\bs$, and $G_1$ as a set of size $\Theta(n^{5/8})$ \footnote{With high probability, this happens in one of of the parallel repetition. In this case, we can get a violation to monotonicity with high probability. }. Furthermore neither of these set intersect with $T_i$, $C_{ij}$ or $T_{i,j,k}$. Note that coordinates in $D_0$ are $1$'s of $z$ while coordinates in $G_1$ are $0$'s of $z$. Since $\bs \in D_0$ we have $g(z)=1$. 
\begin{enumerate}
    \item[]\textbf{Round 3:} Randomly partition $D_0$ into  $n^{1/8}$ sets $\Delta_1, \ldots, \Delta_{n^{1/8}}$ each of size $n^{5/8}$. For each such $\Delta_t$ query the point $w^{(t)}:=z^{G_1 \cup \Delta_{t}}$. 
    
   We will need the following: Observe that if $\bs \not \in \Delta_t$, then $g(w^{(t)})$ must be equal to $1$. Indeed, $\Delta_t,G_1$ do not intersect $T_i,C_{i,j}$ nor $T_{i,j,k}$, so $g(w^{(t)})$ can't become $0$ (maybe $w^{(t)}$ satisfies some new terms or clauses but this can't change the value of $g(w^{(t)})$ to $0$). However, if $\bs \in \Delta_t$, then $g(w^{(i)})=0$ as long as $w^{(t)}$ uniquely satisfies $T_i$, uniquely falsifies $C_{i,j}$ and uniquely satisfies $T_{i,j,k}$ (which happens with constant probability).   

    \item[]\textbf{Round 4:} If in Round 3 we had a unique $t$ with $g(w^{(t)})=0$, let $w=w^{(t)}$; otherwise, skip this this round. First, randomly partition $\Delta_t$ into $n^{1/8}$ sets $F_1, \ldots, F_{n^{1/8}}$ each of size $\sqrt{n}$. Since $|w|=n/2$, we have that $w^{F_j}$ is in the middle layers for each $j \in [n^{1/8}]$. Hence, for each $j$, we query $g(w^{F_j})$. Note that if $\bs \in \Delta_t$, then, when $\bs \in F_j$, we have $g(w^{F_j})=1$ with constant probability. If this happens, we've found $w^{F_j} \prec w$ but $1=g(w^{F_j})>g(w)=0$. 


\end{enumerate}

\subsection{A General $(2\ell+1)$-Round-Adaptive Algorithm for $2\ell$ Levels of Our New Construction}

The algorithm from the previous subsection can easily be generalized into one that works against the $(2\ell)$-level Talagrand construction we gave in \Cref{sec:new-construction}. As before, let $g$ be a function in the support of $\Dno$ with the secret variable $\bs \sim [n]$. 
We will proceed to ``conquer'' the layers inductively using $2\ell$ rounds of queries, after which we will use two rounds to find a violation to monotonicity. We sketch an algorithm which only makes one-sided errors, and finds a violation to monotonicity with probability $\Omega(1)$. We will consider the following even $E_j$ where $j\in [2\ell]$: after we've just performed round $j-1$, we have a point $x^{(j)}$ and sets $C_1^{(j)}$ and $C_0^{(j)}$ with the following properties:
\begin{itemize}
    \item $|x^{(j)}|=n/2$ and $x^{(j)}$ uniquely satisfies (resp. falsifies) a term (resp. clause) at each level $k \leq j$ but nothing in the next level.
    \item $\forall i \in C_1^{(j)}$ we have $x^{(j)}_i = 1$ and $\forall i \in C_0^{(j)}$ we have $x^{(j)}_i = 0$. Furthermore, the sets $C_1^{(j)}$ and $C_0^{(j)}$ do not intersect any of the terms (resp. clauses) that $x^{(j)}$ has satisfied (resp. falsified) uniquely so far. 
    \item $|C_{1}^{(j)}|=\Theta\left(n^{1-\frac{j}{4\ell+2}}\right)$, $|C_{0}^{(j)}|=\Theta\left(n^{1-\frac{j-1}{4\ell+2}}\right)$ and $\bs \in C_0^{(j)}$.
    \item For $j<2\ell-1$, we have that $g(x^{(j)})=j \mod 2$. If $j=2\ell$, then $g(x^{(j)})=1$ (this is because at this point $x^{(2\ell)}$ is at a leaf). 
\end{itemize} 

We will show that $E_1$ happens with $\Omega(1)$ probability for $j=1$ (round $0$). We then show that conditioned on $E_j$ happening, $E_{j+1}$ happens during round $j$ with high probability in one of the parallel repetitions. Finally, assuming $E_{2\ell}$ happened, we will use round $2\ell$ and $2\ell+1$ to find violation to monotonicity with high probability.

\medskip
We proceed with the base case:
Without loss of generality, we can assume that we start with a point $x^{(1)}$ with $\left|x^{(1)}\right|=n/2$ and $x_{\bs}={0}$. Furthermore, we assume that $x^{(1)}$ satisfies some term $T_{i_1}$ uniquely, but does not falsify any $C_{i_1,i_2}$, as this event happens with constant probability. We have $g(x) = 1$. Let $C_0^{(1)}=\{i \mid x^{(1)}_i =0\}$. 

\begin{enumerate}
    \item[]\textbf{Round 0:} 
    Similar to  \Cref{sec:3layers}, we select $t:=n^{\frac{1}{2}- \frac{1}{4\ell+2}}$ random sets $S_1.\cdots, S_{t} \subseteq \{i \in [n] \mid x^{(1)}_i =1\}$ of size $\sqrt{n}$. For each $i\in[t]$, query $g\left(\left(x^{(1)}\right)^{S_i}\right)$. If the output is $1$, add the elements in $S_i$ to $C_1^{(1)}$. It is clear that such a set $S_i$ does not intersect $T_{i_1}$, and the total size of $C_1^{(1)}$ is $\Theta\left(n^{1- \frac{1}{4\ell+2}}\right)$ with high probability by a Chernoff bound. 
\end{enumerate}

\noindent
It is clear that by the end of round $0$, the event $E_1$ happened with constant probability. 

\medskip
\noindent
For $1 \leq j <2\ell$, for each execution of round $j-1$ we execute round $j$ $n^{\frac{1}{4\ell+2}}$ times in parallel \footnote{This means round $j$ is executed $n^{\frac{j}{4\ell+2}}$ times in parallel.}. In the description of Round $j$ bellow, we assume $E_{j}$ happened to argue $E_{j+1}$ happens with high probability in one of the repetitions. However the queries do not depend on whether $E_j$ happened or not.
\begin{enumerate}
\item[]\textbf{Round j:} (a) Pick a random random set $C_{0}^{(j+1)} \subseteq C_0^{(j)}$ with $|C_{0}^{(j+1)}| = |C_{1}^{(j)}|$ and query $x^{(j+1)}:=(x^{(j)})^{C_1 \cup C_0}$. By our assumption on $x^{(j)}$, we have $|x^{(j+1)}|=n/2$. With constant probability, $x^{(j+1)}$ uniquely satisfies (resp. falsifies) a term (resp. clause) at each level $k \leq j+1$ but nothing in the next level. Hence, we have $g(x^{(j+1)})=0$ when $j+1$ is even, and $1$ when $j+1$ is odd. However, when $j+1=2\ell$, then we are at a leaf $u \in [N]^{2\ell}$, and if $\bs \in C_0^{(2\ell)}$, then $g(x^{(2\ell)})=1$ with probability $1/2$ (if $h_{u}$ is the anti-dictatorship $\overline{x}_{\bs}$).

Observe that assuming $\bs \in C_0^{(j)}$ we have $\bs \in C_0^{(j+1)}$ with probability $n^{-\frac{1}{4\ell+2}}$, which is why we repeat this round $n^{\frac{1}{4\ell+2}}$ times.

    \item[] (b) Let $C_1^{(j+1)}=\emptyset$ and repeat the following $n^{\frac{1}{2}-\frac{j+1}{4\ell+2}}$ times: Pick a random subset $R \subseteq C_1^{(j)}$ of size $\sqrt n$. Let $y=(x^{(j+1)})^{R}$. If $g(y)=(j+1)\mod 2$ (in which case $R$ doesn't intersect with the term or clause $x^{(j+1)}$ uniquely satisfies at level $j+1$) add the coordinates in $R$ to $C_1^{(j+1)}$. With high probability we have $|C_1^{(j+1)}|=\Theta\left(n^{1-\frac{j+1}{4\ell+2}}\right)$. Note that $C_1^{(j)}$ doesn't intersect with terms or clauses $x^{(j+1)}$ satisfies. 
\end{enumerate}
\noindent
From the above, it's easy to see that if $E_{j}$ happened in round $j-1$, then with constant probability $E_{j+1}$ happens during round $j$ in one of the parallel repetitions.

\medskip
\noindent
Rounds $2\ell$ and $2\ell+1$ are executed once for each parallel execution of round $2\ell-1$. 
\begin{enumerate}
    \item[] \textbf{Round 2$\boldsymbol{\ell}$:} We assume for simplicity that the event $E_{2\ell}$ happened and we have a point $x^{(2\ell)}$ and sets $C_0^{(2\ell)}, C_1^{(2\ell)}$ respecting the constraints of the event\footnote{Otherwise, there is no guarantee rounds $2\ell$ and $2\ell+1$ find a violation to monotonicity. }. Ignoring the hidden constants in $\Theta$, we assume for simplicity that $|C_0^{(2\ell)}|=n^{1-\frac{2\ell-1}{4\ell+2}}$, $|C_1^{(2\ell)}|=n^{1-\frac{2\ell}{4\ell+2}}$. We proceed as in Round 3 of \Cref{sec:3layers}. Randomly partition $C_0^{(2\ell)}$ into  $n^{\frac{1}{4\ell+2}}$ sets $\Delta_1, \ldots, \Delta_{n^{\frac{1}{4\ell+2}}}$ each of size $n^{1-\frac{2\ell}{4\ell+2}}$. For each such $\Delta_t$ query the point $$w^{(t)}:=\left(x^{(2\ell)}\right)^{C^{(2\ell)}_1 \cup \Delta_{t}}.$$
    

    {Recall that $g(x^{(2\ell)})=1$ and $C_0^{(2\ell)}$ contains the hidden variable $\bs$. So, if $\bs \not \in \Delta_t$, then $g(w^{(t)})$ must be equal to $1$. Indeed, $\Delta_t,C^{2\ell}_1$ do not intersect any of the terms nor clauses $g(w^{(t)})$ satisfies so $g(w^{(t)})$ can't become $0$ (maybe $w^{(t)}$ satisfies some new terms or clauses at different levels but can't change the value of $g(w^{(t)})$ to $0$). However if $\bs \in \Delta_t$, then $g(w^{(t)})=0$ as long as $w^{(t)}$ uniquely satisfies (resp falsifies) the same terms (resp clauses) that $x^{(2\ell)}$ does (which happens with constant probability). }

    \item[]\textbf{Round } \textbf{2$\boldsymbol{\ell}+$1:}
    If in Round $2\ell$ we had a unique $t$ with $g(w^{(t)})=0$, let $w=w^{(t)}$, otherwise skip this this round. First, randomly partition $\Delta_t$ into $n^{\frac{1}{4\ell+2}}$ sets $F_1, \ldots, F_{n^{\frac{1}{4\ell+2}}}$ each of size $\sqrt{n}$. Since $|w|=n/2$, we have that $w^{F_j}$ is in the middle layers for each $j \in \left[n^{\frac{1}{4\ell+2}}\right]$. Hence, for each $j$, we query $g(w^{F_j})$. Note that if $\bs \in \Delta_t$, then when $\bs \in F_j$ we have $g(w^{F_j})=1$ with constant probability. If this happens, we've found $w^{F_j} \prec w$ but $1=g(w^{F_j})>g(w)=0$. 
    
\end{enumerate}

By induction and using the fact $\ell$ is constant, it's easy to see that with probability $\Omega(1)$ the event $E_{2\ell}$ holds for one of the parallel repetitions of round $2\ell-1$. In this case, during round $2\ell,2\ell+1$ our algorithm will find a violation to monotonicity in $g$ with probability $\Omega(1)$. In the above for $j \leq 2\ell-1$, round $j$ is executed in parallel for $n^{\frac{j}{4\ell+2}}$ times, and round $j$ uses $n^{1/2-\frac{j+1}{4\ell+2}}$ queries. Round $2\ell$ and $2\ell+1$ use $n^{\frac{1}{4\ell+2}}$ queries in each parallel repetition but are executed only $n^{\frac{2\ell-1}{4\ell+2}}$ times. So, each round uses $O(n^{\frac{1}{2}-\frac{1}{4\ell+2}})$ queries. Since $\ell$ is a constant, the total number of queries is also $O(n^{\frac{1}{2}-\frac{1}{4\ell+2}})$. 

\subsection{A $3$-Round-Adaptive, $\tilde{O}(n^{1/3})$ Algorithm for the Constant-Level Generalization of \cite{chen2017beyond}}
\label{ssec:algo-can-skip-level-cwx17}

As mentioned in the overview, the naive generalization of \cite{chen2017beyond} to more levels\footnote{Instead of two-levels, we could add more levels by alternating terms and clauses and for each leaf $u$ we sample a secret variable $\bs_u \sim [n]$ independently and uniformly at random. In the $\Dyes$ distribution, we use the dictatorship function $x_{\bs_u}$ at the leaf, and in the $\Dno$ distribution we use the anti-dictatorship function $\overline{x_{\bs_u}}$}, doesn't yield stronger lower bounds then $\tilde{\Omega}(n^{1/3})$ for monotonicity testing. Indeed, unlike in the previous two subsections, an algorithm doesn't need to work level by level to find a violation to monotonicity. 

We sketch an algorithm making $O(n^{1/3})$ queries for the extension of the Talagrand construction of \cite{chen2017beyond} to a constant number of levels. For simplicity, consider \(g\sim \mathcal{D}_{no}\) based on the three-level Talagrand construction (but note that the approach sketched below works for any \(O(1)\)-level construction). The key idea is that the algorithm jumps directly to the penultimate level. From there, the algorithm works the same way as it does for the two-level construction. 

Let $g$ be a function in the support of $\Dno$. Without loss of generality, we can assume that we start with a point $x$ with $|x|=n/2$ such that $x$ satisfies some term $ T_i $ uniquely and falsifies some clause $C_{i,j}$ uniquely, but satisfies no term $T_{i,j,k}$, as this event occurs with constant probability. We denote $A_0:=\{i \in [n] \mid x_i=0\},A_1:=\{i \in [n] \mid x_i=1\}$. We have that $g(x) = 0$. 
\begin{enumerate}
    \item[]\textbf{Round 0:} 
   We select $n^{1/3}$ random sets $S_1.\cdots, S_{n^{1/3}} \subseteq A_0$ of size $\sqrt{n}$. 
   Let $C_0=\emptyset$ and for each $t\in[n^{1/3}]$ query $g(x^{S_t})$. If the output is $0$, we add the elements of $S_t$ to $C_0$.
   
   The total size of $C_0$ is $\Theta(n^{5/6})$ with high probability. It is clear that $C_0$ does not intersect $T_i$ since $C_0 \subseteq A_0$. Furthermore, $g(x^{S_t})$ can be equal to $0$ only if $x^{S_t}$ falsifies $C_{i,j}$. Hence we can see that coordinates in $C_0$ do not appear in $C_{ij}$. 
\end{enumerate}
\noindent
We repeat the following \(n^{1/6}\) (Rounds 1 to 3) times in parallel:
\begin{enumerate}
    \item[] \textbf{Round 1:} (a) Pick a random set $R \subseteq A_0 \setminus C_0$ of size $\sqrt{n}$ and query $y:=x^{R}$.  With constant probability, $y$ satisfies uniquely the term $T_i$, falsifies uniquely  the clause $C_{i,j}$ and satisfies a unique term $T_{i,j,k}$.  Note that by this construction, $T_{i,j,k}$ and $C_0$ are disjoint.  With \(\Theta(n^{-1/6})\) probability, the hidden variable \(\bs_{i,j,k} \in C_0\) (hidden at the leaf corresponding to $T_{i,j,k}$), in which case $g(y)=1$. 
    
    \medskip
    
    (b) Let $C=\emptyset$ and repeat the following $n^{1/6}$ times: Pick a random subset $R \subseteq A_1$ of size $\sqrt n$. Query $g(y^{R})$\footnote{Since, $|y| = \frac{n}{2}+\sqrt n$, and $A_1 \subseteq \{i \in [n] \mid y_i=1\}$ we can flip a $\sqrt n$ coordinates $R$ in $A_1$  every time without getting out of the middle layers.}, if $g(y^R)=1$ (in which case $R$ doesn't intersect with $T_i, T_{i,j,k}$) add the coordinates in $R$ to $C$. With high probability we have $|C|=\Theta(n^{2/3})$. Note that $C$ doesn't intersect with $T_i, C_{i,j}$ nor $T_{i,j,k}$.

\end{enumerate}    

Assuming $\bs_{i,j,k} \in C_0$ and $|C|=\Theta(n^{2/3})$ we can now find a violation efficiently.  

\begin{enumerate}
    \item[] \textbf{Round 2:} Randomly partition $C_0$ into \(n^{1/6}\) subsets \(\Delta_1, \dots, \Delta_{n^{1/6}}\) of size \(n^{2/3}\). For each such $\Delta_t$ query the point $w^{(t)}:=y^{C \cup \Delta_{t}}$.

    Recall that by assumption we have that $g(y)=1$, the hidden variable $\bs_{i,j,k}$ is in $C_0$ and at the corresponding leaf we use the anti-dictatorship function $\overline{x_{\bs_{i,j,k}}}$ . Since $\Delta_t$ and $C$ do not intersect $T_i,C_{i,j}$ nor $T_{i,j,k}$, $g(w^{(t)})$ can't be equal to $0$ (maybe $w^{(t)}$ satisfies some new terms or clauses but this can't change the value of $g(w^{(t)})$ to $0$). However if $\bs_{i,j,k} \in \Delta_t$, then $g(w^{(t)})=0$ as long as $w^{(t)}$ uniquely satisfies $T_i$, uniquely falsifies $C_{i,j}$ and uniquely satisfies $T_{i,j,k}$ (which happens with constant probability).

    \item[] \textbf{Round 3:} 
    If in Round 3 we had a unique $t$ with $g(w^{(t)})=0$, let $w=w^{(t)}$, otherwise skip this this round. First, randomly partition $\Delta_t$ into $n^{1/3}$ sets $F_1, \ldots, F_{n^{1/6}}$ each of size $\sqrt{n}$. As $|w|=n/2 + \sqrt{n}$ and $F_1$ consists of 1 bits of $wx$, we still have that $w^{F_j}$ is in the middle layers for each $j \in [n^{1/3}]$. Hence, for each $j$, we query $g(w^{F_j})$. Note that if $\bs \in \Delta_t$, then when $\bs \in F_j$ we have $g(w^{F_j})=1$ with constant probability. If this happens, we've found $w^{F_j} \prec w$ but $1=g(w^{F_j})>g(w)=0$.
\end{enumerate}

\input{relative-error-appendix}

\end{document}

%% file: relative-error-appendix.tex
\section{Relative-Error Monotonicity and Unateness Testing}\label{sec:relativeerrorsec}
\newcommand{\reldist}{\mathsf{reldist}}

\newcommand{\MQ}{\mathsf{MQ}}

\newcommand{\SAMP}{\mathsf{SAMP}}


\subsection{Background on Relative-Error Testing}\label{sec:relativebackground}

In this appendix, we are interested in the relative-error model of property testing, which was introduced by Chen et al.~\cite{chen2025relative}. The motivation for this new model came from the observation that the standard testing framework is not well suited for testing \emph{sparse} Boolean functions (i.e. functions with $|f^{-1}(1)| \leq p2^{n}$ where $p$ is very small\footnote{For instance imagine a setting where $p=2^{-n/2}$.}) since any such function is $p$ close to the constant-0 function. 
To circumvent this, in the \emph{relative-error} Boolean function property testing model introduced by \cite{chen2025relative} the distance between the function $f: \zo^n \to \zo$ that is being tested and a function $g: \zo^n \to \zo$ is defined to be
\begin{equation}\label{eq:reldist}
\reldist(f,g) := {\frac {|f^{-1}(1) \hspace{0.05cm} \triangle \hspace{0.05cm} g^{-1}(1)|} {|f^{-1}(1)|}}.
\end{equation}
Hence relative distance is measured ``at the scale'' of the function $f$ that is being tested, i.e.~$|f^{-1}(1)|$, rather than at the ``absolute scale'' of $2^n =$ $ |\zo^n|$ that is used in the standard model. Note that if only black-box membership queries to $f$ were allowed, it would take a tester an enormous amount of queries to find a point $x \in \zo^n$ with $f(x)=1$ when $f$ is very sparse. As such, the model also allows the testing algorithm to obtain i.i.d.~uniform elements of $f^{-1}(1)$ by calling a ``random sample'' oracle.  See \Cref{sec:rel-error} for a more detailed description of the relative-error model.
The main result we prove in this appendix is \Cref{thm:relativeerror}, which we restate below for convenience. 
\relativeerror*

\subsubsection{Previous Results on Monotonicity and Unateness Testing}
The work of \cite{chen2025relative} was interested in the relative-error testing of monotone functions. 
The main positive result of \cite{chen2025relative} was a one-sided algorithm which is an $\eps$-relative-error tester for monotonicity, and with high probability makes at most $O(\log(|f^{-1}(1)|/\eps)$ queries, even when the value of $|f^{-1}(1)|$ is not known to the testing algorithm. More recently, \cite{chen2025relativeerrorunatenesstesting} showed that there exists a tester for relative-error unateness testingw using  $\tilde{O}(\log(|f^{-1}(1)|/\eps)$ queries with high probability.

On the lower bound side, \cite{chen2025relative} proved the following result: For any constant $\alpha<1$, there exists a constant $\epsilon>0$ such that any (adaptive) algorithm for testing whether a boolean function $f$ with $|f^{-1}(1)|=\Theta(N)$, where $N \leq 2^{\alpha n}$ needs at least $\tilde{\Omega}(\log( N )^{2/3})$ queries. \cite{chen2025relativeerrorunatenesstesting} observed that the same lower bound applies for unateness testing in the relative-error model. 

In particular, it remained open whether adaptivity can help for monotonicity testing. Furthermore, it is known that unateness testing is harder than monotonicity testing. As mentioned before, \cite{khotminzerSafraOptUpperBound} gave a $\tilde{O}(\sqrt{n}/\epsilon^2)$ upper bound for (non-adaptive) monotonicity testing.  \cite{chen2017beyond} gave a $\Omega(n^{2/3})$ lower bound for adaptive unateness testing, while \cite{ChenWaingartenUnate} gave an (almost) matching upper bound of $\tilde{O}(n^{2/3}/\epsilon^2)$. However, given the upper bounds of \cite{chen2025relative} and \cite{chen2025relativeerrorunatenesstesting} it could very well be that be that in the relative-error model, testing unateness and monotonicity are (almost) equally hard.

\subsubsection{Proof Overview of \Cref{thm:relativeerror}}\label{sec:overview-rel-error}

The ideas behind the lower bound of \cite{chen2025relative} were inspired from the two-level Talagrand functions of \cite{chen2017beyond} (see \Cref{sec:two-level-Talagrand}). Instead of working with the ``middle layers'' the authors introduced ``two-layer functions'' which are functions such that $f(x)=0$ if $|x| <3n/4$ and $f(x)=1$ if $|x|>3n/4+1$. To accommodate the fact we now work with points of ``high'' Hamming weight, \cite{chen2025relative} use a construction similar to two-level Talagrand function but with the following differences: There are $N^{(1)}:=(4/3)^n$ terms $\bT_1, \ldots, \bT_{N^{(1)}}$ on the first level, and for each $i \in \left[N^{1}\right]$ we have $N^{(0)}:=4^n$ clauses $\bC_{i,1}, \ldots, \bC_{i,N^{(0)}}$. Furthermore, the terms and clauses have size $n$ instead of $\sqrt{n}$.

Our idea for the proof of \Cref{thm:relativeerror} follows similarly by adapting our multilevel Talagrand construction to two-layer functions. To draw a function $\bf \sim \Dyes$ or $\bf \sim \Dno$ we will again draw a multiplexer tree $\bM$ and a tuple of function $\bH$. However, the multiplexer tree $\bM$ is drawn slightly differently: We first draw depth $2\ell$ tree where nodes at even depth have $N^{(1)}$ children, and nodes at odd depth have $N^{(0)}$ children. Similarly to \cite{chen2025relative}, the terms and clauses labeling the edges are drawn from $\frak T_{n,n}$ and $\frak C_{n,n}$ (and thus have size $n$ rather than $\sqrt{n}$). 

The proof of \Cref{thm:relativeerror} is similar to that of \Cref{thm:2l-layer-2s-adaptive-lb}. In particular, a reason \cite{chen2025relative} worked with two-level functions is that one can assume the testing algorithm only uses black box queries to $f$ and never queries the oracle returning i.i.d. uniform samples from $f^{-1}(1)$. To prove both the unateness and monotonicity lower bound, we argue that functions drawn from $\Dyes$ are monotone (and thus unate) while function drawn from $\Dno$ are far from unateness in relative distance (and thus from monotonicity). Then by Yao's minimax principle, it suffices to show that no deterministic $q$-query algorithm can distinguish the distribution $\Dyes$ and $\Dno$.


\newcommand{\frakN}{N}
\subsection{Preliminaries}\label{sec:rel-error} 
 In this appendix, we always assume that $\ell$ is a positive integer constant, that $n$ is divisible by $4$ and $(4/3)^n$ is an integer. We will reuse the notation and definitions from \Cref{sec:preliminaries} with the exception that we now use $N:={n \choose 3n/4}$\footnote{We make this choice to stay consistent with the notation used in the previous works of \cite{chen2025relative,chen2025relativeerrorunatenesstesting}}. We write $\frak T'$ for $\frak T_{n,n}$ and $\frak C'$ for $\frak C_{n,n}$ for convenience.  We let $N^{(1)}:=(4/3)^{n}$, $N^{(0)}:=4^n$. We let $L^{(0)}=\{\varepsilon\}$ and given $k \geq 1$ we let $$L^{(k)}:=\left\{ (u_1, \ldots, u_k) \mid u_i \in \left[N^{(i \text{ mod } 2)}\right]\right\}.$$ 


We recall the definition of unateness. 
\begin{definition}
    A function $f:\zo^n \to \zo$ is unate, if there exists $a \in \zo^n$ such that the function $h(x)=f(x \oplus a)$ is monotone (where $x \oplus a$ denotes the bitwise XOR). 
\end{definition}
\newcommand{\Edge}{\mathsf{Edges}}

We have the following easy result:
\begin{lemma}\label{lem:dist_to_unate}
Let $f:\zo^n \to \zo$. For $i \in [n]$ let $\Edge_i^{1}:=\{(x,x^{(i)}) \mid x_i=0, f(x)=0, f(x^{(i)})=1\}$ be the set of strictly monotone edges along direction $i$ and $\Edge_i^{0}:=\{(x,x^{(i)} \mid x_i=0, f(x)=1, f(x^{(i)})=0\}$ be the set of strictly anti-monotone edges along direction $i$. 

We have that:

$$\reldist(f,\textsf{unate})\geq \min\left(|\Edge_i^{1}|,|\Edge_i^{0}|\right)/|f^{-1}(1)|.$$
\end{lemma}
\begin{proof}
    For $b \in \zo$ edges in $\Edge_i^{b}$ are all disjoint.  To make $f$ unate, we either need to make all edges in $\Edge_i^{1}$ anti-monotone, or all edges in $\Edge_i^{0}$ monotone. Hence, we need to change at least the value of $f$ on at least $\min\left(|\Edge_i^{1}|,|\Edge_i^{0}|\right)$ points to make it unate. 
\end{proof}

\subsubsection{The Relative-Error Model}\label{sec:formalmodel}
As defined in \cite{chen2025relative}, a \emph{relative-error} testing algorithm for a class ${\cal C}$ of Boolean functions has oracle access to $\MQ(f)$ (membership queries) and also has access to a $\SAMP(f)$ oracle which, when called, returns a uniformly random element $\bx \sim f^{-1}(1)$.

A relative-error testing algorithm for a class ${\cal C}$ must output ``yes'' with high probability (say at least 2/3) if $f \in {\cal C}$ and must output ``no'' with high probability (say at least 2/3) if $\reldist(f,{\cal C}) \geq \eps$, where
$$\reldist(f,{\cal C}):=\min_{g \in {\cal C}}\hspace{0.05cm}\reldist(f,g) \text{ and $\reldist(f,g)$ is defined in \Cref{eq:reldist}} .
$$

We say that a relative-error testing algorithm is ``non-adaptive'' if after receiving the results of all of its calls to the sampling oracle $\SAMP(f)$, it makes one parallel round of queries to the black-box oracle $\MQ(f)$.

As in the standard model, the tester is called ``one-sided'' if it always accepts when the function $f$ is in $\mathcal{C}$, otherwise it is called ``two-sided''. 

\subsection{Sandwiched-Multilevel Talagrand Functions}
In this section, we revisit the construction of \emph{multilevel Talagrand functions} from \Cref{sec:MTF}, to introduce \emph{sandwiched-multilevel} Talagrand functions. As before, we will use these to obtain the two distributions of functions, $\Dyes$ and $\Dno$,
  which will be used to prove our lower bounds for monotonicity and unateness testing. Indeed, functions in $\Dyes$ will always be monotone while functions in $\Dno$ will be far from unateness with high probability. 
  
\subsubsection{Two-Layer Functions}
Recall that $N={n \choose 3n/4}$. We say a point $x \in \zo^n$ is in \emph{sandwich}-layers if $3n/4 \leq |x| \leq 3n/4+1$. We will also say that $f:\zo^n \to \zo$ \emph{two-layer function} if:
$$f(x)=\begin{cases}
    1 \text{ if } ||x||_1 > 3n/4 + 1 \\
    0 \text{ if } ||x||_1 < 3n/4 
\end{cases}$$
All functions used in the lower bound proof in this appendix will be two-layer functions. {In particular, note that for any two-layer function $f$, we have $|f^{-1}(1)|=\Theta(N)$}. To prove \Cref{thm:relativeerror}, we will prove the following:

\begin{theorem}\label{thm:rel_error_main}
    Let $N={n \choose 3n/4}$. For all $c>0$, there is a constant $\epsilon_c  >0$ such that any two-sided,  adaptive algorithm for testing whether an unknown Boolean function $f: \{0,1\}^n \to \{0,1\}$ with $|f^{-1}(1)|=\Theta(N)$ is monotone (unate) or $\epsilon_{c}$-far from monotone (unate) in relative distance must make $\tilde{\Omega}({n^{1- c}})$. 
\end{theorem}

 In \Cref{thm:rel_error_main} as well as the rest of the appendix, we work with two-layer functions
as described above, where the two layers  are $3n/4$ and $3n/4 + 1$. It is easy to verify that the definition of two-layer functions could be altered to use the two layers $\alpha n$ and $\alpha n + 1$, for any constant $\alpha\in  (1/2, 1)$, and
that \Cref{thm:rel_error_main} would still go through with $N = \Theta({n \choose \alpha n})$.
To see that \Cref{thm:rel_error_main} implies \Cref{thm:relativeerror}, we first note that for any choice of the parameter $N \leq {n \choose 3n/4}$, there exists a positive integer $k \leq n$ such that $\frakN = \Theta\left({k \choose 3k/4}\right)$. The desired $\tilde{\Omega}(\log(\frakN)^{1-c})$
lower bound for relative-error testing of functions with sparsity $\Theta(\frakN)$ can then be obtained from a
routine reduction to \Cref{thm:rel_error_main} (with $n$ set to $k$) by embedding in a suitable subcube of $\zo^n$
using functions $f : \zo^n \to \zo$ of the form
$f(x_1, \ldots, x_n) = (x_{k+1} \land \ldots \land x_n) \land f'
(x_1, \ldots , x_k)$.
Moreover, we can replace $3/4$ by any constant $\alpha \in (1/2, 1)$. Choosing
$\alpha$ to be sufficiently close to $1/2$ extends the lower bound to any $N \leq 2^{\alpha n}$
for any constant $\alpha < 1$.

\subsubsection{Multiplexer Trees and Maps}

%

Our construction is similar to the Multiplexer we used in \Cref{sec:MTM}, we will reuse most of the notation laid out in that section and spell out the differences that we hinted at in \Cref{sec:overview-rel-error}. 
To build a $2\ell$-level \emph{multiplexer tree} $M$, we again build a complete tree of $2\ell$ levels, with the root at level $0$ and leaves at level $2\ell$ and where nodes on level $j<2\ell$ have $N^{(j+1 \text{ mod } 2)}$ children. In particular, there are now $\left(N^{(0)} \cdot N^{(1)}\right)^{\ell}$ leaves in total.

We refer to the root of the tree by the empty tuple $\eps$ and each node at level $j\in [2\ell]$ by a tuple $u=(u_1,\ldots,u_j)\in L^{(j)}$,
with the parent node of $u$ being $\parent(u)=(u_1,\ldots,u_{j-1})\in L^{(j-1)}$.

To finish building the multiplexer tree $M$, we associate each odd-level edge $e$ with a size-$n$ term $T_e\in \frak T'$, and each even edge $e$ with a size-$n$ clause
  $C_e\in \frak C'$.
Formally, a $(2\ell)$-level multiplexer tree is a map $M$ from edges to $\frak T'\cup \frak C'$, such that $M(e)$ is the term $T_e$ of $e$ if it is an odd-level edge and the clause $C_e$ of $e$ if it is an even-level edge.

Every $(2\ell)$-level multiplexer tree $M$ 
  defines a \emph{multiplexer map}
$$\Gamma_{M}:\{0,1\}^n\rightarrow L^{(2\ell)}\cup \{0^*,1^*\},$$
which maps every $x\in \{0,1\}^n$ to either a leaf $u\in L^{(2\ell)}$ of the  tree or one of the two special labels $\{0^*,1^*\}$. 

We reuse the definition of \emph{unique activations} and \emph{unique activation path} we used in \Cref{sec:MTM}.



Using these definitions, we can define the multiplexer map $\Gamma_M$.
For each $x\in \{0,1\}^n$, let $u^0\cdots u^k$ be its unique activation path in the tree.
We set $\Gamma_M(x)=u^{{k}}$ if $u^{k}\in L^{(2\ell)}$ is a leaf; otherwise $\Gamma_M(x)$ is set to $0^*$ or $1^*$ as in \Cref{sec:MTM}.


Before using it to define multilevel Talagrand functions, we record the following simple lemma: 

\begin{lemma}\label{lem:verysimple_reldist}
Let $M$ be a $(2\ell)$-level multiplexer tree and $\Gamma_M$ be the multiplexer map it defines. 
Given any $x\in \{0,1\}^n$ and $i\in [n]$ with $x_i=0$, we have
\begin{itemize}
\item If $\Gamma_M(x)=u\in L^{(2\ell)}$, then $\Gamma_M(x^{\{i\}})$ is either $u$ or $1^*$.
\item If $\Gamma_M(x)=1^*$, then $\Gamma_M(x^{\{i\}})=1^*$.
\end{itemize}
\end{lemma}
The proof is identical to that of \Cref{lem:verysimple}
\subsubsection{Sandwiched-Multilevel Talagrand Functions}


Let $M$ be a $(2\ell)$-level multiplexer tree and $H=(h_u)$ be a tuple of  functions  
  $h_u:\{0,1\}^n\rightarrow \{0,1\}$, one for each leaf $u\in L^{(2\ell)}$ of the tree. 
(So $H$ consists of $(N^{(0)} \cdot N^{(1)})^{\ell}$ functions.)
Together they define the following sandwiched \emph{$(2\ell)$-level Talagrand function}  $f_{M,H}:\{0,1\}^n\rightarrow \{0,1\}$.
For each string $x\in \{0,1\}^n$, we set $f_{M,H}(x)=0$ if $|x|< (3n/4)$; $f_{M,H}(x)=1$ if $|x| > (3n/4)+1$; and 
$$
f_{M,H}(x)=\begin{cases} 0 & \text{if $\Gamma_M(x)=0^*$}\vspace{0.02cm}\\[0.3ex]
1 & \text{if $\Gamma_M(x)=1^*$}\vspace{0.02cm}\\[0.3ex]
h_u(x) & \text{if $\Gamma_M(x)=u\in L^{(2\ell)} $}
\end{cases},
$$
if $x$ is in sandwich layers.

\subsubsection{Distributions $\Dyes$ and $\Dno$}\label{sec:dist_reldist}


We describe the two distributions $\Dyes$ and $\Dno$ over $(2\ell)$-level two-layer functions $f_{M,H}$ that will be used in our lower bound proofs in the appendix.

To draw $\ff\sim \Dyes$, we 
  first draw a multiplexer tree $\bM$ and a tuple of functions $\bH$ as follows: 
\begin{flushleft}\begin{enumerate}
\item We draw $\bM\sim \calM$ as follows: Start with a complete tree of height $(2\ell)$, where nodes on even levels have $N^{(1)}$ children and nodes on odd levels $N^{(0)}$.  
We then draw a term $\bT_e\sim \frak T'$ for each odd-level edge $e$ (i.e., set $M(e)=\bT_e$) and draw   a clause $\bC_e\sim \frak C'$ for each even-level edge (i.e., set $M(e)=\bC_e$), both independently and uniformly at random. 
\item We draw $\bH=(\bh_u)\sim \Hyes$ as follows: For each leaf $u$, $\bh_u$ is set to be the constant-$0$ function with probability $1/2$ and the constant-$1$ function with probability $1/2$, independently. 
\end{enumerate}\end{flushleft} 
Given $\bM\sim\calM$ and $\bH\sim \Hyes$, $\bf$ is set to be the sandwiched $(2\ell)$-level Talagrand function $\bf=f_{\bM,\bH}$.

To draw $\bf\sim \Dno$, we draw   $\bM\sim \calM$ in the same way as in $\Dyes$. 
On the other hand, the tuple of functions $\bH$ is drawn as follows:
\begin{flushleft}\begin{itemize}
\item[$2'$.] We draw $\bH\sim \Hno$ as follows: First we draw a ``\emph{secret variable}'' $\bs\sim [n]$ uniformly at random.
For each leaf $u$,  $\bh_u$ is set to 
  the dictator function $\bh_u(x)=x_{\bs}$ with probability $1/2$ and set to be
  the anti-dictatorship function $\bh_u(x)=\overline{x_{\bs}}$ with probability $1/2$, independently. 
\end{itemize}\end{flushleft}
Given $\bM\sim \calM$ and $\bH\sim \Hno$, $\bf$ is set to be the sandwiched $(2\ell)$-level Talagrand function $\bf=f_{\bM,\bH}$.

We prove two lemmas about $\Dyes$ and $\Dno$, respectively. 
\Cref{lem:monotone_reldist} shows that every function in the support of $\Dyes$ is monotone; \Cref{lem:farmonotone_reldist} shows that $\bf\sim \Dno$ is $\Omega(1)$-far from unate with probability $\Omega(1)$. We note that both hidden constants are exponentially small in $\ell$ {(as with the standard model, 
  this is the obstacle for the current
  construction to obtain an $\tilde{\Omega}(\log(N))$ lower bound)}.

\begin{lemma}\label{lem:monotone_reldist}
Every function in the support of $\Dyes$ is monotone (and thus unate).
\end{lemma}
The proof is similar to that of \Cref{lem:monotone} using \Cref{lem:verysimple_reldist} instead of \Cref{lem:verysimple}. 

\begin{lemma}\label{lem:farmonotone_reldist}
A function $\bf\sim \Dno$ satisfies $\reldist(f, \textsf{unate}) = \Omega(1)$ (and thus $\reldist(f, \textsf{monotone}) = \Omega(1)$) with probability at least $\Omega(1)$.
\end{lemma}
\begin{proof}
Fix an $s\in [n]$. 
We write $\Hno^s$ to denote this distribution of $\bH$ conditioning on $\bs=s$, i.e., each $\bh_u$ is $x_s$ with probability $1/2$ and $\overline{x_s}$ with probability $1/2$. 
It suffices to show that $\bf= f_{\bM,\bH}$ with $\bM\sim\calM$ and $\bH\sim \Hno^s$  has distance $\Omega(1)$ to unateness with probability $\Omega(1)$.

Given $\bM\sim \calM$ and $\bH\sim \Hno^s$, we write $\bX^{(0)}$ to denote the sets of edges $(x,x^\ast)$ in $\{0,1\}^n$ 
such that the following three conditions holds:
\begin{enumerate}
  \item $x_{s}=0$, $x^*=x^{\{s\}}$ and $x$ satisfies $3n/4= |x|$; 
    \item $\Gamma_{\bM}(x) = \Gamma_{\bM}(x^*) = \bu$ for some leaf $\bu\in L^{(2\ell)}$; and 
    \item $\bh_{\bu}(x)$ is the anti-dictatorship function $\overline{x_s}$.
\end{enumerate}
We define $\bX^{(1)}$ similarly, but instead require $\bh_{\bu}(x)$ to be the dictator function ${x_s}$.

Clearly, all strings in edges of $\bX^{+}$ and $\bX^{-}$ are distinct and are along coordinate $s$. Recall that any function in the support of $\Dno$ is a two-layer function and thus has $|f^{-1}(1)|=\Omega(\frakN)$. Hence, by \Cref{lem:dist_to_unate}, it suffices to show that $\min(|\bX^{(0)}|, |\bX^{(1)}|)\ge \Omega(\frakN)$ with probability $\Omega(1)$.

Given that the number of edges that satisfy the first condition is $\Omega(\frakN)$, by linearity of expectation and Markov's inequality,
  it suffices to show that for each 
  edge $(x,x^*)$ satisfying the first condition and $b \in \zo$ we have 
$$\Prob_{\bM\sim \calM,\bH\sim \Hno^s}\big[(x,x^*)\in \bX^{(b)}\big ] = \Omega(1).$$
Fix $b=0$ (the case where $b=1$ is symmetric). Note that the second condition is about $\bM\sim \calM$ and the third condition, conditioning on the second condition, is only about $\bH\sim \Hno^s$ and always holds with probability $1/2$.
So below we show that the second condition holds with probability $\Omega(1)$ when $\bM\sim \calM$.

We partition the above event into $(N^{(1)}\cdot N^{(0)})^{\ell}$ disjoint sub-events, indexed by leaves $u\in L^{(2\ell)}$:
\begin{equation*} 
\sum_{u\in L^{(2\ell)}}\Pr_{\bM\sim \calM} \big[\Gamma_{\bM}(x)=\Gamma_{\bM}(x^*)=u\big].
\end{equation*}
For each $u\in L^{(2\ell)}$, letting $u^0\cdots u^{2\ell}$ denote the path from the root $u^0$ to $u=u^{2\ell}$, the sub-event of $u$ above corresponds to the following $2\ell$ independent conditions: \begin{itemize}
    \item For each $j\in [0:2\ell-1]$, edge $(u^j,u^{j+1})$ is uniquely activated by both $x$ and $x^*$.
\end{itemize}
In particular, fix any even $j$, the probability is at least
$$
\left(\frac{|x|}{n}\right)^{n}
\left(1-\left(\frac{|x^*|}{n}\right)^{n}\right )^{N^{(1)}-1},$$
where the first factor is the probability of the term $\bT_e\sim \frak T'$, where $e=(u^j,u^{j+1})$, is satisfied by $x$ (which implies that it is satisfied by $x^*$ as well); the second factor is the probability of $\bT_{e'}\sim\frak T'$ of every other edge $e'$ of $u^{j}$ is not satisfied by $x^*$ (which implies that they are also not satisfied by $x$). 
Given that both $x$ and $x^*$ are in {sandwich layers}, the probability is at least 
$$
\left(\frac{3n/4}{n}\right)^{n}\left(1-\left(\frac{3n/4+1}{n}\right)^{n}\right)^{N^{(1)}}
=\left(\frac{3}{4}\right)^n 
\left(1-\left(\frac{3}{4}\right)^n\left(1+\frac{4}{3n}\right)^{n}\right)^{N^{(1)}-1}.
$$
Using $N^{(1)}=(4/3)^n$,  $(1+ 4/3n)^{n}=\Theta(1)$ and $(1-\Theta(1/N^{(1)}))^{N^{(1)}-1}=\Theta(1)$, the probability is $\Omega\left(1/N^{(1)}\right)$.

Now, fix any odd $j$, the probability is at least
$$
\left(1-\frac{|x^*|}{n}\right)^{n}
\left(1-\left(1-\frac{|x|}{n}\right)^{n}\right )^{N^{(0)}-1},$$
where the first factor is the probability of the clause $\bC_e\sim \frak C'$, where $e=(u^j,u^{j+1})$, is falsified by $x^*$ (which implies that it is falsified by $x$ as well); the second factor is the probability of $\bC_{e'}\sim\frak C'$ of every other edge $e'$ of $u^{j}$ is satisfied by $x$ (which implies that they are also not satisfied by $x^*$). 
Given that both $x$ and $x^*$ are in {sandwich layers}, the probability is at least 
$$
\left(\frac{n/4-1}{n}\right)^{n}\left(1-\left(\frac{n/4}{n}\right)^{n}\right)^{N^{(0)-1}}
=\left(\frac{1}{4}\right)^n\left(1-\frac{4}{n}\right)^n 
\left(1-\frac{1}{4^n}\right)^{N^{(0)}-1}.
$$
Using $N^{(0)}=4^n$,  $(1- 4/n)^{n}=\Theta(1)$ and $(1-1/N^{(0)})^{N^{(0)}-1}=\Theta(1)$, the probability is $\Omega\left(1/N^{(0)}\right)$.

As a result, 
$$
\sum_{u\in L^{(2\ell)}}\Pr_{\bM\sim \calM} \big[\Gamma_{\bM}(x)=\Gamma_{\bM}(x^*)=u\big] \ge (N^{(0)} \cdot N^{(1)})^{\ell} \cdot  \left(\Omega\left(\frac{1}{N^{(1)}}\right)\right)^{\ell} \cdot \left(\Omega\left(\frac{1}{N^{(0)}}\right)\right)^{\ell}=\Omega(1)$$ as desired, given that $\ell$ is a constant. 
\end{proof}

\subsubsection{Outcomes of Query Points}\label{sec:information-maintained_reldist}

One reason to work with two-layer functions, is that we can assume that the algorithm only uses black box queries to the function $f$ and never queries the $\SAMP$ oracle.

\begin{claim}[Claim 15 of \cite{chen2025relative}]
If $f:\zo^n \to \zo$ is a two-layer function, then for any constant $\tau > 0$, making q
calls to the $\SAMP(f)$ oracle can be simulated, with success probability at least $1 - \tau$ , by making
$O(q)$ calls to the $\MQ$ oracle.
\end{claim}

Given the above we can now apply Yao's minimax principle and prove our lower bounds for monotoncity and unateness testing  by showing that any deterministic, adaptive algorithm $\ALG$ cannot distinguish $\Dyes$ from $\Dno$ when its query complexity is too low. We can assume $\ALG$ only uses black box query and furthermore that all these queries are on points in sandwich layers (since the functions in the support of $\Dyes, \Dno$ are all two-layer functions).

In our lower bound proofs, we again assume that $\ALG$ has access to a ``stronger'' oracle for the unknown sandwiched $(2\ell)$-level Talagrand function $f_{M,H}$. We assume this stronger oracle returns the same information as the one we described in \Cref{sec:information-maintained}




We keep track of the outcome $\rmO$ and use the same definitions as in \Cref{sec:information-maintained,sec:safeoutcomes}. Note that \Cref{fact:simple1,fact:simple2} still hold (but now for the outcome of a sandwiched $(2\ell)$-level Talagrand function) and so does \Cref{lem:hardtotell}.

\begin{fact}
\label{fact:basic-facts-about-coordinate-sets-rel-error}
Let $\rmO=(Q,P,R,\rho)$ be the outcome of some sandwiched $(2\ell)$-level Talagrand function on $Q$. Then 
\begin{enumerate}
\item For any node $u$ with $P_u\ne \emptyset$, we have $$A_{u,0}\cap A_{u,1}=\emptyset\quad\text{and}\quad 
\big|A_{u,0}| \leq n/4 \text{ and } \big|A_{u,1}\big|\le (3n/4)+1.
$$
\item For any nodes $u,v$ such that $u$ is an ancestor of $v$ and $P_u$ and $P_v$ are nonempty, we have 
$$A_{u,0}\subseteq A_{v,0}\quad\text{and}\quad
A_{u,1}\subseteq A_{v,1}.$$
\end{enumerate}
\end{fact}

\subsection{Lower Bounds for Relative-Error Adaptive Monotonicity and Unateness Testing} 




Let $\Dyes$ and $\Dno$ be the two distributions over sandwiched $2\ell$-level Talagrand functions described in \Cref{sec:dist_reldist}.
Let $q$ be the following parameter:
\begin{equation}\label{eq:setq_rel_dists}
{q={n^{1-\frac{1}{2\ell+1}}/\log(n)}}
\end{equation} 
We prove that no $q$-query, deterministic algorithm $\ALG$ can distinguish $\Dyes$ from $\Dno$ under the  stronger oracle described in \Cref{sec:information-maintained_reldist}. To this end, we view $\ALG$ as a depth-$q$ tree as in \Cref{sec:mainlowerbound1}. 

As mentioned before, $\ALG$ only needs to work on functions $f$ in the support of $\Dyes$ and $\Dno$.
For these functions, we always have  $f(x)=1$ if $|x|>3n/4+1$ and $f(x)=0$ if $|x|<3n/4$. Hence we may assume without loss of generality that every query $x\in \{0,1\}^n$ made by $\ALG$ lies in {sandwich layers}, as otherwise $\ALG$ already knows the value of $f(x)$. We also assume the algorithm doesn't use any queries to the $\SAMP$ oracle and only makes $\MQ$ queries. 

Looking ahead, \Cref{thm:rel_error_main} follows from two main  lemmas, \Cref{lem:main1_reldist,lem:main2_reldist}, combined with \Cref{lem:hardtotell} for safe outcomes proved in \Cref{sec:safeoutcomes}. 
Both of them are based on the following adapted notion of \emph{good} outcomes:

\begin{definition}\label{def:good_leaves condition_rel_dist}
Let $\rmO=(Q,P,R,\rho)$ be the outcome of some {$2\ell$-level two-layer Talagrand} function on a query set $Q$.
We say $\rmO$ is a {\emph{good outcome}} if it satisfies the following conditions:
\begin{flushleft}\begin{enumerate}
\item For every odd-level node $u$ with $P_u\ne \emptyset$, we have 
$$
\big|A_{u,1}\big| \geq \frac{3n}{4}-\big|P_{u}\big|\cdot 100\log n.$$
\item For every even-level non-root node $u$ with  $P_u\ne \emptyset$, we have 
$$
\big|A_{u,0}\big| \geq \frac{n}{4}-\big|P_{u}\big|\cdot 100\log n.$$
\item For every leaf $u$ such that  $P_{u}\ne \emptyset$, we have $\rho_{u}(x)=\rho_{u}(y)$ for all $x,y\in P_{u}$. (Note that this is the same condition as in the definition of safe outcomes.)
\end{enumerate}\end{flushleft}
\end{definition}

\Cref{lem:main1_reldist} shows that every good outcome must be safe as well:
\begin{lemma}\label{lem:main1_reldist}
Every good outcome $\rmO=(Q,P,R,\rho)$ with $|Q|\le q$ is also safe. 
\end{lemma}

We define $\Lyes$ and $\Lno$ in an analogous way as we did in \Cref{sec:mainlowerbound1}. Consider the following adaptation of \Cref{lem:main2}:

\begin{restatable}{lemma}{maintworeldist}\label{lem:main2_reldist}
We have 
$$
\Pr_{\brmO\sim \Lyes}\big[\brmO\ \text{is good}\big]\ge 1-o_n(1).
$$
\end{restatable}

\Cref{thm:rel_error_main} follows immediately from the above. 
\begin{proof}[Proof of \Cref{thm:rel_error_main}]
    The proof follows similar to that \Cref{theo:main100} using \Cref{lem:main1_reldist,lem:main2_reldist} instead of \Cref{lem:main1,lem:main2}.
\end{proof}

In the rest of the section, we prove \Cref{lem:main1_reldist} in \Cref{sec:main1_reldist} and  \Cref{lem:main2_reldist} in \Cref{sec:main2_reldist}.


\subsubsection{Proof of \Cref{lem:main1_reldist} }\label{sec:main1_reldist}



Let $\rmO=(Q,P,R,\rho)$ be a good outcome with $|Q|\le q$.
We start with two bounds on  $|A_{u,1}|$ and $|A_{u,0}|$: 
\begin{claim}
\label{claim:1-good-leaf-size-lower-bound_reldist}
For any even-level node $u$ other than the root with $P_u\ne \emptyset$, letting $v=\parent(u)$, we have
$$
\big|A_{u,1}\big| \ge \frac{3n}{4} -  \min\left(\big|P_{u}\big|^2, \big|P_{v}\big|\right)
\cdot 100\log n.
$$
\end{claim}
\begin{proof}
First, by \Cref{fact:simple2} we have $P_u\subseteq P_v$ so $P_v\ne \emptyset$ as well; by \Cref{fact:basic-facts-about-coordinate-sets-rel-error} we have $A_{v,1}\subseteq A_{u,1}$.
Then by the definition of good outcomes (and that $v$ is an odd-level node with $P_v\ne \emptyset$), we have 
    $$\big|A_{u,1}\big| \ge \big|A_{v,1}\big| \ge  \frac{3n}{4} - \big|P_{v}\big|\cdot 100\log n.$$
    
    On the other hand, we also know that for any two strings $x,y\in P_{u}$, we have 
    $$
    \big|\{j\in [n] : x_j = y_j = 0\}\big| \geq \big|A_{u, 0}\big|\geq \frac{n}{4} -  \big|P_{u}\big|\cdot 100 \log n,
    $$
    where the second inequality used the definition of good outcomes (and that $u$ is an even-level node other than the root).
    Given that all points are in {sandwich layers}, we have  
    \begin{align*}
    \big|\{j\in [n] : x_j = 1, y_j = 0\}\big|=\left(n-|y|\right)- \big|\{j\in [n] : x_j = y_j = 0\}\big|\leq \big|P_{u}\big|\cdot 100 \log n.
    \end{align*} 
 As a result, we have 
\begin{align*}
    \big|A_{u, 1}\big| &\geq  |x| - \sum_{y\in P_{u}\setminus \{x\}}\big|\{j: x_j = 1, y_j = 0\}\big|\\ 
    &\geq \frac{3n}{4}-\left(\big|P_u\big|-1\right)\left(\big|P_{u}\big|\cdot 100 \log n\right)\\[0.8ex] &\ge \frac{3n}{4}-\big|P_u\big|^2\cdot 100\log n,
\end{align*}
where we used $|P_u|\ge 1$.
Combining the two inequalities for $|A_{u,1}|$ gives the desired claim. 
\end{proof}
The following claim for odd-level nodes can be proved similarly:
\begin{claim}
\label{claim:0-good-leaf-size-lower-bound_reldist}
For any odd-level node $u$ at level $k\ge 3$ with $P_u\ne \emptyset$, letting $v=\parent(u)$, we have
$$
\big|A_{u,0}\big| \ge \frac{n}{4} - \min\left(\big|P_{u}\big|^2, \big|P_{v}\big|\right)\cdot 100\log n.
$$
\end{claim}


We now let $K':=200\log n$ in the rest of this subsection. 

Recall that for each leaf $u$, the dangerous set $D_u$ at $u$ is the set of coordinates $i\in [n]$ such that points in $P_u$ don't agree on and $B_u$ is the union of dangerous sets $D_w$ over all leaves $w$ in the subtree rooted at $u$. 
So $B_u$ is the same as $D_u$ if $u$ is a leaf, and $B_\eps$ at the root is exactly the union of $D_w$ over all leaves $w$, which we want to bound in size by $o(n)$.
We also have for each internal node $u$ at level $k$ that $B_u=\cup_{a\in \left[N^{(k+1 \text{ mod } 2)}\right]} B_{u\circ a}.$


As a corollary of \Cref{claim:1-good-leaf-size-lower-bound_reldist,claim:0-good-leaf-size-lower-bound_reldist}, we have the following inequality for $|B_u|$:


\begin{corollary}\label{Bij Corollary_rel_dist} 
    For each node $u$ at level $k\ge 2$,  letting $v=\parent(u)$, we have
    $|B_{u}| \leq |P_{v}\big|\cdot K'$. 
\end{corollary}
\begin{proof}
The proof is similar to that of \Cref{Bij Corollary} but using $K'=200 \log n$ and the definition of good outcome of \Cref{def:good_leaves condition_rel_dist} which says, when $u$ is at even level, that
$$
|A_{u,0}|\ge \frac{n}{4}-\big|P_u\big|\cdot 100\log n\quad\text{and}\quad
|A_{v,1}|\ge \frac{3n}{4}-\big|P_v\big| \cdot 100\log n,
$$
and says that
$$
|A_{u,1}|\ge \frac{3n}{4}-\big|P_u\big|\cdot 100\log n\quad\text{and}\quad
|A_{v,0}|\ge \frac{n}{4}-\big|P_v\big| \cdot 100\log n,
$$
when $u$ is at odd level. 
\end{proof}

\begin{corollary}
\label{Claim Bij at last layer rel error}
Let $u$ be a node at level $k$ where $k \not \in  \{0, 2\ell\}$ then;  
    $$\big|B_{u}\big| \leq \sum_{a\in \left[N^{(k+1 \text{ mod } 2)}\right]} \min\Big(\big| P_{u\circ a}\big|^2, \big| P_{u}\big|\Big)\cdot K'.$$
\end{corollary} 
\begin{proof}
Using $B_u=\cup_{a\in \left[N^{(k+1 \text{ mod } 2)}\right]} B_{u\circ a}$,  
we have 
\begin{align*}
\big|B_{u}\big|
    \leq
\sum_{a\in \left[N^{(k+1 \text{ mod } 2)}\right]} 
    \big|B_{u\circ a}\big|.
\end{align*}
For each $a\in [N]$, if $P_{u\circ a}=\emptyset$, then $B_{u\circ a}=\emptyset$ because every dangerous set in the subtree rooted at $u\circ a$ is empty.
Combining this with \Cref{fact: Bij}, we have 
$$
\big|B_u\big|\le \sum_{a\in [N]:P_{u\circ a}\ne \emptyset}\Big(n-\big|A_{u\circ a,0}\big|-\big|A_{u\circ a,1}\big|\Big).
$$

For each $a\in [N]$ with $P_{u\circ a}\ne \emptyset$, it follows by combining \Cref{def:good_leaves condition_rel_dist} and \Cref{claim:1-good-leaf-size-lower-bound_reldist,claim:0-good-leaf-size-lower-bound_reldist} that :
\begin{enumerate}
    \item If $u$ is at an odd level, $|A_{u\circ a,0}|$ is at least $(n/4)-\min(|P_{u\circ a}|^2,|P_u|) \cdot 100\log n$ while $|A_{u\circ a,1}|$ is at least $(3n/4)- 100|P_{u \circ a}|\log n$.
    \item If $u$ is at an even level, $|A_{u\circ a,1}|$ is at least $(3n/4)-\min(|P_{u\circ a}|^2,|P_{u}|) \cdot 100\log n$ while $|A_{u\circ a,0}|$ is at least $(n/4)- 100|P_{u\circ a}|\log n$.
\end{enumerate}

The statement follows by combining these inequalities and that $K'=200\log n$.
\end{proof}

We are now ready to prove \Cref{lem:main1_reldist}, i.e., $|B_\eps|=o(n)$:
\begin{proof}[Proof of \Cref{lem:main1_reldist}]
Using a similar argument to that of \Cref{lem:main1}, we can show

\begin{align*}
\big|B_\eps\big|\le 
\sum_{a\in [N^{(1)}]} \big|B_a\big|
\le 4^{2\ell-1}\sum_{a\in [N^{(1)}]}
\big|P_a\big|^{1+ \frac{1}{2\ell}}\cdot K' \le O\Big(q^{1+\frac{1}{2\ell}}K'\Big),
\end{align*}

Given that we choose $q={n^{1-\frac{1}{2\ell+1}}/\log(n)}$ and 
 $K'=O(\log n)$ finishes the proof that $|B_\eps|=o(n)$.

\end{proof}

\subsubsection{Proof of \Cref{lem:main2_reldist}}\label{sec:main2_reldist}
Finally we prove \Cref{lem:main2_reldist} which we restate below for convenience.
\maintworeldist*

Given that $\brmO$ is drawn from $\Lyes$ here, it suffices to prove that $\brmO\sim \Lyes$ satisfies the first two conditions of \Cref{def:good_leaves condition_rel_dist} with probability at least $1-o_n(1)$. This is  because the third condition is always satisfied (see the comment below \Cref{def:safeoutcomes}).

To prove \Cref{lem:main2_reldist}, it suffices to prove the following lemma and apply a union bound:

\begin{lemma}\label{lem:usealot_reldist}
Let $\rmO=(Q,P,R,\rho)$ be a good outcome labeled at some internal node of $\ALG$, and let $x\in \{0,1\}^n$ be the next query to make labeled {at this node}.
Conditioning on $\bf\sim \Dyes$ reaching
  this node (or equivalently, conditioning on the outcome of $\bf\sim \Dyes$ on $Q$ is being $\rmO$), the probability of $\bf$ reaching a bad outcome after querying $x$ is $o(1/q)$.
\end{lemma}
\begin{proof}
Let $K'=100\log n$ in this proof. 

First, the only possibilities for the updated outcome to become bad after querying $x$ are (note that these events below are only necessary but not sufficient for the updated outcome to be bad):
\begin{flushleft}\begin{enumerate}
\item The query point $x$ is added to some $P_u$ which was empty in $\rmO$ for some odd-level node $u$ and the new $|A_{u,1}|$ becomes lower than $(3n/4)-K'.$ This cannot happen because the new $|A_{u,1}|$ is just $|x|$ and is at least $(3n/4)$ because $x$ is in the {sandwich layers}\footnote{Recall that we can assume without loss of generality that $\ALG$ only queries points in sandwich layers.};
\item The query point $x$ is added to some $P_u$ which was empty in $\rmO$ for some even-level, non-root node $u$ and the new $|A_{u,0}|$ becomes lower than $(n/4)-K'$. This again cannot happen. 
\item The query point $x$ is added to some $P_u$ which was not empty in $\rmO$ for some odd-level node $u$ and the new $|A_{u,1}|$ goes down for more than $K'$. For this to happen, it must be the case that the number of $i\in A_{u,1}$ with $x_i=0$ is at least $K'$.
\item The query point $x$ is added to some $P_u$ which was not empty in $\rmO$ for some even-level, non-root node $u$ and the new $|A_{u,0}|$ goes down for more than $K'$.
For this to happen, it must be the case that the number of $i\in A_{u,0}$ with $x_i=1$ is at least $K'$.
\end{enumerate}\end{flushleft}
We show below that for any odd-level node $u$ such that 
\begin{enumerate}
\item $P_u\ne \emptyset$ in $\rmO$; and  
\item the number of $i\in A_{u,1}$ satisfying $x_i=0$ is at least $K'$,
\end{enumerate}
the probability of $x$ being added to $P_u$ when $\bf\sim \Dyes$ conditioning on reaching $\rmO$ is $o(1/q^2)$. 

The same can be proved, with similar arguments, for even-level nodes (and regarding $A_{u,0}$).

Assuming these, the lemma follows because the number of nonempty $P_u$ in $\rmO$ can be at most $O(\ell |Q|)=O(q)$ by \Cref{fact:simple2} given that $|Q|\le q$ and $\ell$ is a constant.

To this end, fix any odd-level $u$ such that $P_u$ is nonempty and we write $\Delta$ to denote
$$
\Delta:=\big\{i\in A_{u,1}:x_i=0\big\},
$$
with $|\Delta|\ge K'$.
We show that when $\bf\sim \Dyes$ conditioning on it reaching $\rmO$, the probability that~$x$ is added to $P_u$ after it is queried is at most $o(1/q^2)$.
For this purpose, recall from \Cref{fact:simple1} that~the characterization of $\bf\sim \Dyes$ reaching $\rmO$ consists of independent conditions, one condition on the term or clause on each edge and one condition on the function at each leaf.
Regarding~the~term~$\bT_e$ (since $u$ is an odd-level node) at 
  $e=(\parent(u),u)$ in $\bM\sim \calM$:
\begin{flushleft}\begin{enumerate}
\item For $\bf\sim \Dyes$ to reach $\rmO$, the term $\bT_e$ at $e$ can be set to a term $T\in \frak T'$ iff (1) $T(y)=1$ for all $y\in P_u$ and (2) $T(y)=0$ for all $y\in R_e$. Let's denote this event $E_1$ for $\bT_e\sim \frak T'$.
\item For $\bf\sim \Dyes$ to not only reach $\rmO$ but also have $x$ added to $P_u$ after it is queried, $\bT_e$ can be set to a term $T\in \frak T'$ iff (1) $T(y)=1$ for all $y\in P_u\cup \{x\}$ and (2) $T(y)=0$ for all $y\in R_e$. Let's denote this event $E_2$ for $\bT_e\sim \frak T'$.
\end{enumerate}\end{flushleft}
With the definition of $E_1$ and $E_2$ above, it suffices to show that
\begin{equation}\label{eq:lastlast_reldist}
\Pr_{\bT\sim \frak 'T} \big[E_2\big]\le 
o\left(\frac{1}{q^2}\right)\cdot \Pr_{\bT\sim \frak T'} \big[E_1 \big].
\end{equation}

Let $A\subseteq [n]$, $\Delta\subseteq A$ with $|\Delta|\ge K'$ and $R\subseteq \{0,1\}^n$ with $|R|\le n/2$. 
Consider $\bT\sim \frak T'$.
Let $E_1^*$ be the event that 
  (1) all variables in $\bT$ come from $A$ and (2) $\bT(y)=0$ for all $y\in R$;
  let $E_2^*$ be the event that (1) all variables in $\bT$ come from $A\setminus \Delta$ and (2) $\bT(y)=0$ for all $y\in R$.

We prove the following claim under this setting, from which \Cref{eq:lastlast_reldist} follows directly:

\begin{claim}\label{claim:reuse rel error}
We have 
$$
\Pr_{\bT\sim \frak T'}\big[E_2^*\big]\le o\left(\frac{1}{n^{5}}\right)\cdot \Pr_{\bT\sim \frak T'} \big[E_1^*\big].
$$
\end{claim}
\begin{proof}We count  ordered tuples $I=(I_1,\ldots,I_{{n}})\in [n]^{{n}}$ in the following two sets.
\begin{flushleft}\begin{itemize}
\item $U$ contains all $I\in [n]^{{n}}$ such that $I_k\in A$ for all $k\in [{n}]$ and for every $z\in R$,
  there exists at least one $k\in [{n}]$ such that $z_{I_k}=0$; and
\item $V$ contains all $I\in [n]^{{n}}$ such that $I_k\in A \setminus \Delta$ for all $k\in [{n}]$ and for every $z\in R$, there exists at least one $k\in [{n}]$ such that $z_{I_k}=0$.
\end{itemize}\end{flushleft}
It suffices to show that $|V|/|U|\le o(1/n^{5})$. 
To upper bound this ratio, let $t= \log n$ and we use $U'$ to denote the subset of $U$ such that $I\in U$ is in $U'$ if and only if $$\Big|\big\{ k \in  [n] : I_k \in \Delta \big\}\Big| = t.$$Now it suffices  to show that $|V|/|U'|=o(1/n^{5})$ given that $U'\subseteq U$. We define a bipartite graph $G$ between
  $U'$ and $V$: $I'\in U'$ and $I\in V$ have an edge if and only if
  $I'_k=I_k$ for every $k\in [{n}]$ with $I'_k\notin \Delta$.
From the construction, it is clear 
  each $I'\in U'$ has degree at most $|A\setminus \Delta|^t$.

To lower bound the degree of an $I\in V$,
  letting points in $R$ be $z^1,\ldots,z^{|R|}$, we can fix a set of $|R|$ (not necessarily distinct) indices $k_1, \ldots, k_{|R|}$  in $[{n}]$ such that every $z^i$ has  $$\big(z^i\big)_{I_{k_i}}=0.$$ Once these indices are fixed, we can pick any of the $t$ remaining ones and map them to $t$ variables in $\Delta$. 
As a result, the degree of each $I\in V$ is at least:
$${{n-|R|} \choose {t} }\cdot |\Delta|^t.
$$
By counting   edges in $G$ in two different ways and using $|A| \leq n$ and $|R|\le {n}/2$, 
we have 
\[ \frac{|U'|}{|V|}\geq {{n - |R| }\choose t} \cdot \Bigg(\frac{|\Delta|}{|A\setminus \Delta|}\Bigg)^t \geq \Bigg(\frac{n/2}{t}\Bigg)^t\cdot \Bigg(\frac{100t}{n}
\Bigg)^t > \omega(n^{5}).\]
This finishes the proof of the claim.
\end{proof}
This finishes the proof of \Cref{lem:usealot_reldist}.\end{proof}